\documentclass[prd,nopacs,preprintnumbers,twocolumn,amsmath,nofootinbib,amssymb]{revtex4}
\usepackage{graphicx,color,dcolumn,booktabs,bm}
\usepackage{longtable,lscape}
\usepackage{txfonts}
\usepackage{threeparttable}
\usepackage{overpic}
\usepackage{epstopdf}
\usepackage{indentfirst}
\usepackage{feynmf}   
\usepackage{slashed}  
\usepackage{cancel}
\usepackage{cases}
\usepackage{color,xcolor}
\usepackage{soul,ulem}
\usepackage{float}
\usepackage{makecell}
\usepackage{multirow}
\usepackage{graphicx,color,dcolumn,booktabs,bm}
\usepackage{epsfig,dsfont,amssymb,amsmath,amsfonts,amsbsy,mathrsfs}

\graphicspath{{Figures/}} %

\usepackage{hyperref}
\hypersetup{colorlinks,citecolor=blue,anchorcolor=red,menucolor=red,linkcolor=red,filecolor=red,runcolor=red,urlcolor=blue,frenchlinks=red}

\makeatletter
\@addtoreset{equation}{section}
\makeatother
\renewcommand{\theequation}{\arabic{section}.\arabic{equation}}
\allowdisplaybreaks

\newcolumntype{I}{!{\vrule width 0.9pt}}

\begin{document}

\title{Transition form factors and  angular distributions of the $\bm{\Lambda_b\to\Lambda(1520)(\to N\bar{K})\ell^+\ell^-}$ decay supported by baryon spectroscopy}
\author{Yu-Shuai Li$^{1,2}$}\email{liysh20@lzu.edu.cn}
\author{Su-Ping Jin$^{3}$}\email{jinsuping@nankai.edu.cn}
\author{Jing Gao$^{3}$}\email{9820210055@nankai.edu.cn}
\author{Xiang Liu$^{1,2,4,5}$}\email{xiangliu@lzu.edu.cn}
\affiliation{$^1$School of Physical Science and Technology, Lanzhou University, Lanzhou 730000, China\\
$^2$Research Center for Hadron and CSR Physics, Lanzhou University and Institute of Modern Physics of CAS, Lanzhou 730000, China\\
$^3$School of Physics, Nankai University, Tianjin 300071, China\\
$^4$Lanzhou Center for Theoretical Physics, Key Laboratory of Theoretical Physics of Gansu Province, and Frontiers Science Center for Rare Isotopes, Lanzhou University, Lanzhou 730000, China\\
$^5$Key Laboratory of Quantum Theory and Applications of MoE, Lanzhou University,
Lanzhou 730000, China}

\begin{abstract}
We calculate the weak transition form factors of the $\Lambda_b\to\Lambda(1520)$ transition, and further calculate the angular distributions of the rare decays $\Lambda_b\to\Lambda(1520)(\to N\bar{K})\ell^{+}\ell^{-}$ ($N\bar{K}=\{pK^-,n\bar{K}^0\}$) with unpolarized $\Lambda_b$ and massive leptons. The form factors are calculated by the three-body light-front quark model with the support of numerical wave functions of $\Lambda_b$ and $\Lambda(1520)$ from solving the semirelativistic potential model associated with the Gaussian expansion method. By fitting the mass spectrum of the observed single bottom and charmed baryons, the parameters of the potential model are fixed, so this strategy can avoid the uncertainties arising from the choice of a simple harmonic oscillator wave function of the baryons. With more data accumulated in the LHCb experiment, our result can help for exploring the  $\Lambda_b\to\Lambda(1520)\ell^+\ell^-$ decay and deepen our understanding on the $b\to s\ell^+\ell^-$ processes.
\end{abstract}

\maketitle

\section{introduction}
\label{sec1}

The flavor-changing neutral-current (FCNC) processes, including the high-profile $b\to s\ell^+\ell^-$ process, can play a crucial role in indirect searches for  physics beyond the Standard Model (SM). These transitions are forbidden at the tree level and can only operate through loop diagrams in the SM, and are therefore highly sensitive to potential new physics (NP) effects, such as the much-discussed $R_{D^{(*)}}=\mathcal{B}(B\to D^{(*)}\tau\nu_\tau)/\mathcal{B}(B\to D^{(*)}e(\mu)\nu_{e(\mu)})$~\cite{Belle:2015qfa,LHCb:2015gmp,Belle:2016dyj,Belle:2019rba}. These processes thus provided a unique platform to deepen our understanding of both quantum chromodynamics (QCD) and the dynamics of weak processes, and to help  hunt for NP signs. Therefore, the rare decays of $b\to s$ have attracted the attention of both theorists and experimentalists~\cite{Aliev:2002ww,CDF:2011buy,LHCb:2015tgy,Das:2018iap,LHCb:2018jna,Li:2022tbh,Altmannshofer:2022hfs}.

For example, the rare decay $\Lambda_b\to\Lambda\ell^+\ell^-$ has been theoretically studied  by various approaches, including lattice QCD (LQCD)~\cite{Detmold:2012vy,Detmold:2016pkz}, QCD sum rules \cite{Chen:2001sj}, light-cone sum rule~\cite{Aslam:2008hp,Wang:2008sm,Wang:2009hra,Aliev:2010uy,Wang:2015ndk}, covariant quark model~\cite{Gutsche:2013pp}, nonrelativistic quark model~\cite{Mott:2011cx,Mott:2015zma}, and the Bethe-Salpeter approach~\cite{Liu:2019igt}, etc., and was first measured by the CDF Collaboration~\cite{CDF:2011buy} and later by the LHCb Collaboration~\cite{LHCb:2015tgy,LHCb:2018jna}. In addition to the differential branching ratio, such abundant phenomenologies of various angular distributions have also been studied. Compared with the measured data, the angular distribution of $\Lambda_b\to\Lambda\ell^+\ell^-$ was studied in Refs.~\cite{Boer:2014kda,Yan:2019tgn} with unpolarized $\Lambda_b$ baryon, and with polarized $\Lambda_b$ baryon in Ref.~\cite{Blake:2017une}. Furthermore, the authors studied the $b\to s\mu^+\mu^-$ Wilson coefficients in Ref.~\cite{Blake:2019guk} using the measured full angular distribution of the rare decay $\Lambda_b\to\Lambda(\to p\pi)\mu^{+}\mu^{-}$ by the LHCb Collaboration~\cite{LHCb:2018jna}.

With the previous experiences on the decay to the ground state $\Lambda$, it is therefore worth to further testing the $b\to s\ell^+\ell^-$ transition in the baryon sector decaying to the excited hyperon with quantum number being $J^P=3/2^-$. The form factors of the weak transition were calculated by the quark model~\cite{Mott:2011cx,Mott:2015zma}, LQCD~\cite{Meinel:2020owd,Meinel:2021mdj}, and the heavy quark expansion~\cite{Bordone:2021bop}. The angular analysis was performed  in Ref.~\cite{Descotes-Genon:2019dbw} and Ref.~\cite{Das:2020cpv} for massless and massive leptons, respectively. The authors of Ref.~\cite{Hiller:2021zth} studied the kinematic endpoint relations for $\Lambda_b\to\Lambda(1520)\ell^{+}\ell^{-}$ decays and provided the corresponding angular distributions.
Amhis {\it et al.}~\cite{Amhis:2022vcd} used the dispersive techniques to provide a model-independent parameterization of the form factors of $\Lambda_b\to\Lambda(1520)$ and further investigated the FCNC decay $\Lambda_b\to\Lambda(1520)\ell^{+}\ell^{-}$ with the LQCD data. In addition, Xing {\it et al.} also studied the multibody decay $\Lambda_b\to\Lambda_{J}^{*}(\to pK^{-})J/\psi(\to \ell^{+}\ell^{-})$~\cite{Xing:2022uqu}. In addition, Amhis {\it et al.} studied the angular distributions of $\Lambda_b\to\Lambda(1520)\ell^+\ell^-$ and talked about the potential to identify NP effects~\cite{Amhis:2020phx}. Obviously, the $\Lambda_b\to\Lambda(1520)$ is less studied. Following this line, we further study the $\Lambda_b\to\Lambda(1520)(\to N\bar{K})\ell^+\ell^-$ with the $N\bar{K}=\{pK^-,n\bar{K}^0\}$ process and investigate the corresponding angular observables.

From a theoretical point of view, apart from the consideration of new operators beyond the SM, the calculation of the weak transition form factors is a key issue. In addition, how to solve the three-body system
for the  $\Lambda_b$ baryon and $\Lambda^{*}$ hyperon involved is also a challenge. In previous work on baryon weak decays~\cite{Guo:2005qa,Zhu:2018jet,Zhao:2018zcb,Chua:2018lfa,Chua:2019yqh}, the quark-diquark scheme has been widely adopted as an approximate treatment. Meanwhile, the spatial wave functions of hadrons are often approximated  as  simple harmonic oscillator (SHO) wave functions~\cite{Guo:2005qa,Zhu:2018jet,Zhao:2018zcb,Chua:2018lfa,Chua:2019yqh,Ke:2019smy,Ke:2021pxk}, which makes the results dependent on the relevant parameters. To avoid the correlative uncertainties of the above approximations, in this work we calculate the $\Lambda_b\to\Lambda^{*}$ form factors by the three-body light-front quark model. Moreover, in the realistic calculation, we take the numerical spatial wave functions as input, where the semirelativistic potential model combined with the Gaussian expansion method (GEM)~\cite{Hiyama:2003cu,Yoshida:2015tia,Hiyama:2018ivm,Yang:2019lsg} is adopted. By fitting the mass spectrum of the observed single bottom and charmed baryons, the parameters of the semirelativistic potential model can be fixed. Compared with the SHO wave function approximation, our strategy can avoid the uncertainties arising from the selection of the spatial wave functions of the baryons.

The structure of this paper is as follows.  After the Introduction, we derive the helicity amplitudes of $\Lambda_b\to\Lambda^{*}(\to N\bar{K})\ell^+\ell^-$ ($N\bar{K}=\{pK^-,n\bar{K}^0\}$) processes and define some angular observables with unpolarized $\Lambda_b$ baryons and massive leptons in Sec.~\ref{sec2}. The formulas for the weak transition form factors are derived in the three-body light-front quark model in Sec.~\ref{sec3}. And then, to obtain the spatial wave functions of the involved baryons, the applied semirelativistic potential model and GEM are briefly introduced in Sec.~\ref{sec4}. In Sec.~\ref{sec5}, we present our numerical results, including both the relevant form factors and the physical observables in $\Lambda_b\to\Lambda^{*}(\to N\bar{K})\ell^+\ell^-$ decays. Finally, this paper ends with a short summary in Sec.~\ref{sec6}.

\section{The angular distribution of $\mathbf{\Lambda_b\to\Lambda^{*}(\to N\bar{K})\ell^+\ell^-}$}
\label{sec2}

In this paper, we use a model-independent approach with the effective Hamiltonian\cite{Grinstein:1987vj,Buchalla:1995vs}
\begin{equation}
\mathcal{H}_{\text{eff}}(b\to s\ell^+\ell^-)=-\frac{4G_F}{\sqrt{2}}V_{tb}V_{ts}^\ast\sum_{i=1}^{10}\mathcal{C}_i(\mu)\mathcal{O}_i(\mu)
\end{equation}
to study the $b\to s\ell^+\ell^-$ process, where $G_F=1.16637\times10^{-5}\text{GeV}^{-2}$ is the Fermi coupling constant and $|V_{tb}V_{ts}^*|=0.04088$~\cite{Detmold:2016pkz} is the product of the Cabibbo-Kobayashi-Maskawa  matrix elements. Furthermore, the Wilson coefficients $\mathcal{C}_i(\mu)$ describe the short-distance physics, while the four fermion operators $\mathcal{O}_i(\mu)$ describe the long-distance physics, where $\mathcal{O}_{1,2}$ are the current-current operators, $\mathcal{O}_{3-6}$ are the QCD penguin operators, $\mathcal{O}_{7,8}$ denote the electromagnetic and chromomagnetic penguin operators respectively, and $\mathcal{O}_{9,10}$ stand for the semileptonic operators.

In our calculation, we follow the treatment given in Refs.~\cite{Das:2018iap,Liu:2019igt}, adding the factorable quark-loop contributions from $\mathcal{O}_{1-6}$ and $\mathcal{O}_{8}$ to the effective Wilson coefficients $\mathcal{C}_7^{\text{eff}}$ and $\mathcal{C}_9^{\text{eff}}$. The effective Hamiltonian can be written as
\begin{equation}
\begin{split}
\mathcal{H}_{\text{eff}}(b\to s\ell^+\ell^-)=&-\frac{4G_F}{\sqrt{2}}V_{tb}V_{ts}^*\frac{\alpha_e}{4\pi}
\Bigg{\{}\bar{s}\Bigg{[}\mathcal{C}_9^{\text{eff}}(\mu,q^2)\gamma^\mu P_L\\
&-\frac{2m_b}{q^2}\mathcal{C}_{7}^{\text{eff}}(\mu)i\sigma^{\mu\nu}q_\nu P_R\Bigg{]}
b(\bar{\ell}\gamma_\mu\ell)\\
&+\mathcal{C}_{10}(\mu)(\bar{s}\gamma^{\mu}P_Lb)(\bar{\ell}\gamma_{\mu}\gamma^5\ell)\Bigg{\}},
\label{eq:Hamiltonian}
\end{split}
\end{equation}
where $P_{R(L)}=(1\pm\gamma^5)/2$ and $\sigma^{\mu\nu}=i[\gamma^{\mu},\gamma^{\nu}]/2$. The electromagnetic coupling constant is $\alpha_e=1/137$. For the leading logarithmic approximation, we take $m_b=4.80\ \text{GeV}$~\cite{Yan:2000dc,Azizi:2012vy} and the Wilson coefficients as $\mathcal{C}_7^\text{eff}(m_b)=-0.313$ and $\mathcal{C}_{10}(m_b)=-4.669$ in the calculation~\cite{Yan:2000dc,Li:2004vh,Ahmed:2011sa,Azizi:2012vy}. In addition, the short-distance contributions from the soft-gluon emission and the one-loop contributions of the four-quark operators $\mathcal{O}_{1}$-$\mathcal{O}_{6}$, and the long-distance effects due to the charmonium resonances, $J/\psi$ and $\psi(2S)$ are taken into account, where we adopt the $\mathcal{C}_{9}^{\text{eff}}(\mu,q^2)$ as~\cite{Ali:1994bf,Buras:1994dj,Yan:2000dc}
\begin{equation}
\mathcal{C}_{9}^{\text{eff}}(\mu,q^2)=\mathcal{C}_{9}(\mu)+Y_{\text{pert}}(\hat{s})+Y_{\text{res}}(q^2).
\end{equation}
The $Y_{\text{pert}}$ term can be written as
\begin{equation}
\begin{split}
Y_{\text{pert}}(\hat{s})=&g(\hat{m_c},\hat{s})\mathcal{C}(\mu)\\
&-\frac{1}{2}g(1,\hat{s})(4\mathcal{C}_3(\mu)+4\mathcal{C}_4(\mu)+3\mathcal{C}_5(\mu)+\mathcal{C}_6(\mu))\\
&-\frac{1}{2}g(0,\hat{s})(\mathcal{C}_3(\mu)+3\mathcal{C}_4(\mu))\\
&+\frac{2}{9}(3\mathcal{C}_3(\mu)+\mathcal{C}_4(\mu)+3\mathcal{C}_5(\mu)+\mathcal{C}_6(\mu)),
\end{split}
\end{equation}
where $\hat{m}_{c}=m_c/m_b$, $\hat{s}=q^2/m_{b}^2$, $\mathcal{C}(\mu)=3\mathcal{C}_{1}(\mu)+\mathcal{C}_{2}(\mu)+3\mathcal{C}_{3}(\mu)+\mathcal{C}_{4}(\mu)+3\mathcal{C}_{5}(\mu)+\mathcal{C}_{6}(\mu)$, and~\cite{Yan:2000dc}
\begin{equation}
\begin{split}
g(z,\hat{s})=&-\frac{8}{9}\ln{z}+\frac{8}{27}+\frac{4}{9}x-\frac{2}{9}(2+x)\sqrt{\vert 1-x \vert}\\
&\times\Bigg{\{}
\begin{array}{ll}
\ln\vert\frac{1+\sqrt{1-x}}{1-\sqrt{1-x}}\vert-i\pi &\text{for}\ x\equiv4z^2/\hat{s}<1\\
2\arctan{\frac{1}{\sqrt{x-1}}} &\text{for}\ x\equiv4z^2/\hat{s}>1
\end{array},\\
g(0,\hat{s})=&\frac{8}{27}-\frac{8}{9}\ln{\frac{m_b}{\mu}}-\frac{4}{9}\ln{\hat{s}}+\frac{4}{9}i\pi.
\end{split}
\end{equation}
The Wilson coefficients are used as $\mathcal{C}_{1}(m_b)=-0.248$, $\mathcal{C}_{2}(m_b)=1.107$, $\mathcal{C}_{3}(m_b)=0.011$, $\mathcal{C}_{4}(m_b)=-0.026$, $\mathcal{C}_{5}(m_b)=0.007$, and $\mathcal{C}_{6}(m_b)=-0.031$~\cite{Yan:2000dc}. Besides, $m_{c}=1.4\ \text{GeV}$~\cite{Yan:2000dc}.
The $Y_{\text{res}}$ term can be parametrized by using the Breit-Wigner ansatz (it is a model-dependent treatment, and one can refer to Refs.~\cite{Khodjamirian:2010vf,Khodjamirian:2012rm} for more detailed discussions) as~\cite{Azizi:2012vy}
\begin{equation}
Y_{\text{res}}(q^2)=\frac{3\pi}{\alpha_{e}^{2}}C^{(0)}
\sum_{V_{i}=J/\psi,\psi(2S)}\kappa_{Vi}\frac{\Gamma(V_{i}\to\ell^{+}\ell^{-})m_{V_{i}}}{m_{V_{i}}^2-q^{2}-im_{V_{i}}\Gamma_{V_{i}}},
\end{equation}
where $C^{(0)}=0.362$, $\kappa_{J/\psi}=1$, and $\kappa_{\psi(2S)}=2$. The masses and total widths associated with the relevant charmonium resonances are taken to be $3.096\ \text{GeV}$ and $92.9\ \text{keV}$ for $J/\psi$, and $3.686\ \text{GeV}$ and $294\ \text{keV}$ for $\psi(2S)$~\cite{ParticleDataGroup:2020ssz}. The decay widths are taken as $\Gamma(J/\psi\to\ell^{+}\ell^{-})=5.53\ \text{keV}$ and $\Gamma(\psi(2S)\to\ell^{+}\ell^{-})=2.33\ \text{keV}$~\cite{ParticleDataGroup:2020ssz}.

Since the quarks are confined in hadron, the weak transition matrix element cannot be calculated in the framework of perturbative QCD. They are conventionally parametrized in terms of eight (axial-)vector  and six (pseudo-)tensor type dimensionless form factors~\cite{Leibovich:1997az,Pervin:2005ve,Feldmann:2011xf,Mott:2011cx,Boer:2014kda,Descotes-Genon:2019dbw,Das:2020cpv}. In this work, we adopt the helicity-based form as~\cite{Descotes-Genon:2019dbw,Das:2020cpv}
\begin{widetext}
\begin{equation}
\begin{split}
\langle\Lambda^\ast(k,s_{\Lambda^\ast})|\bar{s}\gamma^{\mu}b|\Lambda_b(p,s_{\Lambda_b})\rangle=
&\bar{u}_\alpha(k,s_{\Lambda^\ast})
\Bigg{\{}p^{\alpha}\Bigg{[}
f_{t}^V(q^2)(m_{\Lambda_b}-m_{\Lambda^\ast})\frac{q^{\mu}}{q^2}\\
&+f_{0}^V(q^2)\frac{m_{\Lambda_b}+m_{\Lambda^\ast}}{s_+}
\left(p^{\mu}+k^{\mu}-(m_{\Lambda_b}^2-m_{\Lambda^\ast}^2)\frac{q^{\mu}}{q^2}\right)\\
&+f_{\bot}^V(q^2)\left(\gamma^{\mu}-\frac{2m_{\Lambda^\ast}}{s_+}p^{\mu}-\frac{2m_{\Lambda_b}}{s_+}k^{\mu}\right)\Bigg{]}\\
&+f_g^V(q^2)\Bigg{[}g^{\alpha\mu}+m_{\Lambda^\ast}\frac{p^\alpha}{s_-}\Bigg{(}\gamma^\mu-\frac{2k^\mu}{m_{\Lambda^\ast}}+\frac{2(m_{\Lambda^\ast}p^\mu+m_{\Lambda_b}k^\mu)}{s_+}\Bigg{)}\Bigg{]}
\Bigg{\}}u(p,s_{\Lambda_b}),
\label{eq:ffs01}
\end{split}
\end{equation}
\begin{equation}
\begin{split}
\langle\Lambda^\ast(k,s_{\Lambda^\ast})|\bar{s}\gamma^{\mu}\gamma^5b|\Lambda_b(p,s_{\Lambda_b})\rangle=
&-\bar{u}_\alpha(k,s_{\Lambda^\ast})\gamma^5
\Bigg{\{}p^\alpha\Bigg{[}
f_{t}^A(q^2)(m_{\Lambda_b}+m_{\Lambda^\ast})\frac{q^{\mu}}{q^2}\\
&+f_{0}^A(q^2)\frac{m_{\Lambda_b}-m_{\Lambda^\ast}}{s_-}
\left(p^{\mu}+k^{\mu}-(m_{\Lambda_b}^2-m_{\Lambda^\ast}^2)\frac{q^{\mu}}{q^2}\right)\\
&+f_{\bot}^A(q^2)\left(\gamma^{\mu}+\frac{2m_{\Lambda^\ast}}{s_-}p^{\mu}-\frac{2m_{\Lambda_b}}{s_-}k^{\mu}\right)
\Bigg{]}\\
&+f_g^A(q^2)\Bigg{[}g^{\alpha\mu}-m_{\Lambda^\ast}\frac{p^\alpha}{s_+}\Bigg{(}\gamma^{\mu}+\frac{2k^{\mu}}{m_{\Lambda^\ast}}-\frac{2(m_{\Lambda^\ast}p^{\mu}-m_{\Lambda_b}k^{\mu})}{s_-}\Bigg{)}\Bigg{]}
\Bigg{\}}u(p,s_{\Lambda_b}),
\label{eq:ffs02}
\end{split}
\end{equation}
\begin{equation}
\begin{split}
\langle\Lambda^\ast(k,s_{\Lambda^\ast})|\bar{s}i\sigma^{\mu\nu}q_{\nu}b|\Lambda_b(p,s_{\Lambda_b})\rangle=
&-\bar{u}_{\alpha}(k,s_{\Lambda^\ast})
\Bigg{\{}p^{\alpha}\Bigg{[}
f_{0}^{T}(q^2)\frac{q^2}{s_+}\left(p^{\mu}+k^{\mu}-(m_{\Lambda_b}^2-m_{\Lambda^\ast}^2)\frac{q^{\mu}}{q^2}\right)\\
&+f_{\bot}^{T}(q^2)(m_{\Lambda_b}+m_{\Lambda^\ast})\left(\gamma^{\mu}-\frac{2m_{\Lambda^\ast}}{s_+}p^{\mu}-\frac{2m_{\Lambda_b}}{s_+}k^{\mu}\right)
\Bigg{]}\\
&+f_g^T(q^2)\Bigg{[}g^{\alpha\mu}+m_{\Lambda^\ast}\frac{p^\alpha}{s_-}\Bigg{(}\gamma^\mu-\frac{2k^\mu}{m_{\Lambda^\ast}}+\frac{2(m_{\Lambda^\ast}p^\mu+m_{\Lambda_b}k^\mu)}{s_+}\Bigg{)}\Bigg{]}
\Bigg{\}}u(p,s_{\Lambda_b}),
\label{eq:ffs03}
\end{split}
\end{equation}
\begin{equation}
\begin{split}
\langle\Lambda^\ast(k,s_{\Lambda^\ast})|\bar{s}i\sigma^{\mu\nu}q_{\nu}\gamma^5b|\Lambda_b(p,s_{\Lambda_b})\rangle=
&-\bar{u}_{\alpha}(k,s_{\Lambda^\ast})\gamma^5
\Bigg{\{}p^{\alpha}\Bigg{[}
f_{0}^{T5}(q^2)\frac{q^2}{s_-}\left(p^{\mu}+k^{\mu}-(m_{\Lambda_b}^2-m_{\Lambda^\ast}^2)\frac{q^{\mu}}{q^2}\right)\\
&+f_{\bot}^{T5}(q^2)(m_{\Lambda_b}-m_{\Lambda^\ast})\left(\gamma^{\mu}+\frac{2m_{\Lambda^\ast}}{s_-}p^{\mu}-\frac{2m_{\Lambda_b}}{s_-}k^{\mu}\right)
\Bigg{]}\\
&+f_g^{T5}(q^2)\Bigg{[}g^{\alpha\mu}-m_{\Lambda^\ast}\frac{p^\alpha}{s_+}\Bigg{(}\gamma^{\mu}+\frac{2k^{\mu}}{m_{\Lambda^\ast}}-\frac{2(m_{\Lambda^\ast}p^{\mu}-m_{\Lambda_b}k^{\mu})}{s_-}\Bigg{)}\Bigg{]}
\Bigg{\}}u(p,s_{\Lambda_b}).
\label{eq:ffs04}
\end{split}
\end{equation}
This form defined above is convenient for calculating the corresponding helicity amplitudes, where $q^2$ is the transferred momentum square and $s_{\pm}=(m_{\Lambda_b}\pm m_{\Lambda^\ast})^2-q^2$.

\subsection{The helicity amplitudes of the $\mathbf{\Lambda_b\to\Lambda^\ast\ell^+\ell^-}$ decay}

To calculate the $\Lambda_b\to\Lambda^\ast\ell^+\ell^-$ process, we define the corresponding helicity amplitudes of the  $\Lambda_b(s_{\Lambda_b})\to\Lambda^\ast(s_{\Lambda^\ast})$ transition as
\begin{equation}
H^{(V,A,T,T5)}(s_{\Lambda_b},s_{\Lambda^\ast},\lambda_W)=
\epsilon_{\mu}^{*}(\lambda_W)\langle\Lambda^\ast(s_{\Lambda^\ast})\vert\bar{s}
\big{\{}
\gamma^{\mu},\gamma^{\mu}\gamma^5,i\sigma^{\mu\nu}q_{\nu},i\sigma^{\mu\nu}q_{\nu}\gamma^5
\big{\}}
b\vert\Lambda_b(s_{\Lambda_b})\rangle,
\label{eq:helicityH}
\end{equation}
where $\epsilon^{\mu}(\lambda_W=t,\pm,0)$ are the polarization vectors of the virtual gauge boson in the $\Lambda_b$ rest frame, $s_{\Lambda_b}$ and $s_{\Lambda^\ast}$ are the polarizations of $\Lambda_b$ and $\Lambda^\ast$, respectively. For the vector current, the complete helicity amplitudes $H^{V}(s_{\Lambda_b},s_{\Lambda^\ast},\lambda_W)$ read~\cite{Descotes-Genon:2019dbw}
\begin{eqnarray}
H^V(s_{\Lambda_b},s_{\Lambda^\ast},t)&=&\epsilon_{\mu}^{*}(t)\langle\Lambda^\ast(k,s_{\Lambda^\ast})|\bar{s}\gamma^{\mu}b|\Lambda_b(p,s_{\Lambda_b})\rangle\nonumber\\
&=&f_t^V(q^2)\frac{m_{\Lambda_b}-m_{\Lambda^\ast}}{\sqrt{q^2}}\bar{u}_{\alpha}(k,s_{\Lambda^\ast})p^{\alpha}u(p,s_{\Lambda_b}),\\
H^V(s_{\Lambda_b},s_{\Lambda^\ast},0)&=&\epsilon_{\mu}^{*}(0)\langle\Lambda^\ast(k,s_{\Lambda^\ast})|\bar{s}\gamma^{\mu}b|\Lambda_b(p,s_{\Lambda_b})\rangle\nonumber\\
&=&2f_0^V(q^2)\frac{m_{\Lambda_b}+m_{\Lambda^\ast}}{s_+}k\cdot\epsilon^*(0)\bar{u}_{\alpha}(k,s_{\Lambda^\ast})p^{\alpha}u(p,s_{\Lambda_b}),\\
H^V(s_{\Lambda_b},s_{\Lambda^\ast},\pm)&=&\epsilon_{\mu}^{*}(\pm)\langle\Lambda^\ast(k,s_{\Lambda^\ast})|\bar{s}\gamma^{\mu}b|\Lambda_b(p,s_{\Lambda_b})\rangle\nonumber\\
&=&\Big{(}f_{\bot}^V(q^2)+f_g^V(q^2)\frac{m_{\Lambda^\ast}}{s_-}\Big{)}\bar{u}_{\alpha}(k,s_{\Lambda^\ast})p^{\alpha}\slashed{\epsilon}^{*}(\pm)u(p,s_{\Lambda_b})\nonumber\\
&&+f_g^V(q^2)\bar{u}_{\alpha}(k,s_{\Lambda^\ast})\epsilon^{*\alpha}(\pm)u(p,s_{\Lambda_b}).
\end{eqnarray}
Analogous expressions for the helicity amplitudes of the axial-vector, tensor, and pseudotensor currents are written as
\begin{eqnarray}
H^A(s_{\Lambda_b},s_{\Lambda^\ast},t)&=&\epsilon_{\mu}^{*}(t)\langle\Lambda^\ast(k,s_{\Lambda^\ast})|\bar{s}\gamma^{\mu}\gamma^5b|\Lambda_b(p,s_{\Lambda_b})\rangle\nonumber\\
&=&-f_t^A(q^2)\frac{m_{\Lambda_b}+m_{\Lambda^\ast}}{\sqrt{q^2}}\bar{u}_{\alpha}(k,s_{\Lambda^\ast})p^{\alpha}\gamma^5u(p,s_{\Lambda_b}),\\
H^A(s_{\Lambda_b},s_{\Lambda^\ast},0)&=&\epsilon_{\mu}^{*}(0)\langle\Lambda^\ast(k,s_{\Lambda^\ast})|\bar{s}\gamma^{\mu}\gamma^5b|\Lambda_b(p,s_{\Lambda_b})\rangle\nonumber\\
&=&-2f_0^A(q^2)\frac{m_{\Lambda_b}-m_{\Lambda^\ast}}{s_-}k\cdot\epsilon^*(0)\bar{u}_{\alpha}(k,s_{\Lambda^\ast})p^{\alpha}\gamma^5u(p,s_{\Lambda_b}),\\
H^A(s_{\Lambda_b},s_{\Lambda^\ast},\pm)&=&\epsilon_{\mu}^{*}(\pm)\langle\Lambda^\ast(k,s_{\Lambda^\ast})|\bar{s}\gamma^{\mu}\gamma^5b|\Lambda_b(p,s_{\Lambda_b})\rangle\nonumber\\
&=&\Big{(}f_{\bot}^A(q^2)-f_g^A(q^2)\frac{m_{\Lambda^\ast}}{s_+}\Big{)}\bar{u}_{\alpha}(k,s_{\Lambda^\ast})p^{\alpha}\slashed{\epsilon}^{*}(\pm)\gamma^5u(p,s_{\Lambda_b})\nonumber\\
&&-f_g^A(q^2)\bar{u}_{\alpha}(k,s_{\Lambda^\ast})\epsilon^{*\alpha}(\pm)\gamma^5u(p,s_{\Lambda_b}),
\end{eqnarray}
\begin{eqnarray}
H^T(s_{\Lambda_b},s_{\Lambda^\ast},0)&=&\epsilon_{\mu}^{*}(0)\langle\Lambda^\ast(k,s_{\Lambda^\ast})|\bar{s}i\sigma^{\mu\nu}q_{\nu}b|\Lambda_b(p,s_{\Lambda_b})\rangle\nonumber\\
&=&-2f_0^T(q^2)\frac{q^2}{s_+}k\cdot\varepsilon^{*}(0)\bar{u}_{\alpha}(k,s_{\Lambda^\ast})p^{\alpha}u(p,s_{\Lambda_b}),\\
H^T(s_{\Lambda_b},s_{\Lambda^\ast},\pm)&=&\epsilon_{\mu}^{*}(\pm)\langle\Lambda^\ast(k,s_{\Lambda^\ast})|\bar{s}i\sigma^{\mu\nu}q_{\nu}b|\Lambda_b(p,s_{\Lambda_b})\rangle\nonumber\\
&=&-\Big{(}f_{\bot}^T(q^2)(m_{\Lambda_b}+m_{\Lambda^\ast})+f_g^T(q^2)\frac{m_{\Lambda^\ast}}{s_-}\Big{)}\bar{u}_{\alpha}(k,s_{\Lambda^\ast})p^{\alpha}\slashed{\epsilon}^{*}(\pm)u(p,s_{\Lambda_b})\nonumber\\
&&-f_g^T(q^2)\bar{u}_{\alpha}(k,s_{\Lambda^\ast})\epsilon^{*\alpha}(\pm)u(p,s_{\Lambda_b}),
\end{eqnarray}
\begin{eqnarray}
H^{T5}(s_{\Lambda_b},s_{\Lambda^\ast},0)&=&\epsilon_{\mu}^{*}(0)\langle\Lambda^\ast(k,s_{\Lambda^\ast})|\bar{s}i\sigma^{\mu\nu}q_{\nu}\gamma^5b|\Lambda_b(p,s_{\Lambda_b})\rangle\nonumber\\
&=&-2f_0^{T5}(q^2)\frac{q^2}{s_-}k\cdot\epsilon^{*}(0)\bar{u}_{\alpha}(k,s_{\Lambda^\ast})p^{\alpha}\gamma^5u(p,s_{\Lambda_b}),\\
H^{T5}(s_{\Lambda_b},s_{\Lambda^\ast},\pm)&=&\epsilon_{\mu}^{*}(\pm)\langle\Lambda^\ast(k,s_{\Lambda^\ast})|\bar{s}i\sigma^{\mu\nu}q_{\nu}\gamma^5b|\Lambda_b(p,s_{\Lambda_b})\rangle\nonumber\\
&=&\Big{(}f_{\bot}^{T5}(q^2)(m_{\Lambda_b}-m_{\Lambda^\ast})-f_g^{T5}(q^2)\frac{m_{\Lambda^\ast}}{s_+}\Big{)}\bar{u}_{\alpha}(k,s_{\Lambda^\ast})p^{\alpha}\slashed{\epsilon}^{*}(\pm)\gamma^5u(p,s_{\Lambda_b})\nonumber\\
&&-f_g^{T5}(q^2)\bar{u}_{\alpha}(k,s_{\Lambda^\ast})\epsilon^{*\alpha}(\pm)\gamma^5u(p,s_{\Lambda_b}),
\end{eqnarray}
respectively. Using the kinematic conventions presented in Appendix~\ref{app02.1}, the nonzero terms for the above helicity amplitudes of the vector, axial-vector, tensor, and pseudotensor currents are~\cite{Descotes-Genon:2019dbw}
\begin{equation}
\begin{split}
H^V(+1/2,+1/2,t)&=H^V(-1/2,-1/2,t)=f_t^V(q^2)\frac{m_{\Lambda_b}-m_{\Lambda^\ast}}{\sqrt{q^2}}\frac{s_+\sqrt{s_-}}{\sqrt{6}m_{\Lambda^\ast}},\\
H^V(+1/2,+1/2,0)&=H^V(-1/2,-1/2,0)=-f_0^V(q^2)\frac{m_{\Lambda_b}+m_{\Lambda^\ast}}{\sqrt{q^2}}\frac{s_-\sqrt{s_+}}{\sqrt{6}m_{\Lambda^\ast}},\\
H^V(+1/2,-1/2,+)&=H^V(-1/2,+1/2,-)=-f_\bot^V(q^2)\frac{s_-\sqrt{s_+}}{\sqrt{3}m_{\Lambda^\ast}},\\
H^V(-1/2,-3/2,+)&=H^V(+1/2,+3/2,-)=f_g^V(q^2)\sqrt{s_+},
\label{eq:nonzeroHV}
\end{split}
\end{equation}
\begin{equation}
\begin{split}
H^A(+1/2,+1/2,t)&=-H^A(-1/2,-1/2,t)=f_t^A(q^2)\frac{m_{\Lambda_b}+m_{\Lambda^\ast}}{\sqrt{q^2}}\frac{s_-\sqrt{s_+}}{\sqrt{6}m_{\Lambda^\ast}},\\
H^A(+1/2,+1/2,0)&=-H^A(-1/2,-1/2,0)=-f_0^A(q^2)\frac{m_{\Lambda_b}-m_{\Lambda^\ast}}{\sqrt{q^2}}\frac{s_+\sqrt{s_-}}{\sqrt{6}m_{\Lambda^\ast}},\\
H^A(+1/2,-1/2,+)&=-H^A(-1/2,+1/2,-)=f_\bot^A(q^2)\frac{s_+\sqrt{s_-}}{\sqrt{3}m_{\Lambda^\ast}},\\
H^A(-1/2,-3/2,+)&=-H^A(+1/2,+3/2,-)=-f_g^A(q^2)\sqrt{s_-},
\label{eq:nonzeroHA}
\end{split}
\end{equation}
\begin{equation}
\begin{split}
H^T(+1/2,+1/2,0)&=H^T(-1/2,-1/2,0)=f_0^T(q^2)\sqrt{q^2}\frac{s_-\sqrt{s_+}}{\sqrt{6}m_{\Lambda^\ast}},\\
H^T(+1/2,-1/2,+)&=H^T(-1/2,+1/2,-)=f_\bot^T(q^2)(m_{\Lambda_b}+m_{\Lambda^\ast})\frac{s_-\sqrt{s_+}}{\sqrt{3}m_{\Lambda^\ast}},\\
H^T(-1/2,-3/2,+)&=H^T(+1/2,+3/2,-)=-f_g^T(q^2)\sqrt{s_+},
\label{eq:nonzeroHT}
\end{split}
\end{equation}
\begin{equation}
\begin{split}
H^{T5}(+1/2,+1/2,0)&=-H^{T5}(-1/2,-1/2,0)=-f_0^{T5}(q^2)\sqrt{q^2}\frac{s_+\sqrt{s_-}}{\sqrt{6}m_{\Lambda^\ast}},\\
H^{T5}(+1/2,-1/2,+)&=-H^{T5}(-1/2,+1/2,-)=f_\bot^{T5}(q^2)(m_{\Lambda_b}-m_{\Lambda^\ast})\frac{s_+\sqrt{s_-}}{\sqrt{3}m_{\Lambda^\ast}},\\
H^{T5}(-1/2,-3/2,+)&=-H^{T5}(+1/2,+3/2,-)=-f_g^{T5}(q^2)\sqrt{s_-},
\label{eq:nonzeroHT5}
\end{split}
\end{equation}
respectively.

Similarly, we define the leptonic helicity amplitudes as
\begin{equation}
\begin{split}
L^{(V,A)}(s_{\ell^-},s_{\ell^+},\lambda_W)&=\bar{\epsilon}^{\mu}(\lambda_W)\langle\ell^-\ell^+\vert\bar{\ell}^-
\big{\{}\gamma_\mu,\gamma_\mu\gamma^5\big{\}}
\ell^+\vert0\rangle\\
&=\bar{\epsilon}^{\mu}(\lambda_W)\bar{u}(p_{\ell},s_{\ell^-})
\big{\{}\gamma_{\mu},\gamma_{\mu}\gamma^5\big{\}}
v(-p_{\ell},s_{\ell^+}),
\label{eq:helicityL}
\end{split}
\end{equation}
where $\bar{\epsilon}^{\mu}(\lambda_W=t,\pm,0)$ are the polarization vectors of the virtual gauge boson in the dilepton rest frame. Using the kinematic conventions presented in Appendix~\ref{app02.2}, the nonzero terms are obtained as~\cite{Das:2020cpv}
\begin{equation}
\begin{split}
L^V(\pm1/2,\pm1/2,0)&=\pm2m_{\ell}\cos\theta_\ell,~~~~~~~~~~~~~~~~~~~~~~
L^V(\pm1/2,\mp1/2,0)=-\sqrt{q^2}\sin\theta_\ell,\\
L^V(+1/2,+1/2,\pm)&=\mp\sqrt{2}m_\ell\sin\theta_\ell,~~~~~~~~~~~~~~~~~~
L^V(-1/2,-1/2,\mp)=\mp\sqrt{2}m_\ell\sin\theta_\ell,\\
L^V(\pm1/2,\mp1/2,\pm)&=\mp\frac{\sqrt{q^2}}{\sqrt{2}}(1+\cos\theta_\ell),~~~~~~~~~~~
L^V(\pm1/2,\mp1/2,\mp)=\mp\frac{\sqrt{q^2}}{\sqrt{2}}(1-\cos\theta_\ell),\\
L^A(\pm1/2,\pm1/2,t)&=-2m_\ell,~~~~~~~~~~~~~~~~~~~~~~~~~~~~~~~
L^A(\pm1/2,\mp1/2,0)=\mp\sin\theta_\ell\sqrt{q^2}\beta_\ell,\\
L^A(\pm1/2,\mp1/2,\pm)&=-\frac{\sqrt{q^2}}{\sqrt{2}}(1+\cos\theta_\ell)\beta_\ell,~~~~~~~
L^A(\pm1/2,\mp1/2,\mp)=-\frac{\sqrt{q^2}}{\sqrt{2}}(1-\cos\theta_\ell)\beta_\ell,
\label{eq:nonzeroL}
\end{split}
\end{equation}
where $\beta_\ell\equiv\sqrt{1-4m_{\ell}^2/q^2}$.

\subsection{The helicity amplitudes of the $\mathbf{\Lambda^*\to N\bar{K}}$ decay}

We use the effective Lagrangian approach to describe the strong decay process $\Lambda^*\to N\bar{K}$. The concerned effective Lagrangian is
\begin{equation}
\mathcal{L}_{\Lambda^{\ast}KN}=g_{\Lambda^{\ast}KN}\bar{N}\gamma_{5}\Lambda^{\ast}_{\alpha}\partial^{\alpha}K,
\end{equation}
where $g_{\Lambda^{\ast}KN}$ is the coupling constant. So the decay amplitude for the $\Lambda^\ast\to N\bar{K}$ process can be expressed as
\begin{equation}
\mathcal{M}_{\Lambda^\ast\to N\bar{K}}(s_{\Lambda^\ast},s_{N})=
g_{\Lambda^{\ast}KN}\bar{u}_N(s_N)\gamma_5u_{\Lambda^\ast,\alpha}(s_{\Lambda^\ast})k_{2}^{\alpha},
\end{equation}
where $k_{2}^{\alpha}$ is the four momentum of the $K$ meson, $u_{\Lambda^\ast,\alpha}$ is the Rarita-Schwinger spinor describing the hyperon $\Lambda^\ast$, while $u_N$ is the Dirac spinor describing the nucleon. The interference terms between matrix elements with different $\Lambda^\ast$ polarizations can be written as
\begin{equation}
\begin{split}
\Gamma_{\Lambda^\ast\to N\bar{K}}\big{(}s_{\Lambda^\ast}^a,s_{\Lambda^\ast}^b\big{)}=&
\frac{\sqrt{r_-r_+}}{16\pi m_{\Lambda^\ast}^3}
\sum_{s_N}
\bigg{[}\mathcal{M}_{\Lambda^\ast\to N\bar{K}}\big{(}s_{\Lambda^\ast}^b,s_N\big{)}\bigg{]}^{\ast}\\
&\times\mathcal{M}_{\Lambda^\ast\to N\bar{K}}\big{(}s_{\Lambda^\ast}^a,s_N\big{)},
\label{eq:helicityLambda}
\end{split}
\end{equation}
where $r_{\pm}=(m_{\Lambda^\ast}\pm m_{N})^2-m_{K}^2$, and then the decay width of $\Lambda^\ast\to N\bar{K}$ can be obtained by
\begin{equation}
\Gamma\big{(}\Lambda^\ast\to N\bar{K}\big{)}=\frac{1}{4}
\sum_{s_{\Lambda^\ast}}
\Gamma_{\Lambda^\ast\to N\bar{K}}\big{(}s_{\Lambda^\ast},s_{\Lambda^\ast}\big{)},
\end{equation}
where the factor 4 comes from averaging over the polarization of $\Lambda^\ast$.

With respect to the forms of Rarita-Schwinger spinors and Dirac spinors presented in Appendix~\ref{app02.3}, we obtain~\cite{Descotes-Genon:2019dbw,Das:2020cpv}
\begin{equation}
\Gamma_{\Lambda^\ast\to N\bar{K}}\big{(}s_{\Lambda^\ast}^a,s_{\Lambda^\ast}^b\big{)}=
\frac{\Gamma\big{(}\Lambda^\ast\to N\bar{K}\big{)}}{4}
\begin{pmatrix}
6\sin^2(\theta_{\Lambda^\ast}) & 2\sqrt{3}e^{-i\phi}\sin(2\theta_{\Lambda^\ast})
& -2\sqrt{3}e^{-2i\phi}\sin^2(\theta_{\Lambda^\ast}) & 0 \\
2\sqrt{3}e^{i\phi}\sin(2\theta_{\Lambda^\ast}) & 3\cos(2\theta_{\Lambda^\ast})+5
& 0 & -2\sqrt{3}e^{-2i\phi}\sin^2(\theta_{\Lambda^\ast}) \\
-2\sqrt{3}e^{2i\phi}\sin^2(\theta_{\Lambda^\ast}) & 0 &
3\cos(2\theta_{\Lambda^\ast})+5 & -2\sqrt{3}e^{-i\phi}\sin(2\theta_{\Lambda^\ast}) \\
0 & -2\sqrt{3}e^{2i\phi}\sin^2(\theta_{\Lambda^\ast}) &
-2\sqrt{3}e^{i\phi}\sin(2\theta_{\Lambda^\ast}) & 6\sin^2(\theta_{\Lambda^\ast}) \\
\end{pmatrix},
\label{eq:nonzeroLambda}
\end{equation}
with rows and columns corresponding to the polarizations of $s_{\Lambda^\ast}^a,s_{\Lambda^\ast}^b=-3/2,-1/2,1/2,3/2$ from top to bottom and from left to right. We emphasize that $\Gamma\big{(}\Lambda^\ast\to N\bar{K}\big{)}=\mathcal{B}_{\Lambda^\ast}\times\Gamma_{\Lambda^\ast}$, where $\mathcal{B}_{\Lambda^\ast}\equiv\mathcal{B}\big{(}\Lambda^\ast\to N\bar{K}\big{)}$ is the corresponding branching ratio and $\Gamma_{\Lambda^\ast}$ is the inclusive decay width of the $\Lambda^\ast$ hyperon.

\subsection{The total amplitudes of $\Lambda_b\to\Lambda^{*}(\to N\bar{K})\ell^{+}\ell^{-}$ process}

The invariant amplitude of $\Lambda_b\to\Lambda^\ast(\to N\bar{K})\ell^+\ell^-$ is~\cite{Yan:2019tgn}
\begin{equation}
\begin{split}
\mathcal{M}(s_{\Lambda_b},s_N,s_{\ell^-},s_{\ell^+})=&
\langle N(s_N)\bar{K}\ell^-(s_{\ell^-})\ell^+(s_{\ell^+})\vert\mathcal{H}_{\text{eff}}\vert\Lambda_b(s_{\Lambda_b})\rangle\\
=&\sum_{s_{\Lambda^\ast}}\frac{i}{k^2-m_{\Lambda^\ast}^2}
\mathcal{M}_{\Lambda^\ast\to N\bar{K}}(s_{\Lambda^\ast},s_{N})
\langle\Lambda^\ast(s_\Lambda^\ast)\ell^-(s_{\ell^-})\ell^+(s_{\ell^+})
\vert\mathcal{H}_{\text{eff}}\vert\Lambda_b(s_{\Lambda_b})\rangle\\
=&\sum_{s_{\Lambda^{*}}}\frac{iN}{2(k^2-m_{\Lambda^{*}}^2)}
\mathcal{M}_{\Lambda^{*}\to N\bar{K}}(s_{\Lambda^{*}},s_{N})g^{\mu\nu}
\langle\Lambda^{*}(s_{\Lambda^{*}})\vert j_{\mu} \vert\Lambda_b(s_{\Lambda_b})\rangle
\langle\ell^{-}(s_{\ell^{-}})\ell^{+}(s_{\ell^{+}})\vert j_{\nu} \vert0\rangle\\
=&\sum_{s_{\Lambda^{*}}}\frac{iN}{2(k^2-m_{\Lambda^{*}}^2)}
\mathcal{M}_{\Lambda^{*}\to N\bar{K}}(s_{\Lambda^{*}},s_{N})
\bigg{[}C_{9}^{\text{eff}}H^{V}(s_{\Lambda_{b}},s_{\Lambda^{*}},t)L^{V}(s_{\ell^{-}},s_{\ell^{+}},t)
-C_{9}^{\text{eff}}H^{A}(s_{\Lambda_{b}},s_{\Lambda^{*}},t)L^{V}(s_{\ell^{-}},s_{\ell^{+}},t)\\
&+C_{10}H^{V}(s_{\Lambda_{b}},s_{\Lambda^{*}},t)L^{A}(s_{\ell^{-}},s_{\ell^{+}},t)
-C_{10}H^{A}(s_{\Lambda_{b}},s_{\Lambda^{*}},t)L^{A}(s_{\ell^{-}},s_{\ell^{+}},t)\\
&-\frac{2m_{b}}{q^2}C_{7}^{\text{eff}}H^{T}(s_{\Lambda_{b}},s_{\Lambda^{*}},t)L^{V}(s_{\ell^{-}},s_{\ell^{+}},t)
-\frac{2m_{b}}{q^2}C_{7}^{\text{eff}}H^{T5}(s_{\Lambda_{b}},s_{\Lambda^{*}},t)L^{V}(s_{\ell^{-}},s_{\ell^{+}},t)\\
&-\sum_{\lambda=\pm,0}\Big{(}
C_{9}^{\text{eff}}H^{V}(s_{\Lambda_{b}},s_{\Lambda^{*}},\lambda)L^{V}(s_{\ell^{-}},s_{\ell^{+}},\lambda)
-C_{9}^{\text{eff}}H^{A}(s_{\Lambda_{b}},s_{\Lambda^{*}},\lambda)L^{V}(s_{\ell^{-}},s_{\ell^{+}},\lambda)\\
&+C_{10}H^{V}(s_{\Lambda_{b}},s_{\Lambda^{*}},\lambda)L^{A}(s_{\ell^{-}},s_{\ell^{+}},\lambda)
-C_{10}H^{A}(s_{\Lambda_{b}},s_{\Lambda^{*}},\lambda)L^{A}(s_{\ell^{-}},s_{\ell^{+}},\lambda)\\
&-\frac{2m_{b}}{q^2}C_{7}^{\text{eff}}H^{T}(s_{\Lambda_{b}},s_{\Lambda^{*}},\lambda)L^{V}(s_{\ell^{-}},s_{\ell^{+}},\lambda)
-\frac{2m_{b}}{q^2}C_{7}^{\text{eff}}H^{T5}(s_{\Lambda_{b}},s_{\Lambda^{*}},\lambda)L^{V}(s_{\ell^{-}},s_{\ell^{+}},\lambda)
\Big{)}\bigg{]},
\end{split}
\end{equation}
where $N\equiv\frac{4G_F}{\sqrt{2}}V_{tb}V_{ts}^{*}\frac{\alpha_e}{4\pi}$, and the factor $1/2$ comes from the definition of $P_{L(R)}$. Besides, the relation $g^{\mu\nu}=\epsilon_{t}^{*\mu}\epsilon_{t}^{\nu}-\sum_{\lambda=\pm,0}\epsilon_{\lambda}^{*\mu}\epsilon_{\lambda}^{\nu}$ is implied.  The helicity amplitudes defined in Eqs.~\eqref{eq:helicityH}, \eqref{eq:helicityL}, \eqref{eq:helicityLambda}, and \eqref{eq:nonzeroLambda} are implied. Finally, with the nonzero helicity amplitudes presented in Eqs.~(\ref{eq:nonzeroHV})-(\ref{eq:nonzeroHT5}), \eqref{eq:nonzeroL} and \eqref{eq:nonzeroLambda}, and the expression of the differential width by considering the narrow-width approximation shown in Eq.~\eqref{appeq:decayrate}, the differential decay width can be obtained.

As analyzed in Refs.~\cite{Descotes-Genon:2019dbw,Das:2020cpv}, the angular distribution for the four-body decay $\Lambda_b\to\Lambda^\ast(\to N\bar{K})\ell^+\ell^-$ can be reduced as
\begin{equation}
\begin{split}
\frac{d^4\Gamma}{dq^2 d\cos\theta_{\Lambda^\ast} d\cos\theta_\ell d\phi}=&\frac{3}{8\pi}\sum_{i}L_{i}(q^2)f_{i}(q^2,\theta_{\ell},\theta_{\Lambda^{*}},\phi)\\
=&\frac{3}{8\pi}\Big{[}
\Big{(}L_{1c}\cos\theta_\ell+L_{1cc}\cos^2\theta_\ell+L_{1ss}\sin^2\theta_\ell\Big{)}\cos^2\theta_{\Lambda^*}\\
&+\Big{(}L_{2c}\cos\theta_\ell+L_{2cc}\cos^2\theta_\ell+L_{2ss}\sin^2\theta_\ell\Big{)}\sin^2\theta_{\Lambda^*}\\
&+\Big{(}L_{3ss}\sin^2\theta_\ell\cos^2\phi+L_{4ss}\sin^2\theta_\ell\sin\phi\cos\phi\Big{)}\sin^2\theta_{\Lambda^*}\\
&+\Big{(}L_{5s}\sin\theta_\ell+L_{5sc}\sin\theta_\ell\cos\theta_\ell\Big{)}\sin\theta_{\Lambda^*}\cos\theta_{\Lambda^*}\cos\phi\\
&+\Big{(}L_{6s}\sin\theta_\ell+L_{6sc}\sin\theta_\ell\cos\theta_\ell\Big{)}\sin\theta_{\Lambda^*}\cos\theta_{\Lambda^*}\sin\phi
\Big{]}.
\label{eq:AbgularDistribution2}
\end{split}
\end{equation}
\end{widetext}
The complete expressions for the series angular coefficients $L_{i}$ can be found in  Appendix G of Ref.~\cite{Das:2020cpv}.

\subsection{Physical observable in the four-body process}

By integrating over the angles in the regions $\theta_\ell\in[0,\pi]$, $\theta_\Lambda\in[0,\pi]$, and $\phi\in[0,2\pi]$, the relevant physical observables are listed as follows:
\begin{enumerate}
\item[{$\bullet$}]
    The differential width is
    \begin{equation}
    \begin{split}
    \frac{d\Gamma}{dq^2}&=\!\int_{-1}^{1}\!d\cos\theta_\ell\!\int_{-1}^{1}\!d\cos\theta_{\Lambda^\ast}\!\int_{0}^{2\pi}\!d\phi
    \frac{d\Gamma}{dq^2 d\cos\theta_{\Lambda^\ast} d\cos\theta_\ell d\phi}\\
    &=\frac{1}{3}(L_{1cc}+2L_{1ss}+2L_{2cc}+4L_{2ss}+2L_{3ss}).
    \label{eq:branchingrate}
    \end{split}
    \end{equation}
\item[{$\bullet$}]
    The lepton-side forward-backward symmetry $A_{FB}^{\ell}$ is
    \begin{equation}
    \begin{split}
    A_{FB}^{\ell}&\!=\!
    \frac{\bigg{(}\!\int_{-1}^{0}\!-\!\int_{0}^{1}\!\bigg{)}d\cos\theta_\ell\!\int_{-1}^{1}\!d\cos\theta_{\Lambda^\ast}\!\int_{0}^{2\pi}\!d\phi
    \frac{d\Gamma}{dq^2d\cos\theta_{\Lambda^\ast} d\cos\theta_\ell d\phi}}
    {\int_{-1}^{1}d\cos\theta_\ell\int_{-1}^{1}d\cos\theta_{\Lambda^\ast}\int_{0}^{2\pi}d\phi
    \frac{d\Gamma}{dq^2 d\cos\theta_{\Lambda^\ast} d\cos\theta_\ell d\phi}}\\
    &=\frac{3}{2}\frac{L_{1c}+2L_{2c}}{L_{1cc}+2L_{1ss}+2L_{2cc}+4L_{2ss}+2L_{3ss}}.
    \end{split}
    \end{equation}
\item[{$\bullet$}]
    The hadron-side forward-backward symmetry $A_{FB}^{\Lambda^\ast}$ and the lepton-hadron-side forward-backward symmetry $A_{FB}^{\ell\Lambda^\ast}$ will undoubtedly disappear, since the decay $\Lambda^\ast\to N\bar{K}$ is a strong process~\cite{Descotes-Genon:2019dbw}. This can be tested in future experiments.
\item[{$\bullet$}]
    The transverse and longitudinal polarization fractions of the dilepton system are defined as~\cite{Descotes-Genon:2019dbw}
    \begin{eqnarray}
    F_T&\!=\!&\frac{2(L_{1cc}+2L_{2cc})}{L_{1cc}+2(L_{1ss}+L_{2cc}+2L_{2ss}+L_{3ss})},\\
    F_L&\!=\!&1\!-\!\frac{2(L_{1cc}+2L_{2cc})}{L_{1cc}+2(L_{1ss}+L_{2cc}+2L_{2ss}+L_{3ss})},
    \end{eqnarray}
    respectively.
\item[{$\bullet$}]
    We also define the normalized angular observables as
    \begin{eqnarray}
    \langle A\rangle_{[q_{\text{min}}^2,q_{\text{max}}^2]}=
    \bigg{[}\int_{q_{\text{min}}^2}^{q_{\text{max}}^2}A[q^2]\frac{d\Gamma}{dq^2}dq^2\bigg{]}\bigg{/}
    \bigg{[}\int_{q_{\text{min}}^2}^{q_{\text{max}}^2}\frac{d\Gamma}{dq^2}dq^2\bigg{]}
    \label{eq:AveragedPhysicalObservable}
    \end{eqnarray}
    with $A[q^2]=A_{FB}^{\ell}[q^2]$, $F_{T}[q^2]$, or $F_{L}[q^2]$.
\end{enumerate}

\section{the light-front quark model for calculating weak transition form factors}
\label{sec3}

In this section, we will calculate the form factors involved in the three-body light-front quark model. First, the vertex function of a baryon $\mathcal{B}$ with spin $J$ and momentum $P$ is~\cite{Cheung:1995ub,Cheng:1996if,Geng:1997ws,Cheng:2004cc,Ke:2019smy,Geng:2020gjh,Geng:2022xpn,Ke:2021pxk}
\begin{equation}
\begin{split}
\vert\mathcal{B}(P,J&,J_z)\rangle=\int\frac{d^3\tilde{p}_{1}}{2(2\pi)^3}\frac{d^3\tilde{p}_{2}}{2(2\pi)^3}\frac{d^3\tilde{p}_{3}}{2(2\pi)^3}2(2\pi)^3\\
&\times\sum_{\lambda_1,\lambda_2,\lambda_3}\Psi^{J,J_z}(\tilde{p}_i,\lambda_i)C^{\alpha\beta\gamma}
\delta^{3}(\!\tilde{P}\!-\!\tilde{p}_1\!-\!\tilde{p}_2\!-\!\tilde{p}_3\!)\\
&\times~F_{q_1q_2q_3}~\vert q_{1\alpha}(\tilde{p}_1,\lambda_1)\rangle~\vert q_{2\beta}(\tilde{p}_2,\lambda_2)\rangle~\vert q_{3\gamma}(\tilde{p}_3,\lambda_3)\rangle,
\end{split}
\end{equation}
where the $C^{\alpha\beta\gamma}$ and $F_{q_1q_2q_3}$ are the color and flavor factors, respectively, and $\lambda_i$ and $p_i$ ($i$=1,2,3) are the helicities and light-front momenta of the on-mass-shell quarks, respectively, defined as
\begin{equation}
\tilde{p}_i=(p_i^+, \vec{p}_{i\bot}),
\quad p_i^+=p_i^0+p_i^3,
\quad \vec{p}_{i\bot}=(p_i^1, p_i^2).
\end{equation}
To describe the motion of the constituents, the intrinsic variables $(x_{i},~\vec{k}_{i})$ ($i=1,2,3$) are as follows
\begin{equation}
p_{i}^{+}=x_{i}^{}P^{+},~~
\vec{p}_{i\bot}=x_{i}\vec{P}_{i\bot}+\vec{k}_{i\bot},~~
\sum_{i=1}^{3}\vec{k}_{i\bot}=0,~~
\sum_{i=1}^{3}x_{i}=1,
\end{equation}
where $x_{i}$ represents the light-front momentum fractions bounded by $0<x_{i}<1$.

The vertex function should be normalized by
\begin{equation}
\langle\mathcal{B}(P^{\prime},J,J_z^{\prime})\vert\mathcal{B}(P,J,J_z)\rangle=
2(2\pi)^3P^{+}\delta^3(\tilde{P}-\tilde{P}^{\prime})\delta_{J_z,J_z^{\prime}},
\label{eq:normalization1}
\end{equation}
and
\begin{equation}
\begin{split}
\int&\Bigg{(}\prod_{i=1}^{3}\frac{dx_id^2\vec{k}_{i\bot}}{2(2\pi)^3}\Bigg{)}2(2\pi)^3\delta\Big{(}1-\sum_{i}x_i\Big{)}\\
&\times\delta^2\Big{(}\sum_{i}\vec{k}_{i\bot}\Big{)}\psi^{*}(x_i,\vec{k}_{i\bot})\psi(x_i,\vec{k}_{i\bot})=1.
\label{eq:normalization2}
\end{split}
\end{equation}

As proposed by Refs.~\cite{Korner:1994nh,Hussain:1995xs,Tawfiq:1998nk}, the spin-spatial wave functions for the $\Lambda$-type baryon with $J^{P}=1/2^{+}$ and $3/2^{-}$ are written as
\begin{equation}
\begin{split}
\Psi^{1/2,J_z}(\tilde{p}_i,\lambda_i)=&A_0\bar{u}(p_1,\lambda_1)[(\slashed{P}+M_0)\gamma^5]v(p_2,\lambda_2)\\
&\times\bar{u}_{Q}(p_3,\lambda_3)u(P,J_z)\psi(x_i,\vec{k}_{i\bot}),
\end{split}
\end{equation}
\begin{equation}
\begin{split}
\Psi^{3/2,J_z}(\tilde{p}_i,\lambda_i)=&B_0\bar{u}(p_1,\lambda_1)[(\slashed{P}+M_0)\gamma^5]v(p_2,\lambda_2)\\
&\times\bar{u}_{Q}(p_3,\lambda_3)K^{\alpha}u_{\alpha}(P,J_z)\psi(x_i,\vec{k}_{i\bot}),
\end{split}
\end{equation}
respectively, where
\begin{equation*}
\begin{split}
A_0&=\frac{1}{\sqrt{16P^+M_0^3(e_1+m_1)(e_2+m_2)(e_3+m_3)}},\\
B_0&=\frac{\sqrt{3}}{\sqrt{16P^+M_0^3(e_1+m_1)(e_2+m_2)(e_3-m_3)(e_3+m_3)^2}}
\end{split}
\end{equation*}
are the corresponding normalized factors determined by Eq.~\eqref{eq:normalization1}. The $K=\big{[}(m_1+m_2)p_3-m_3(p_1+p_2)\big{]}/\big{(}m_1+m_2+m_3\big{)}$ is the momentum of the $\lambda$ mode.

In the context of the three-body light-front quark model, the general expression for the weak transition matrix element is written as
\begin{equation}
\begin{split}
\langle\Lambda&^\ast(P^{\prime},J_z^{\prime})\vert\bar{s}\Gamma^{\mu}_{i}b\vert\Lambda_b(P,J_z)\rangle\\
=&\!\int\!\Big{(}\frac{dx_1d^2\vec{k}_{1\bot}}{2(2\pi)^3}\Big{)}
\Big{(}\frac{dx_2d^2\vec{k}_{2\bot}}{2(2\pi)^3}\Big{)}
\frac{\psi_b(x_i,\vec{k}_{i\bot})\psi_s^{\ast}(x_i^{\prime},\vec{k}_{i\bot}^{\prime})}{(16/\sqrt{3})\sqrt{x_3x_3^{\prime}M_0^3M_0^{\prime3}}}\\
&\!\times\!\frac{\text{Tr}[(\slashed{P}^{\prime}-M_0^{\prime})\gamma^5(\slashed{p}_1+m_1)(\slashed{P}+M_0)\gamma^5(\slashed{p}_2-m_2)]}
{\!\sqrt{\!(\!e_1\!+\!m_1\!)(\!e_2\!+\!m_2\!)(\!e_3\!+\!m_3\!)
(\!e_1^\prime\!+\!m_1^\prime\!)(\!e_2^\prime\!+\!m_2^\prime\!)(\!e_3^\prime\!-\!m_3^\prime\!)(\!e_3^{\prime}\!+\!m_3^{\prime}\!)^2\!}}\\
&\!\times\!\bar{u}_{\alpha}(P^{\prime},J_z^{\prime})K^{\prime\alpha}(\slashed{p}_3^{\prime}+m_3^{\prime})\Gamma^{\mu}_{i}(\slashed{p}_3+m_3)u(P,J_z),
\label{eq:ffs}
\end{split}
\end{equation}
where the $\Gamma^{\mu}_i=\big{\{}\gamma^{\mu},\gamma^{\mu}\gamma^{5},i\sigma^{\mu\nu}q_{\nu},i\sigma^{\mu\nu}q_{\nu}\gamma^{5}\big{\}}$. $P=p_1+p_2+p_3$ and $P^\prime=p_1+p_2+p_3^\prime$ are the light-front momenta for initial and final baryons, respectively, considering $p_1 \equiv p_1^\prime$ and $p_2 \equiv p_2^\prime$ in the spectator scheme. The following kinematics of the constituent quarks as
\begin{equation}
\begin{split}
p_{i}^{(\prime)+}&=x_{i}^{(\prime)}P^{(\prime)+},~~~~
\vec{p}_{i\bot}^{(\prime)}=x_{i}^{(\prime)}\vec{P}_{i\bot}^{(\prime)}+\vec{k}_{i\bot}^{(\prime)},\\
p_{3}^+-p_{3}^{\prime+}&=q^+,~~~~~~~~~~~~~
\vec{p}_{3\bot}-\vec{p}_{3\bot}^{\prime}=\vec{q}_{\bot},\\
\sum_{i=1}^{3}\vec{k}_{i\bot}^{(\prime)}&=0,~~~~~~~~~~~~~~
\sum_{i=1}^{3}x_{i}^{(\prime)}=1
\end{split}
\end{equation}
have been used to simplify the above matrix element.

In addition, the $\psi_b$ and $\psi_s$ are the spatial wave functions of $\Lambda_b$ and $\Lambda^\ast$, respectively. Their forms are written as
\begin{equation}
\begin{split}
\psi(x_i,\vec{k}_{i})=&N_{\psi}\sqrt{\frac{e_1e_2e_3}{x_1x_2x_3M_{0}}}\phi_{\rho}\Big{(}\frac{m_1\vec{k}_{2}-m_2\vec{k}_{1}}{m_1+m_2}\Big{)}\\
&\times\phi_{\lambda}\Big{(}\frac{(m_1+m_2)\vec{k}_3-m_3(\vec{k}_1+\vec{k}_2)}{m_1+m_2+m_3}\Big{)}
\end{split}
\end{equation}
in this paper, where $\vec{k}_{i}=(\vec{k}_{i\bot},k_{iz})$ with
\begin{equation}
k_{iz}=\frac{x_{i}M_{0}}{2}-\frac{m_{i}^{2}+\vec{k}_{i\bot}^{2}}{2x_{i}M_{0}}.
\end{equation}
By the way, $\phi_{\rho(\lambda)}$ is the spatial wave function of $\rho(\lambda)$ mode. The factor $N_{\psi}=(4\pi^{3/2})^2$ for the ground state $\Lambda_b$ and the factor $N_{\psi}=(4\pi^{3/2})^2/\sqrt{3}$ for the $P$-wave state $\Lambda(1520)$ are determined by Eq.~\eqref{eq:normalization2}. The additional factor $/1\sqrt{3}$ for the $P$-wave state comes from  different angular components of the spatial wave functions described by the spherical harmonic functions compared to the ground state.

In previous work~\cite{Chua:2018lfa,Ke:2019smy,Chua:2019yqh}, the spatial wave function of the baryon is usually adopted as a SHO form with an oscillator parameter $\beta$, which causes the $\beta$ dependence of the form factors. To avoid this uncertainty, we  adopt the numerical spatial wave function obtained by solving the three-body Schr\"{o}dinger equation with the semirelativistic potential model. The detailed discussions are presented in Sec.~\ref{sec4}.

The next content discusses how to extract the form factors in the light-front quark model. Here, we consider the $q^+=0$ and $\vec{q}_{\bot}\neq0$ condition. In order to extract the four form factors in vector current, one can multiply the $\bar{u}(P,J_z)\Gamma_{i}^{V,\mu\beta}u_{\beta}(P^{\prime},J_z^{\prime})$ on both sides of Eq.~\eqref{eq:ffs} with specific setting $\Gamma_i^{\mu}=\gamma^{\mu}$ and sum over the polarizations of the initial and final states. And then the left side can be replaced by Eq.~\eqref{eq:ffs01}, and the right side can be calculated out by carrying out the traces and then the integration. The Lorentz structures are $\Gamma_{i}^{V,\mu\beta}=\big{\{}g^{\beta\mu},P^{\beta}\gamma^{\mu},P^{\beta}P^{\prime\mu},P^{\beta}P^{\mu}\big{\}}$. The complete expressions for the form factors in vector current are obtained by solving
\begin{widetext}
\begin{equation}
\begin{split}
\text{Tr}[(G_{\Lambda^{\ast}})^{\beta\alpha}.[{\rm form\ factors\ in\ Eq.~\eqref{eq:ffs01}}].(\slashed{P}+M_{0}).\Gamma_{(1,2,3,4),\mu}^{V,\beta}]
=&\int\bigg{(}\frac{dx_1d^2\vec{k}_{1\bot}}{2(2\pi)^3}\bigg{)}\bigg{(}\frac{dx_2d^2\vec{k}_{2\bot}}{2(2\pi)^3}\bigg{)}\frac{\psi_b(x_i,\vec{k}_{i\bot})\psi_s^{\ast}(x_i^{\prime},\vec{k}^{\prime}_{i\bot})}{\sqrt{x_3x_3^{\prime}}}A_0B_0^{\prime}\text{Tr}[\cdots]\\
&\times\text{Tr}\big{[}(G_{\Lambda^{*}})_{\beta\alpha}K^{\prime\alpha}(\slashed{p}_{3}^{\prime}+m_{3}^{\prime})\gamma^{\mu}
(\slashed{p}_{3}+m_{3})(\slashed{P}+M_0)\Gamma_{(1,2,3,4),\mu}^{V,\beta}\big{]},
\label{eq:ffs05}
\end{split}
\end{equation}
where
\begin{eqnarray}
A_{0}&=&1\big{/}{\sqrt{16M_0^3(e_1+m_1)(e_2+m_2)(e_3+m_3)}},\\
B_{0}^{\prime}&=&\sqrt{3}\big{/}{\sqrt{16M_{0}^{\prime3}(e_1^\prime+m_1^\prime)(e_2^\prime+m_2^\prime)(e_3^\prime-m_3^\prime)(e_3^{\prime}+m_3^{\prime})^2}},\\
\text{Tr}[\cdots]&=&\text{Tr}[(\slashed{P}^{\prime}-M_0^{\prime})\gamma^5(\slashed{p}_1+m_1)(\slashed{P}-M_0)\gamma^5(\slashed{p}_2-m_2)],\\
(G_{\Lambda^\ast})^{\mu\nu}&=&-(\slashed{P}^{\prime}+M_0^{\prime})
\Big{[}g^{\mu\nu}-\frac{1}{3}\gamma^{\mu}\gamma^{\nu}-\frac{2}{3M_0^{\prime2}}P^{\prime\mu}P^{\prime\nu}-\frac{1}{3M_0^{\prime}}\big{(}\gamma^{\mu}P^{\prime\nu}-\gamma^{\nu}P^{\prime\mu}\big{)}\Big{]}.
\end{eqnarray}
Analogously, the form factors in the axial-vector, tensor, and pseudotensor currents can be obtained by using the structures $\bar{u}(P,J_z)\Gamma^{A,\mu\beta}_{i}u_{\beta}(P^{\prime},J_z^{\prime})$, $\bar{u}(P,J_z)\Gamma^{T,\mu\beta}_{i}u_{\beta}(P^{\prime},J_z^{\prime})$ and $\bar{u}(P,J_z)\Gamma^{T5,\mu\beta}_{i}u_{\beta}(P^{\prime},J_z^{\prime})$ with setting $\Gamma^{\mu}_{i}=\gamma^{\mu}\gamma^5$, $i\sigma^{\mu\nu}q_{\nu}$ and $i\sigma^{\mu\nu}q_{\nu}\gamma^5$ in Eq.~\eqref{eq:ffs}, respectively. The Lorentz structures are $\Gamma_{i}^{A,\mu\beta}=\big{\{}g^{\beta\mu}\gamma^{5},P^{\beta}\gamma^{\mu}\gamma^{5},P^{\beta}P^{\prime\mu}\gamma^{5},P^{\beta}P^{\mu}\gamma^{5}\big{\}}$, $\Gamma_{i}^{T,\mu\beta}=\big{\{}g^{\beta\mu},P^{\beta}\gamma^{\mu},P^{\beta}P^{\prime\mu}\big{\}}$ and $\Gamma_{i}^{T5,\mu\beta}=\big{\{}g^{\beta\mu}\gamma^{5},P^{\beta}\gamma^{\mu}\gamma^{5},P^{\beta}P^{\prime\mu}\gamma^{5}\big{\}}$.
The complete expressions of the form factors can be obtained by solving
\begin{equation}
\begin{split}
\text{Tr}[(G_{\Lambda^{\ast}})^{\beta\alpha}.[{\rm form~ factors~ in~ Eq}.~\eqref{eq:ffs02}].(\slashed{P}+M_{0}).\Gamma_{(1,2,3,4),\mu}^{A,\beta}]
=&\int\bigg{(}\frac{dx_1d^2\vec{k}_{1\bot}}{2(2\pi)^3}\bigg{)}\bigg{(}\frac{dx_2d^2\vec{k}_{2\bot}}{2(2\pi)^3}\bigg{)}\frac{\psi_b(x_i,\vec{k}_{i\bot})\psi_s^{\ast}(x_i^{\prime},\vec{k}^{\prime}_{i\bot})}{\sqrt{x_3x_3^{\prime}}}
A_0B_0^{\prime}\text{Tr}[\cdots]\\
&\times\text{Tr}\big{[}(G_{G_{\Lambda^{*}}})_{\beta\alpha}K^{\prime\alpha}(\slashed{p}_{3}^{\prime}+m_{3}^{\prime})\gamma^{\mu}\gamma^{5}
(\slashed{p}_{3}+m_{3})(\slashed{P}+M_0)\Gamma_{(1,2,3,4),\mu}^{A,\beta}\big{]},\\
\label{eq:ffs06}
\end{split}
\end{equation}
\begin{equation}
\begin{split}
\text{Tr}[(G_{\Lambda^{\ast}})^{\beta\alpha}.[{\rm form~ factors~ in~ Eq}.~\eqref{eq:ffs03}].(\slashed{P}+M_{0}).\Gamma_{(1,2,3),\mu}^{T,\beta}]
=&\int\bigg{(}\frac{dx_1d^2\vec{k}_{1\bot}}{2(2\pi)^3}\bigg{)}\bigg{(}\frac{dx_2d^2\vec{k}_{2\bot}}{2(2\pi)^3}\bigg{)}\frac{\psi_b(x_i,\vec{k}_{i\bot})\psi_s^{\ast}(x_i^{\prime},\vec{k}^{\prime}_{i\bot})}{\sqrt{x_3x_3^{\prime}}}
A_0B_0^{\prime}\text{Tr}[\cdots]\\
&\times\text{Tr}\big{[}(G_{G_{\Lambda^{*}}})_{\beta\alpha}K^{\prime\alpha}(\slashed{p}_{3}^{\prime}+m_{3}^{\prime})i\sigma^{\mu\nu}q_{\nu}
(\slashed{p}_{3}+m_{3})(\slashed{P}+M_0)\Gamma_{(1,2,3),\mu}^{T,\beta}\big{]},
\label{eq:ffs07}
\end{split}
\end{equation}
\begin{equation}
\begin{split}
\text{Tr}[(G_{\Lambda^{\ast}})^{\beta\alpha}.[{\rm form~ factors~ in~ Eq}.~\eqref{eq:ffs04}].(\slashed{P}+M_{0}).\Gamma_{(1,2,3),\mu}^{T5,\beta}]
=&\int\bigg{(}\frac{dx_1d^2\vec{k}_{1\bot}}{2(2\pi)^3}\bigg{)}\bigg{(}\frac{dx_2d^2\vec{k}_{2\bot}}{2(2\pi)^3}\bigg{)}\frac{\psi_b(x_i,\vec{k}_{i\bot})\psi_s^{\ast}(x_i^{\prime},\vec{k}^{\prime}_{i\bot})}{\sqrt{x_3x_3^{\prime}}}
A_0B_0^{\prime}\text{Tr}[\cdots]\\
&\times\text{Tr}\big{[}(G_{G_{\Lambda^{*}}})_{\beta\alpha}K^{\prime\alpha}(\slashed{p}_{3}^{\prime}+m_{3}^{\prime})i\sigma^{\mu\nu}q_{\nu}\gamma^{5}
(\slashed{p}_{3}+m_{3})(\slashed{P}+M_0)\Gamma_{(1,2,3),\mu}^{T5,\beta}\big{]}.
\label{eq:ffs08}
\end{split}
\end{equation}
\end{widetext}
This approach has been used to evaluate the form factors of triple heavy baryon transitions from $3/2\to1/2$ cases~\cite{Wang:2022ias,Zhao:2022vfr}.

\section{The semirelativistic potential model for calculating baryon wave function}
\label{sec4}

In this section, we will derive the wave function using the GEM with semirelativistic potential model. In general, to obtain the wave function and mass of a baryon, we need to solve the three-body Schr\"{o}dinger equation,
\begin{equation}
\mathcal{H} \Psi_{\mathbf{J},\mathbf{M_J}}=E \Psi_{\mathbf{J},\mathbf{M_J}},
\end{equation}
where $\mathcal{H}$ is the Hamiltonian and $E$ is the corresponding eigenvalue. It can be solved by using the Rayleigh-Ritz variational principle.

Unlike a meson system, a baryon in the traditional quark model is a typical three-body system. In our calculation, the semirelativistic potentials used in Refs.~\cite{Capstick:1985xss,Li:2021qod,Li:2021kfb} are applied. The  Hamiltonian in question~\cite{Li:2021qod,Li:2021kfb}
\begin{equation}
\mathcal{H}=K+\sum_{i<j}(S_{ij}+G_{ij}+V^{\text{so(s)}}_{ij}+V^{\text{so(v)}}_{ij}+V^{\text{ten}}_{ij}+V^{\text{con}}_{ij})
\end{equation}
includes the kinetic energy $K$, the spin-independent linear confinement piece $S$, the Coulomb-like potential $G$, and the higher-order terms containing the scalar-type spin-orbit interaction $V^{\text{so}(s)}$, the vector-type spin-orbit interaction $V^{\text{so}(v)}$, the tensor potential $V^{\text{tens}}$, and the spin-dependent contact potential $V^{\text{con}}$. The concrete expressions are given as~\cite{Capstick:1985xss,Li:2021qod,Li:2021kfb}
\begin{eqnarray}
K&=&\sum_{i=1,2,3}\sqrt{m_i^2+p_i^2},\\
S_{ij}&=&-\frac{3}{4}\left(br_{ij}\left[\frac{e^{-\sigma^2r_{ij}^2}}{\sqrt{\pi}\sigma r_{ij}}+\left(1+\frac{1}{2\sigma^2r_{ij}^2}\right)\frac{2}{\sqrt{\pi}}\right.\right.\nonumber\\&&\left.\left.
\times\int_{0}^{\sigma r_{ij}}e^{-x^2}dx\right]\right)\mathbf{F_i}\cdot\mathbf{F_j}+\frac{c}{3},\\
G_{ij}&=&\sum_{k}\frac{\alpha_k}{r_{ij}}\left[\frac{2}{\sqrt{\pi}}\int_{0}^{\tau_k r_{ij}}e^{-x^2}dx\right]\mathbf{F_i}\cdot\mathbf{F_j}
\end{eqnarray}
for the spin-independent terms with
\begin{equation}
\sigma^2=\sigma_0^2\left[\frac{1}{2}+\frac{1}{2}\left(\frac{4m_im_j}{(m_i+m_j)^2}\right)^4+s^2\left(\frac{2m_im_j}{m_i+m_j}\right)^2\right],
\end{equation}
and the $\langle\mathbf{F_i}\cdot\mathbf{F_j}\rangle=-2/3$ for the quark-quark interaction, and
\begin{eqnarray}
V^{\text{so}(s)}_{ij}\!&\!=\!&-\frac{\mathbf{r_{ij}}\times\mathbf{p_{i}}\cdot\mathbf{S_i}}{2m_i^2}\frac{1}{r_{ij}}
\frac{\partial S_{ij}}{\partial r_{ij}}+\frac{\mathbf{r_{ij}}\times\mathbf{p_{j}}\cdot\mathbf{S_j}}{2m_j^2}\frac{1}{r_{ij}}\frac{\partial S_{ij}}{\partial r_{ij}},\\
V^{\text{so}(v)}_{ij}\!&\!=\!&\frac{\mathbf{r_{ij}}\times\mathbf{p_{i}}\cdot\mathbf{S_i}}{2m_i^2}\frac{1}{r_{ij}}\frac{\partial G_{ij}}{\partial r_{ij}}
-\frac{\mathbf{r_{ij}}\times\mathbf{p_{j}}\cdot\mathbf{S_j}}{2m_j^2}\frac{1}{r_{ij}}\frac{\partial G_{ij}}{\partial r_{ij}}
\nonumber\\&&-\frac{\mathbf{r_{ij}}\times\mathbf{p_{j}}\cdot\mathbf{S_i}-\mathbf{r_{ij}}\times\mathbf{p_{i}}\cdot\mathbf{S_j}}{m_i~m_j}\frac{1}{r_{ij}}\frac{\partial G_{ij}}{\partial r_{ij}},\\
V^{\text{tens}}_{ij}\!&\!=\!&-\frac{1}{m_im_j}\left[\left(\mathbf{S_i}\cdot\mathbf{\hat r_{ij}}\right)\left(\mathbf{S_j}\cdot \mathbf{\hat r_{ij}}\right)-\frac{\mathbf{S_i}\cdot\mathbf{S_j}}{3}\right]
\nonumber\\&&\times\left(\frac{\partial^2G_{ij}}{\partial r_{ij}^2}-\frac{\partial G_{ij}}{r_{ij}\partial r_{ij}}\right),\\
V^{\text{con}}_{ij}\!&\!=\!&\frac{2\mathbf{S_i}\cdot\mathbf{S_j}}{3m_i m_j}\nabla^2G_{ij}
\end{eqnarray}
for the spin-dependent terms, where $m_i$ is the mass of the $i$th constituent quark, and $\mathbf{S_i}$ is the corresponding spin operator.

Next, a general transformation based on the center of mass of the interacting quarks and the momentum is set up to compensate for the loss of relativistic effect in the nonrelativistic limit~\cite{Godfrey:1985xj,Capstick:1985xss}
\begin{equation}
\begin{split}
&G_{ij}\to\left(1+\frac{p^2}{E_iE_j}\right)^{1/2} G_{ij}\left(1+\frac{p^2}{E_iE_j}\right)^{1/2},\\
&\frac{V^{k}_{ij}}{m_im_j}\to\left(\frac{m_im_j}{E_iE_j}\right)^{1/2+\epsilon_k}\frac{V^k_{ij}}{m_im_j}\left(\frac{m_im_j}{E_iE_j}\right)^{1/2+\epsilon_k},
\end{split}
\end{equation}
where $E_i=\sqrt{p^2+m_i^2}$ is the energy of the $i$th constituent quark, the subscript $k$ is used to distinguish the contact, tensor, vector spin-orbit, and scalar spin-orbit terms, and the $\epsilon_k$ is used to denote the relevant modification parameters, which are collected in Table~\ref{tab:parametersofGI}.

\begin{table}
\centering
\caption{The parameters used in the semirelativistic potential model. The quark masses are also chosen to be $m_{u}=220\ \text{MeV}$, $m_{d}=220\ \text{MeV}$, $m_{s}=419\ \text{MeV}$, $m_{c}=1628\ \text{MeV}$, and $m_{b}=4977\ \text{MeV}$~\cite{Godfrey:1985xj,Capstick:1985xss}.}
\label{tab:parametersofGI}
\renewcommand\arraystretch{1.05}
\begin{tabular*}{80mm}{c@{\extracolsep{\fill}}ccc}
\toprule[1pt]
\toprule[0.7pt]
Parameters     & Values   & Parameters   & Values\\
\toprule[0.7pt]
$b~(\text{GeV}^2)$        &$0.1466\pm0.0007$      &$\epsilon^{\text{so}(s)}$   &$0.5000\pm0.0762$\\
$c~(\text{GeV})  $        &$-0.3490\pm0.0050$     &$\epsilon^{\text{so}(v)}$   &$-0.1637\pm0.0131$\\
$\sigma_0~(\text{GeV})$   &$1.7197\pm0.0304$      &$\epsilon^{\text{tens}}$    &$-0.3790\pm0.5011$\\
$s$                       &$0.5278\pm0.0718$      &$\epsilon^{\text{con}}$     &$-0.1612\pm0.0015$\\
\bottomrule[0.7pt]
\bottomrule[1pt]
\end{tabular*}
\end{table}

\begin{figure}[htbp]\centering
  \includegraphics[width=51mm]{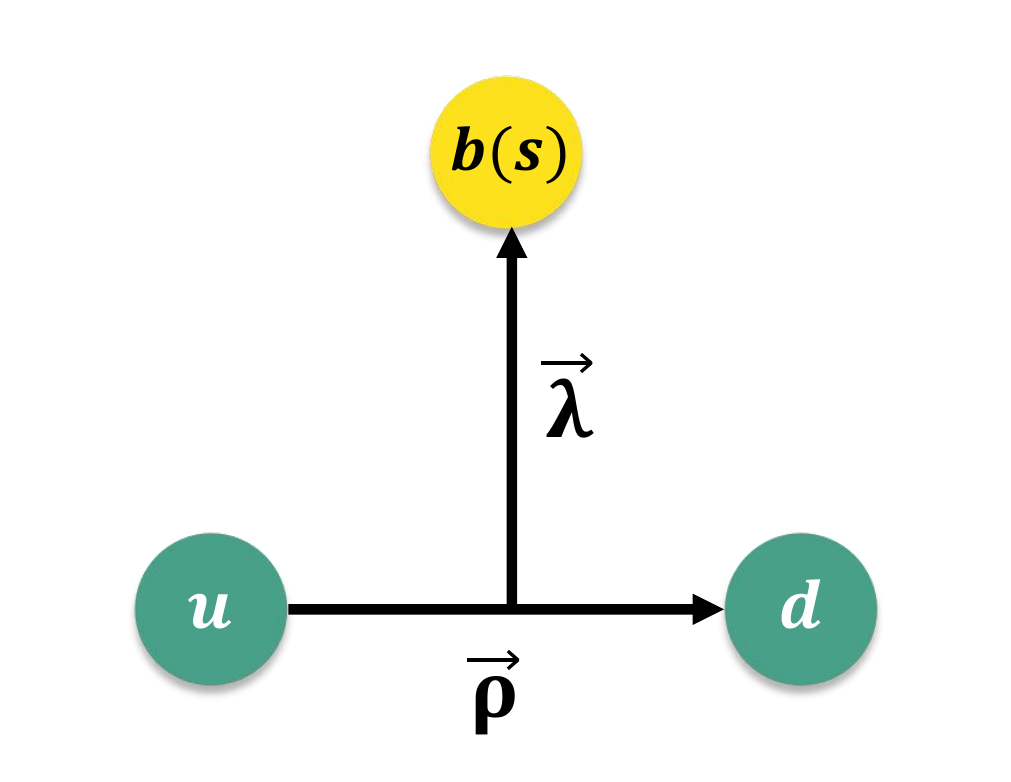}\\
  \caption{The definition of the  internal Jacobi coordinates $\vec{\rho}$ and $\vec{\lambda}$, where we use green spheres to represent the $u$ and $d$ quarks and yellow spheres to represent the $b$ (or $s$) quark.}
  \label{fig:Jacobi}
\end{figure}

The total wave function of the baryon is composed of color, spin, spatial and flavor wave functions, {\it i.e.},
\begin{equation}
\Psi_{\mathbf{J},\mathbf{M_J}}=
\chi^{\text{color}}
\left\{{\chi^\text{spin}}_{\mathbf{S},\mathbf{M_S}}
\psi^{\text{spatial}}_{\mathbf{L},\mathbf{M_L}}(\vec{\rho},\vec{\lambda})\right\}_{\mathbf{J},\mathbf{M_J}}
\psi^{\text{flavor}},
\end{equation}
where $\chi^{\text{color}}=(rgb-rbg+gbr-grb+brg-bgr)/\sqrt{6}$ is the universal color wave function for the baryon. For the affected $\Lambda_b$ and $\Lambda^\ast$, their flavor wave functions are chosen as $\psi^{\text{flavor}}=(ud-du)Q/\sqrt{2}$ where $Q=b~\text{or}~s$. Also, \textbf{S} is the total spin and \textbf{L} is the total orbital angular momentum. $\psi^{\text{spatial}}_{\mathbf{L},\mathbf{M_L}}$ is the spatial wave function, which is composed of the $\rho$ mode and $\lambda$ mode
\begin{equation}
\psi^{\text{partial}}_{\mathbf{L},\mathbf{M_L}}(\vec{\rho},\vec{\lambda})=\left\{\phi_{\pmb{l_{\rho}},\pmb{ml_{\rho}}}(\vec{\rho})
\phi_{\pmb{l_{\lambda}},\pmb{ml_{\lambda}}}(\vec{\lambda})\right\}_{\mathbf{L},\mathbf{M_L}},
\end{equation}
where the subscripts $\pmb{l_{\rho}}$ and $\pmb{l_{\lambda}}$ represent the orbital angular momentum quanta for the $\rho$ and $\lambda$ modes, respectively, and the internal Jacobi coordinates are chosen to be
\begin{equation}
\begin{split}
\vec{\rho}&=\vec{r}_2-\vec{r}_1,\\
\vec{\lambda}&=\vec{r}_3-\frac{m_1\vec{r}_1+m_2\vec{r}_2}{m_1+m_2}.
\end{split}
\end{equation}
As shown in Fig.~\ref{fig:Jacobi}, the $\Lambda_b(\Lambda^\ast)$ is considered as a bound state with the $u$ and $d$ quarks bound to form the $\rho$ mode and then  bounded to the $b\ (\text{or}\ s)$ quark to form the $\lambda$ mode.

In this calculation, the Gaussian basis~\cite{Hiyama:2003cu,Yoshida:2015tia,Yang:2019lsg},
\begin{equation}
\begin{split}
\phi_{nlm}^{G}(\vec{r})&=\phi^{G}_{nl}(r)~Y_{lm}(\hat{r})\\
&=\sqrt{\frac{2^{l+2}(2\nu_{n})^{l+3/2}}{\sqrt{\pi}(2l+1)!!}}\lim_{\varepsilon\to0}\frac{1}{(\nu_{n}\varepsilon)^l}
\sum_{k=1}^{k_{\text{max}}}C_{lm,k}e^{-\nu_{n}(\vec{r}-\varepsilon\vec{D}_{lm,k})^2},
\label{Gaussian basis}
\end{split}
\end{equation}
is used to expand the spatial wave functions $\phi_{\pmb{l_{\rho}},\pmb{ml_{\rho}}}$ and $\phi_{\pmb{l_{\lambda}},\pmb{ml_{\lambda}}}$ ($n=1,2,\cdots,n_{\text{max}}$), where the freedom parameter $n_{\text{max}}$ should be chosen from positive integers, and then the Gaussian size parameter $\nu_{n}$ can be settled as~\cite{Luo:2022cun}
\begin{equation}
\nu_{n}=1/r^2_{n}, ~~~r_{n}=r_{\text{min}}~a^{n-1},
\end{equation}
where $$a=\left(\frac{r_{\text{max}}}{r_{\text{min}}}\right)^{\frac{1}{n_{\text{max}}-1}}.$$In our calculation the values of $\rho_{\text{min}}$ and $\rho_{\text{max}}$ are chosen to be $0.2$~fm and $2.0$~fm, respectively, and the parameter $n_{\rho_{\text{max}}}=6$. For the $\lambda$ mode, we also use the same Gaussian-sized parameters.

\begin{table*}
\centering
\caption{Experimentally observed masses of charmed and bottom baryons used to fit the potential model parameters, where only the central values are given.}
\label{tab:spectrum}
\renewcommand\arraystretch{1.05}
\begin{tabular*}{160mm}{c@{\extracolsep{\fill}}ccccccc}
\toprule[1pt]
\toprule[0.5pt]
States              &$J^{P}$  &This work (GeV)   &Expt. (GeV)~\cite{ParticleDataGroup:2020ssz}   &States        &$J^{P}$    &This work (GeV)   &Expt. (GeV)~\cite{ParticleDataGroup:2020ssz}\\
\toprule[0.5pt]
$\Lambda_c$         &$\frac{1}{2}^{+}$       &$2.286$      &$2.286$       &$\Lambda_b$         &$\frac{1}{2}^{+}$        &$5.621$      &$5.619$\\
$\Lambda_c(2595)$   &$\frac{1}{2}^{-}$       &$2.595$      &$2.595$       &$\Lambda_b(5912)$   &$\frac{1}{2}^{-}$        &$5.896$      &$5.912$\\
$\Lambda_c(2625)$   &$\frac{3}{2}^{-}$       &$2.627$      &$2.625$       &$\Lambda_b(5920)$   &$\frac{3}{2}^{-}$        &$5.909$      &$5.919$\\
$\Lambda_c(2765)$   &$?^{?}$                 &$2.768$      &$2.765$       &$\Lambda_b(6070)$   &$\frac{1}{2}^{+}$        &$6.046$      &$6.072$\\
$\Lambda_c(2860)$   &$\frac{3}{2}^{+}$       &$2.872$      &$2.856$       &$\Lambda_b(6146)$   &$\frac{3}{2}^{+}$        &$6.133$      &$6.146$\\
$\Lambda_c(2880)$   &$\frac{5}{2}^{+}$       &$2.894$      &$2.881$       &$\Lambda_b(6152)$   &$\frac{5}{2}^{+}$        &$6.144$      &$6.152$\\
$\Sigma_c$          &$\frac{1}{2}^{+}$       &$2.446$      &$2.453$       &$\Sigma_b$          &$\frac{1}{2}^{+}$        &$5.809$      &$5.811$\\
$\Sigma_c(2520)$    &$\frac{3}{2}^{+}$       &$2.519$      &$2.518$       &$\Sigma_b^{*}$      &$\frac{3}{2}^{+}$        &$5.835$      &$5.832$\\
$\Xi_c$             &$\frac{1}{2}^{+}$       &$2.478$      &$2.467$       &$\Xi_b$             &$\frac{1}{2}^{+}$        &$5.809$      &$5.794$\\
$\Xi_c(2790)$       &$\frac{1}{2}^{-}$       &$2.787$      &$2.792$       &$\Xi_b(6100)$       &$\frac{3}{2}^{-}$        &$6.093$      &$6.100$\\
$\Xi_c(2815)$       &$\frac{3}{2}^{-}$       &$2.814$      &$2.816$       &$\Xi_b(6327)$~\cite{LHCb:2021ssn}       &$?^{?}$           &$6.316$      &$6.327$\\
$\Xi_c(2970)$       &$?^{?}$                 &$2.953$      &$2.970$       &$\Xi_b(6333)$~\cite{LHCb:2021ssn}       &$?^{?}$           &$6.324$      &$6.332$\\
$\Xi_c(3055)$       &$?^{?}$                 &$3.059$      &$3.055$       &$\Xi_b^{\prime}(5935)$    &$\frac{1}{2}^{+}$  &$5.939$      &$5.935$\\
$\Xi_c(3080)$       &$?^{?}$                 &$3.077$      &$3.080$       &$\Xi_b(5945)$       &$\frac{3}{2}^{+}$        &$5.963$      &$5.949$\\
$\Xi_c^{\prime}$    &$\frac{1}{2}^{+}$       &$2.583$      &$2.577$       &$\Omega_b$          &$\frac{1}{2}^{+}$        &$6.043$      &$6.046$\\
$\Xi_c(2645)$       &$\frac{3}{2}^{+}$       &$2.648$      &$2.645$       &   &   &   &\\
$\Omega_c$          &$\frac{1}{2}^{+}$       &$2.693$      &$2.695$       &   &   &   & \\
$\Omega_c(2770)$    &$\frac{3}{2}^{+}$       &$2.755$      &$2.765$       &   &   &   &\\
\bottomrule[0.5pt]
\bottomrule[1pt]
\end{tabular*}
\end{table*}

\begin{table*}[htbp]\centering
\caption{The comparison of the masses of $\Lambda_b$ and $\Lambda(1520)$ from our calculation and the PDG~\cite{ParticleDataGroup:2020ssz} data, and the radial components of the spatial wave functions of the concerned $\Lambda_b $ and $\Lambda(1520)$ from the semirelativistic potential model and GEM. The Gaussian bases $(n_{\rho},n_{\lambda})$ listed in the fourth column are arranged as $[(1,1),(1,2),\cdots,(1,n_{\lambda_{\text{max}}}),(2,1),(2,2), \cdots,(2,n_{\lambda_{\text{max}}}),\cdots,(n_{\rho_{\text{max}}},1), (n_{\rho_{\text{max}}},2),\cdots,(n_{\rho_{\text{max}}},n_{\lambda_{\text{max}}})]$.}
\label{tab:wavefunctions}
\renewcommand\arraystretch{1.05}
\begin{tabular*}{172mm}{c@{\extracolsep{\fill}}cccc}
\toprule[1pt]
\toprule[0.5pt]
States  &This work (GeV)  &Experiment (MeV)~\cite{ParticleDataGroup:2020ssz}  &Eigenvectors\\
\midrule[0.5pt]
\multirow{9}*{\shortstack{$\Lambda_b$}}        &\multirow{9}*{$5.621\pm0.005$}    &\multirow{9}*{$5619.60\pm0.17$}
&$\big{[}0.0068\pm0.0007,0.0442\pm0.0014,0.0732\pm0.0016,0.0032\pm0.0003,$\\
&&&$0.0011\pm0.0001,-0.0004\pm0.0000,0.0270\pm0.0012,0.0204\pm0.0010,$\\
&&&$0.0273\pm0.0022,0.0067\pm0.0004,-0.0027\pm0.0001,0.0007\pm0.0000,$\\
&&&$-0.0170\pm0.0002,0.2541\pm0.0058,0.2427\pm0.0006,0.0005\pm0.0002,$\\
&&&$0.0060\pm0.0001,-0.0017\pm0.0000,-0.0037\pm0.0003,-0.0426\pm0.0010,$\\
&&&$0.4052\pm0.0028,0.0253\pm0.0025,-0.0023\pm0.0007,0.0004\pm0.0002,$\\
&&&$0.0071\pm0.0001,-0.0052\pm0.0008,0.0105\pm0.0008,0.1224\pm0.0015,$\\
&&&$-0.0246\pm0.0001,0.0054\pm0.0000,-0.0020\pm0.0000,0.0010\pm0.0003,$\\
&&&$-0.0112\pm0.0003,-0.0139\pm0.0001,0.0086\pm0.0001,-0.0017\pm0.0000\big{]}$\\
\midrule[0.5pt]
\multirow{9}*{\shortstack{$\Lambda(1520)$}}        &\multirow{9}*{$1.561\pm0.007$}    &\multirow{9}*{$1517.5\pm0.4$}
&$\big{[}0.0000\pm0.0001,-0.0096\pm0.0004,-0.0488\pm0.0017,-0.0576\pm0.0010$\\
&&&$-0.0011\pm0.0001,-0.0004\pm0.0000,-0.0049\pm0.0004,0.0041\pm0.0002$\\
&&&$-0.0295\pm0.0012,-0.0279\pm0.0020,0.0011\pm0.0003,-0.0006\pm0.0002$\\
&&&$-0.0010\pm0.0001,-0.0510\pm0.0019,-0.1771\pm0.0036,-0.1890\pm0.0014$\\
&&&$-0.0036\pm0.0004,-0.0008\pm0.0003,0.0003\pm0.0001,0.0222\pm0.0005$\\
&&&$-0.2146\pm0.0003,-0.2766\pm0.0040,-0.0001\pm0.0013,-0.0036\pm0.0006$\\
&&&$-0.0025\pm0.0001,0.0028\pm0.0001,0.0135\pm0.0012,-0.1653\pm0.0011$\\
&&&$-0.0174\pm0.0008,0.0019\pm0.0005,0.0010\pm0.0000,-0.0020\pm0.0000$\\
&&&$0.0035\pm0.0004,0.0277\pm0.0002,-0.0061\pm0.0005,0.0010\pm0.0003\big{]}$\\
\toprule[0.5pt]
\toprule[1pt]
\end{tabular*}
\end{table*}

In this paper, we fit the single charmed and bottom baryon spectrum to fix the phenomenological parameters in the semirelativistic potential model. The experimentally observed masses of charmed and bottom baryons are collected in Table~\ref{tab:spectrum}. The $\chi^{2}$ method, i.e., finding the minimum $\chi^{2}$ value, is used for the fitting. In our fit, the $\chi^{2}$ value is defined as
\begin{equation}
\chi^{2}=\frac{1}{n(n-1)}\sum_{i}^{n}\Bigg{(}\frac{m_{i}^{\text{Exp}}-m_{i}^{\text{The}}}{\sigma_{i}}\Bigg{)}^2,
\end{equation}
where $m_{i}^{\text{Exp}}$ and $m_{i}^{\text{The}}$ are experimental and theoretical values of the mass of the $i$th baryon, respectively. {The errors $\sigma_{i}=1\ \text{MeV}$\footnote{Checking the PDG~\cite{ParticleDataGroup:2020ssz}, we find that the uncertainties of the measured masses of the charmed and bottom baryons are around a few MeV. In order to make the baryons act in the same proportions in our fitting, we choose a universe value of 1 MeV as the uncertainty.} are universal for all baryons.} In this fitting, the $\chi^2$ is given as 2.84. The fitted parameters are collected in Table~\ref{tab:parametersofGI}. Meanwhile, our results for the masses of the charmed and bottom baryons are presented in Table~\ref{tab:spectrum}.

With the above preparations, we can calculate the spatial wave functions of $\Lambda_b$ and $\Lambda(1520)$. Their masses and radial components of spatial wave functions are shown in Table~\ref{tab:wavefunctions}. It is obvious that the calculated mass of $\Lambda_b$ is consistent with the Particle Data Group (PDG)~\cite{ParticleDataGroup:2020ssz} averaged value, while that of $\Lambda(1520)$ is about $40$ MeV higher than the PDG value.

\section{Numerical results}
\label{sec5}

\subsection{The weak transition form factors}

With the input of the numerical wave functions of $\Lambda_b$ and $\Lambda^\ast$, and the complete expressions of the form factors obtained by solving Eq.~\eqref{eq:ffs05} and Eqs.~(\ref{eq:ffs06})-(\ref{eq:ffs08}), we present our numerical results of the form factors of the $\Lambda_b\to\Lambda^\ast$ transition. Since the form factors calculated in the light-front quark model are valid in the spacelike region ($q^2<0$), we have to extrapolate them to the timelike region ($q^2>0$).

{Before we do the extrapolation, we need to talk about some constraints on the form factors at the $q^{2}=q^{2}_{\text{max}}$ point. To make sure that the helicity amplitudes in Eqs.~(\ref{eq:nonzeroHV})-(\ref{eq:nonzeroHT5}) have no singularities and are nonzero values in the $q^{2}\to q^{2}_{\text{max}}$ limit, we get the constraints in this limit as
\begin{equation}
\begin{split}
f_{t}^{V}=&O\Big{(}\frac{1}{\sqrt{s_{-}}}\Big{)},
~f_{0}^{V}=O\Big{(}\frac{1}{s_{-}}\Big{)},
~f_{\bot}^{V}=O\Big{(}\frac{1}{s_{-}}\Big{)},
~f_{g}^{V}=O(1),\\
f_{t}^{A}=&O\Big{(}\frac{1}{s_{-}}\Big{)},
~f_{0}^{A}=O\Big{(}\frac{1}{\sqrt{s_{-}}}\Big{)},
~f_{\bot}^{A}=O\Big{(}\frac{1}{\sqrt{s_{-}}}\Big{)},
~f_{g}^{A}=O\Big{(}\frac{1}{\sqrt{s_{-}}}\Big{)},\\
f_{0}^{T}=&O\Big{(}\frac{1}{s_{-}}\Big{)},
~f_{\bot}^{T}=O\Big{(}\frac{1}{s_{-}}\Big{)},
~f_{g}^{T}=O(1),\\
f_{0}^{T5}=&O\Big{(}\frac{1}{\sqrt{s_{-}}}\Big{)},
~f_{\bot}^{T5}=O\Big{(}\frac{1}{\sqrt{s_{-}}}\Big{)},
~f_{g}^{T5}=O\Big{(}\frac{1}{\sqrt{s_{-}}}\Big{)}.
\end{split}
\end{equation}
The form factors that show less singular behavior in the $q^{2}\to q^{2}_{\text{max}}$ limit are also reasonable. This would lead the helicity amplitudes to be zero in $q^{2}=q^{2}_{\text{max}}$. The above features have been discussed in Ref.~\cite{Descotes-Genon:2019dbw}. However, the above requirement is not strict enough, since it gives a broad limit. {This will make nonunique extrapolations of the form factors.}

Since the LQCD calculation of $\Lambda_b\to\Lambda(1520)$ form factors has been done in Refs.~\cite{Meinel:2020owd,Meinel:2021mdj}, and their results work well in the kinematic region near $q_{\text{max}}^{2}$, we will talk about the characters of the form factors in the LQCD. The LQCD calculation has been completed in Refs.~\cite{Meinel:2020owd,Meinel:2021mdj}. The authors obtained finite values of the form factors of  $\Lambda_b\to\Lambda(1520)$  in their definition (i.e., $f_{0,+,\bot,\bot^{\prime}}$, $g_{0,+,\bot,\bot^{\prime}}$, $h_{+,\bot,\bot^{\prime}}$, and $\tilde{h}_{+,\bot,\bot^{\prime}}$) in the  $q^{2}=q_{\text{max}}^{2}$ limit. Their definition of the form factors can be converted to ours by \cite{Meinel:2020owd}
\begin{equation}
\begin{split}
f_{t}^{V}=&\frac{m_{\Lambda^{*}}}{s_{+}}f_{0},~~
f_{0}^{V}=\frac{m_{\Lambda^{*}}}{s_{-}}f_{+},~~~
f_{\bot}^{V}=\frac{m_{\Lambda^{*}}}{s_{-}}f_{\bot},~~~
f_{g}^{V}=f_{\bot^{\prime}},\\
f_{t}^{A}=&\frac{m_{\Lambda^{*}}}{s_{-}}g_{0},~~~
f_{0}^{A}=\frac{m_{\Lambda^{*}}}{s_{+}}g_{+},~~~
f_{\bot}^{A}=\frac{m_{\Lambda^{*}}}{s_{+}}g_{\bot},~~~
f_{g}^{A}=-g_{\bot^{\prime}},\\
f_{0}^{T}=&\frac{m_{\Lambda^{*}}}{s_{-}}h_{+},~~~
f_{\bot}^{T}=\frac{m_{\Lambda^{*}}}{s_{-}}h_{\bot},~~~
f_{g}^{T}=(m_{\Lambda_{b}}+m_{\Lambda^{*}})h_{\bot^{\prime}},\\
f_{0}^{T5}=&\frac{m_{\Lambda^{*}}}{s_{+}}\tilde{h}_{+},~~~
f_{\bot}^{T5}=\frac{m_{\Lambda^{*}}}{s_{+}}\tilde{h}_{\bot},~~~
f_{g}^{T5}=-(m_{\Lambda_{b}}-m_{\Lambda^{*}})\tilde{h}_{\bot^{\prime}}.
\label{eq:relations}
\end{split}
\end{equation}
This shows that in the $q^{2}=q_{\text{max}}^{2}$ limit, the LQCD results~\cite{Meinel:2021mdj} show
\begin{equation}
\begin{split}
f_{t}^{V}&=O(1),~~~f_{0}^{V}=O\Big{(}\frac{1}{s_{-}}\Big{)},~~~f_{\bot}^{V}=O\Big{(}\frac{1}{s_{-}}\Big{)},~~~f_{g}^{V}=O(1),\\
f_{0}^{A}&=O(1),~~~f_{\bot}^{A}=O(1),~~~f_{g}^{A}=O(1),\\
f_{0}^{T}&=O\Big{(}\frac{1}{s_{-}}\Big{)},~~~f_{\bot}^{T}=O\Big{(}\frac{1}{s_{-}}\Big{)},~~~f_{g}^{T}=O(1),\\
f_{0}^{T5}&=O(1),~~~f_{\bot}^{T5}=O(1),~~~f_{g}^{T5}=O(1).
\end{split}
\end{equation}
These characters fulfill the requirements. Also we have  $g_{0}(q^{2})=\frac{a_{1}^{g_{0}}}{1-q^{2}/(m_{\text{pole}}^{f})^{2}}(\omega-1)$~\cite{Meinel:2021mdj} with $\omega=(m_{\Lambda_{b}}^{2}+m_{\Lambda^{*}}^{2}-q^{2})/(2m_{\Lambda_b}m_{\Lambda^{*}})$, where $a_{1}^{g_{0}}$ is a nonzero value. According to Eq.~\eqref{eq:relations}, the $f_{t}^{A}(q^{2})=\frac{m_{\Lambda^{*}}}{s_{-}}\frac{a_{1}^{g_{0}}}{1-q^{2}/(m_{\text{pole}}^{f})^{2}}(\omega-1)$, and this implies, in the $q^{2}=q_{\text{max}}^{2}$ limit, that $f_{t}^{A}=O(1)$. This also satisfies the requirement.

In order to align with the LQCD results, we take the following strategy for the analytical continuation:
\begin{enumerate}
\item[{\bf 1.}]
To do the extrapolations of the form factors $f_{t,g}^{V}$, $f_{t,0,\bot,g}^{A}$, $f_{g}^{T}$, and $f_{0,\bot,g}^{T5}$, the $z$-series form~\cite{Boyd:1995cf,Bourrely:2008za,Khodjamirian:2011ub,Amhis:2022vcd}
\begin{equation}
f(q^2)=\frac{1}{1-q^2/(m_{\text{pole}}^f)^2}\left[a_0^f+a_1^fz^f(q^2)+a_2^f(z^f(q^2))^2\right]
\label{eq:fitness}
\end{equation}
is adopted where $a_{0}^{f}$, $a_{1}^{f}$, and $a_{2}^{f}$ are free parameters needed to  fit in the spacelike region, and
\begin{equation}
\begin{split}
z^f(q^2)&=\frac{\sqrt{t_+^f-q^2}-\sqrt{t_+^f-t_0}}{\sqrt{t_+^f-q^2}+\sqrt{t_+^f-t_0}},\\
t_{\pm}^f&=(m_{B}\pm m_{K})^2.
\end{split}
\end{equation}
The parameter $t_0$ is set to
\begin{equation}
0\leqslant t_0=t_{+}\bigg{(}1-\sqrt{1-\frac{t_{-}}{t_+}}\bigg{)}\leqslant t_{-}.
\end{equation}
The $m^{f}_{\text{pole}}$ is collected in Table~\ref{tab:polemass}.

\begin{table}[htbp]\centering
\caption{The pole masses of the form factors in Eq.~\eqref{eq:fitness}, where the $0^-$, $1^-$, and $1^+$ masses are taken from the PDG~\cite{ParticleDataGroup:2020ssz}, while the $0^+$ mass is taken from the LQCD calculation~\cite{Lang:2015hza}.}
\label{tab:polemass}
\renewcommand\arraystretch{1.05}
\begin{tabular*}{76mm}{c@{\extracolsep{\fill}}cc}
\toprule[1pt]
\toprule[0.5pt]
$f$                                 & $J^{P}$   &$m^{f}_{\text{pole}}(\text{GeV})$ \\
\toprule[0.5pt]
$f_t^V$                             & $0^{+}$   & $5.711$ \\
$f_g^V$, $f_{g}^T$   & $1^{-}$   & $5.415$ \\
$f_t^A$                             & $0^{-}$   & $5.367$ \\
$f_0^A$, $f_{\bot}^A$, $f_{g}^A$, $f_{0}^{T5}$, $f_{\bot}^{T5}$, $f_{g}^{T5}$    & $1^{+}$   & $5.828$ \\
\bottomrule[0.5pt]
\bottomrule[1pt]
\end{tabular*}
\end{table}
\item[{\bf 2.}]
For the form factors $f_{0,\bot}^{V}$ and $f_{0,\bot}^{T}$, we use the form as
\begin{equation}
f(q^2)=\frac{1}{1-q^2/m_{-}^2}\left[a_0^f+a_1^fz^f(q^2)+a_2^f(z^f(q^2))^2\right],
\label{eq:fitness2}
\end{equation}
where $m_{-}=m_{\Lambda_b}-m_{\Lambda^{*}}$.
\end{enumerate}
}

\begin{figure*}[htbp]\centering
  \begin{tabular}{lll}
  \includegraphics[width=54mm]{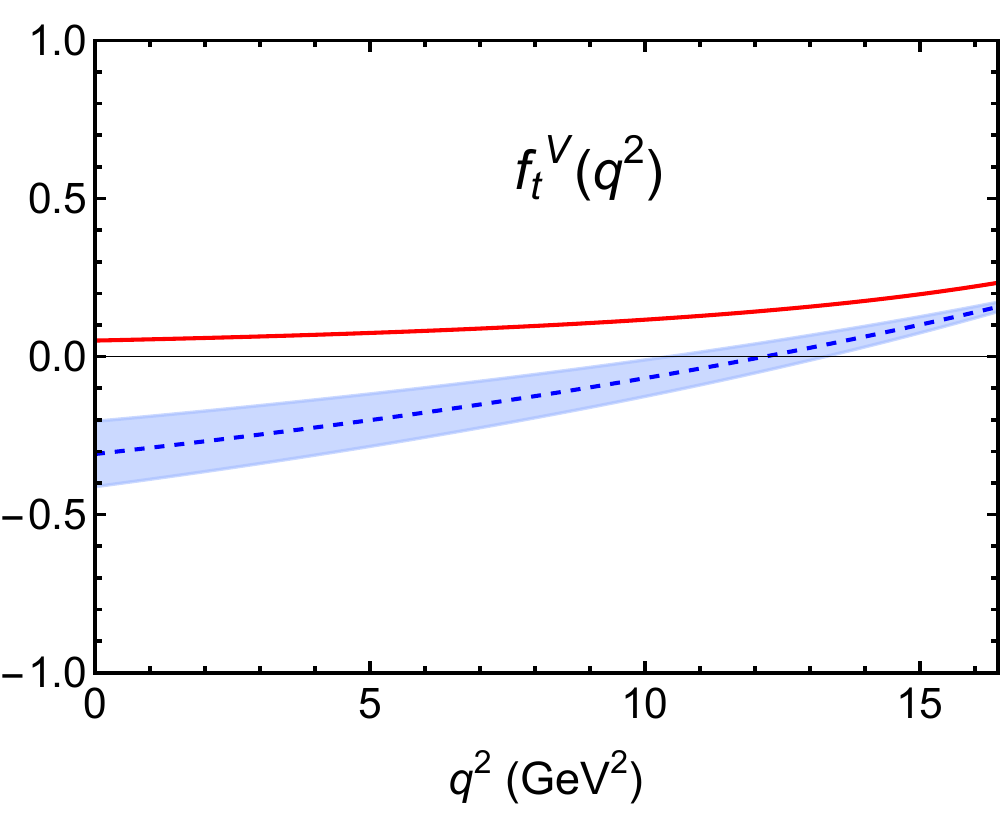}
  \includegraphics[width=54mm]{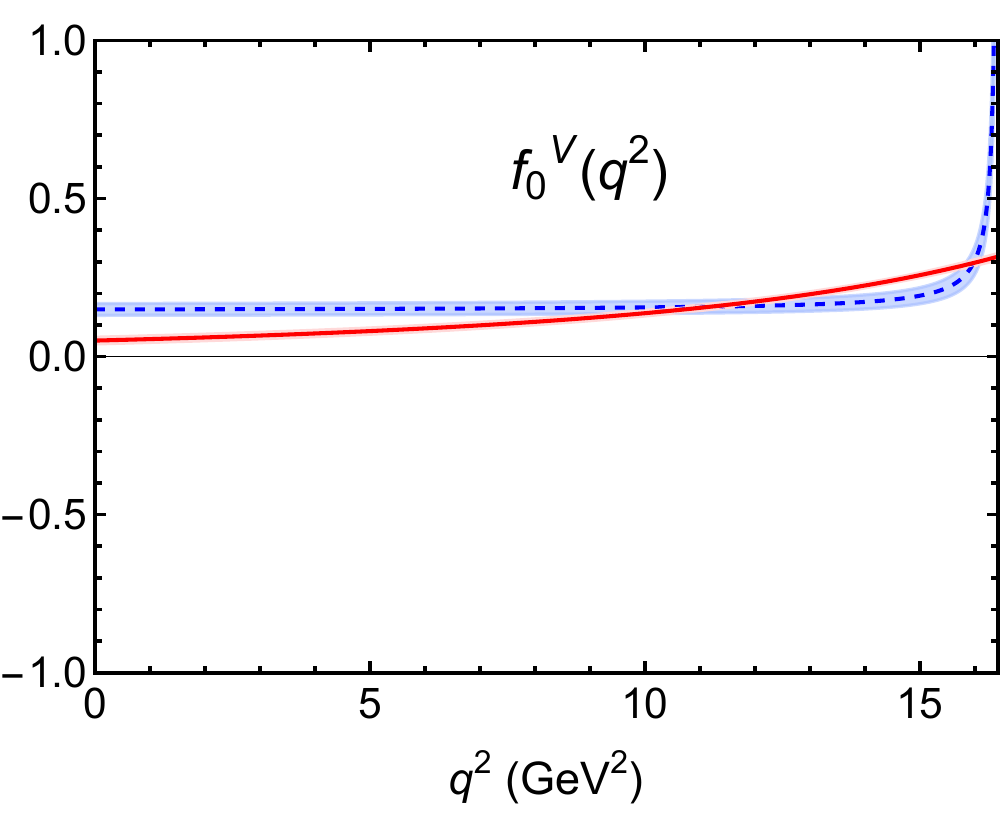}
  \includegraphics[width=54mm]{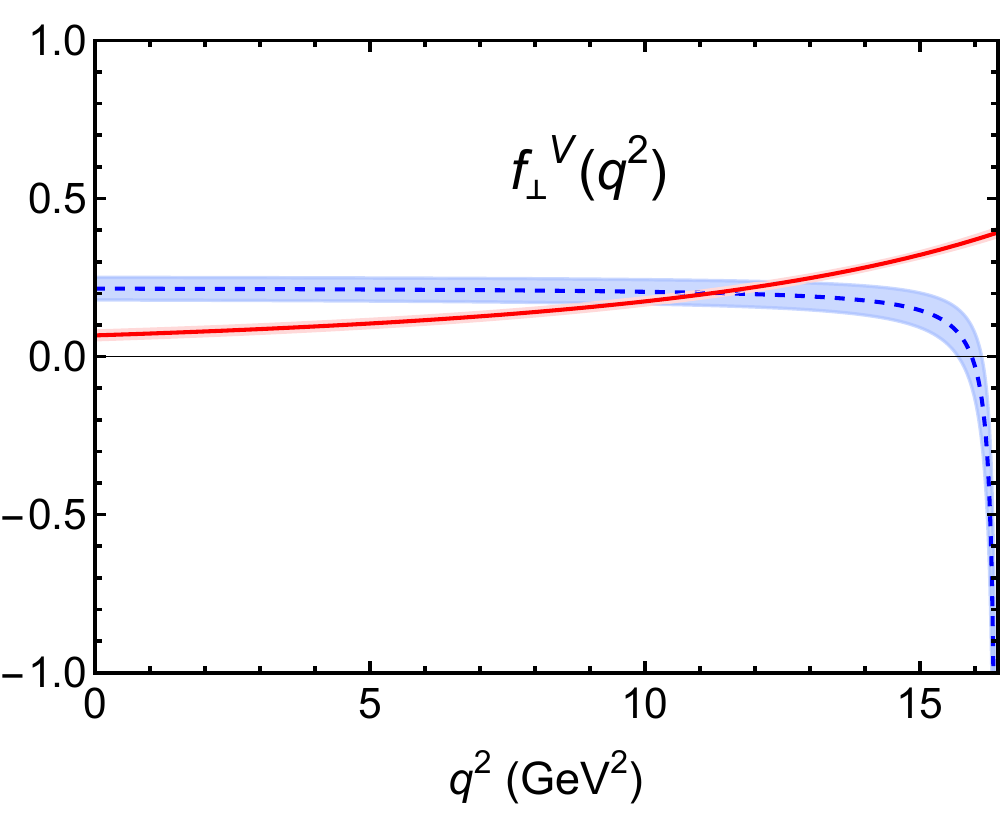}\\
  \includegraphics[width=54mm]{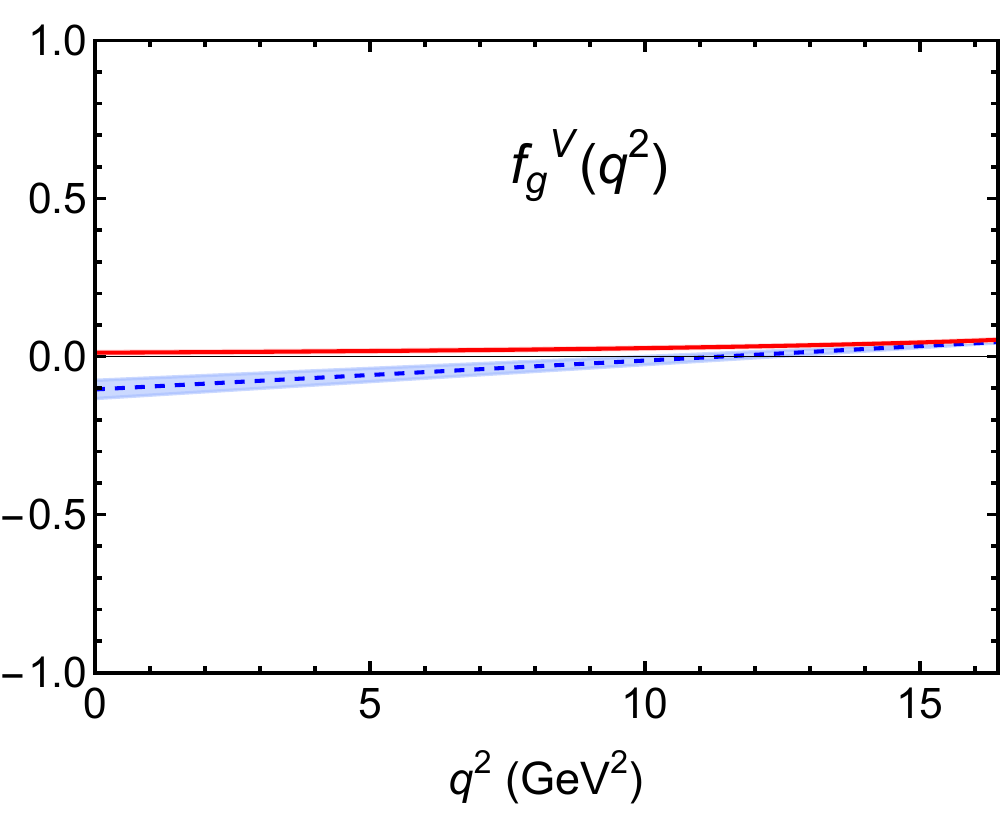}
  \includegraphics[width=54mm]{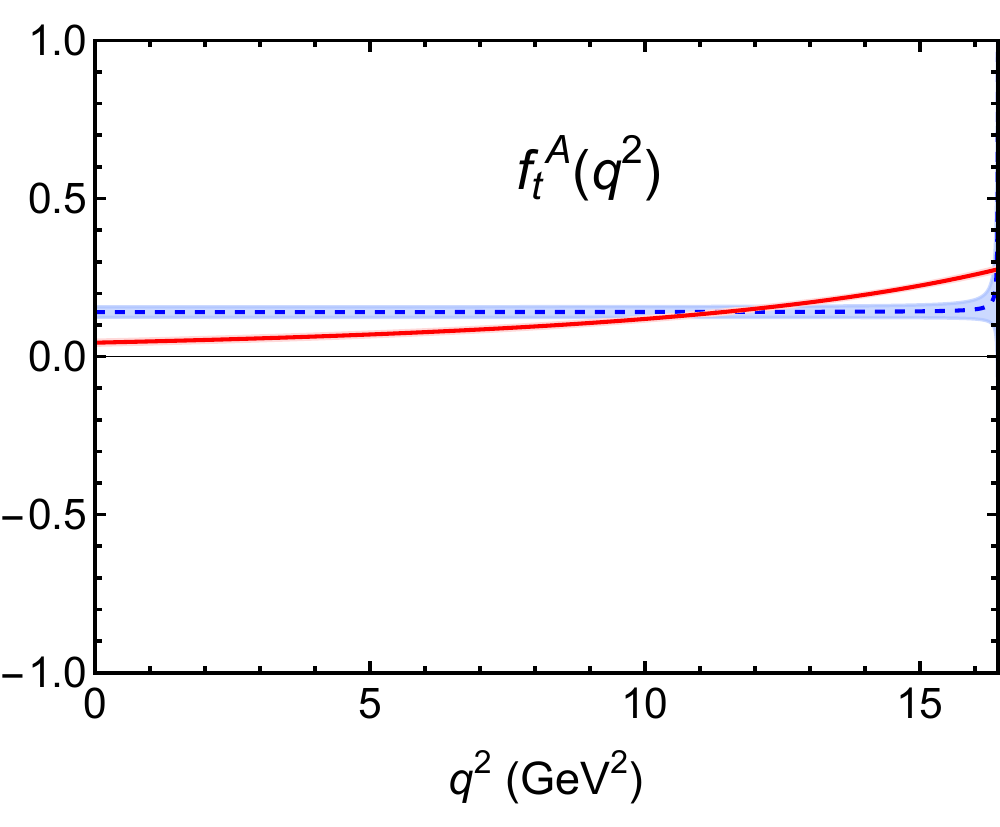}
  \includegraphics[width=54mm]{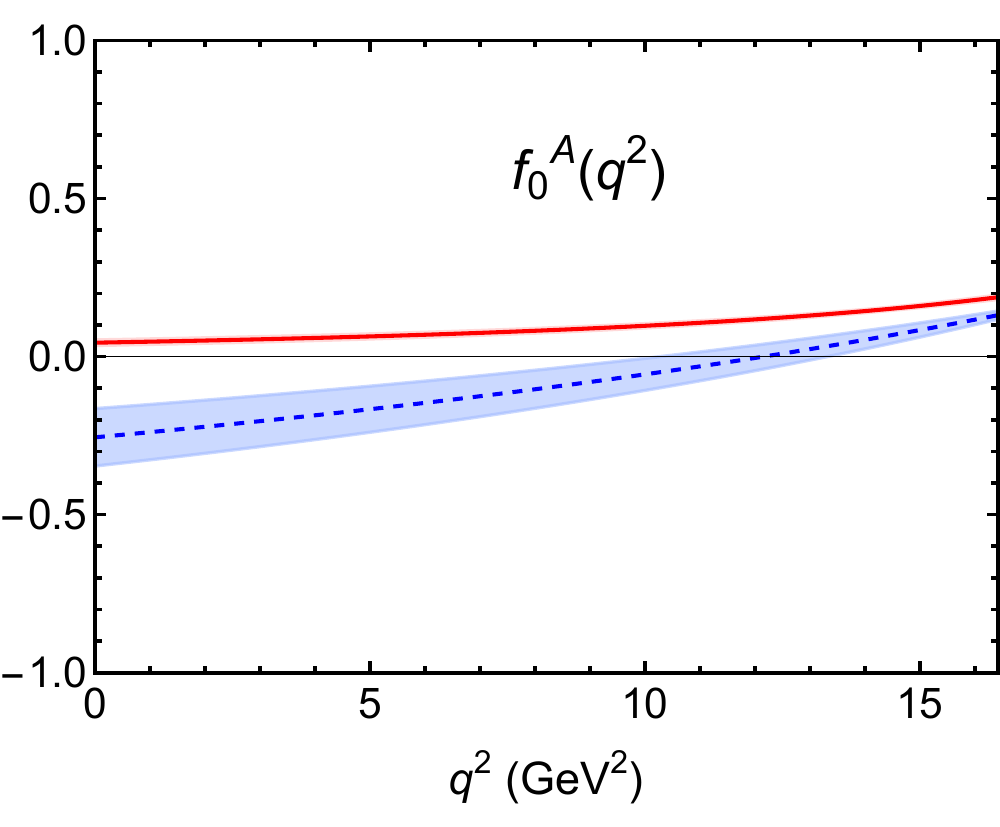}\\
  \includegraphics[width=54mm]{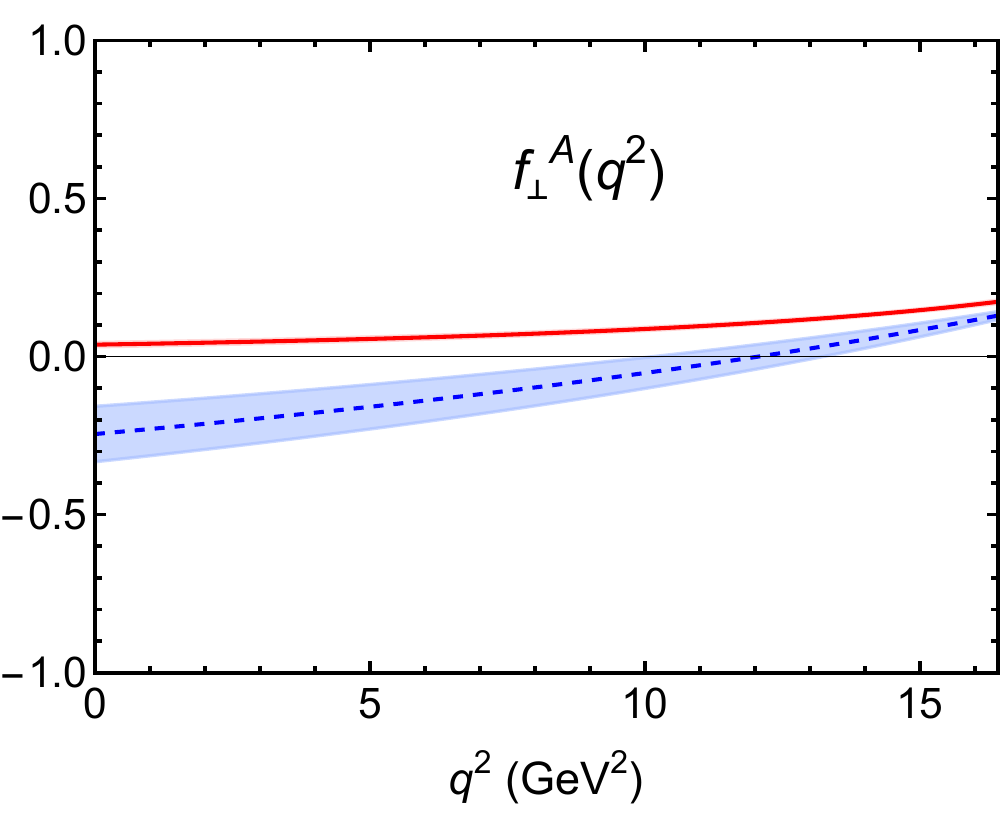}
  \includegraphics[width=54mm]{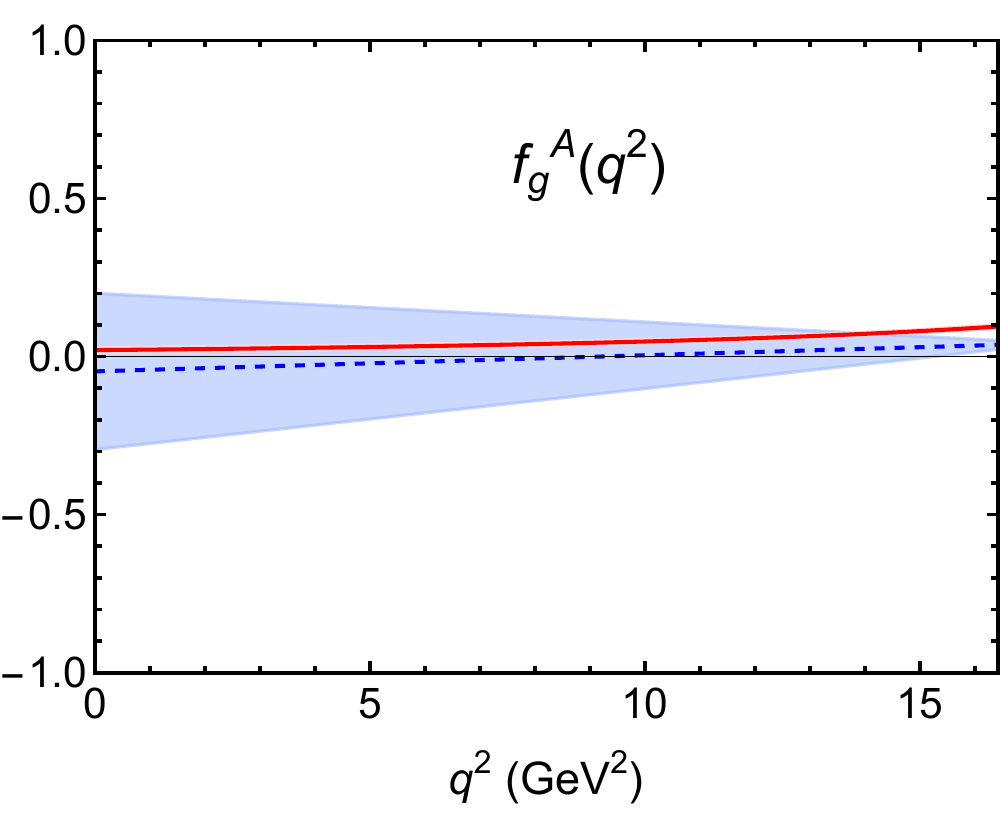}
  \includegraphics[width=54mm]{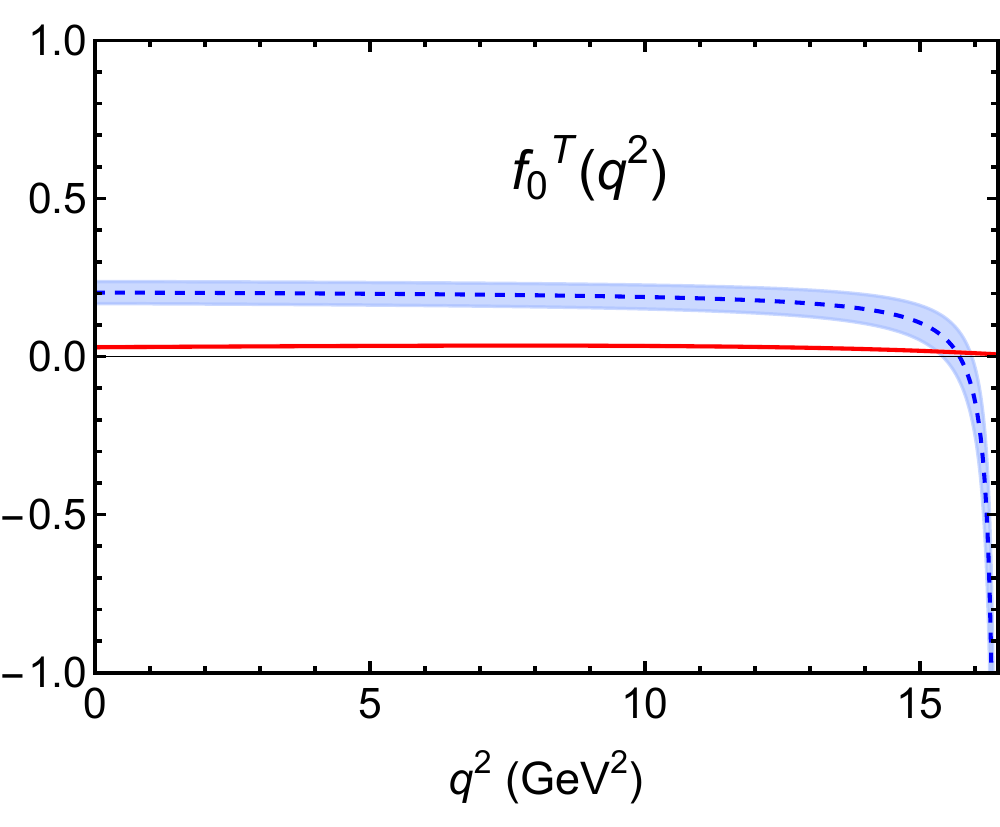}\\
  \includegraphics[width=54mm]{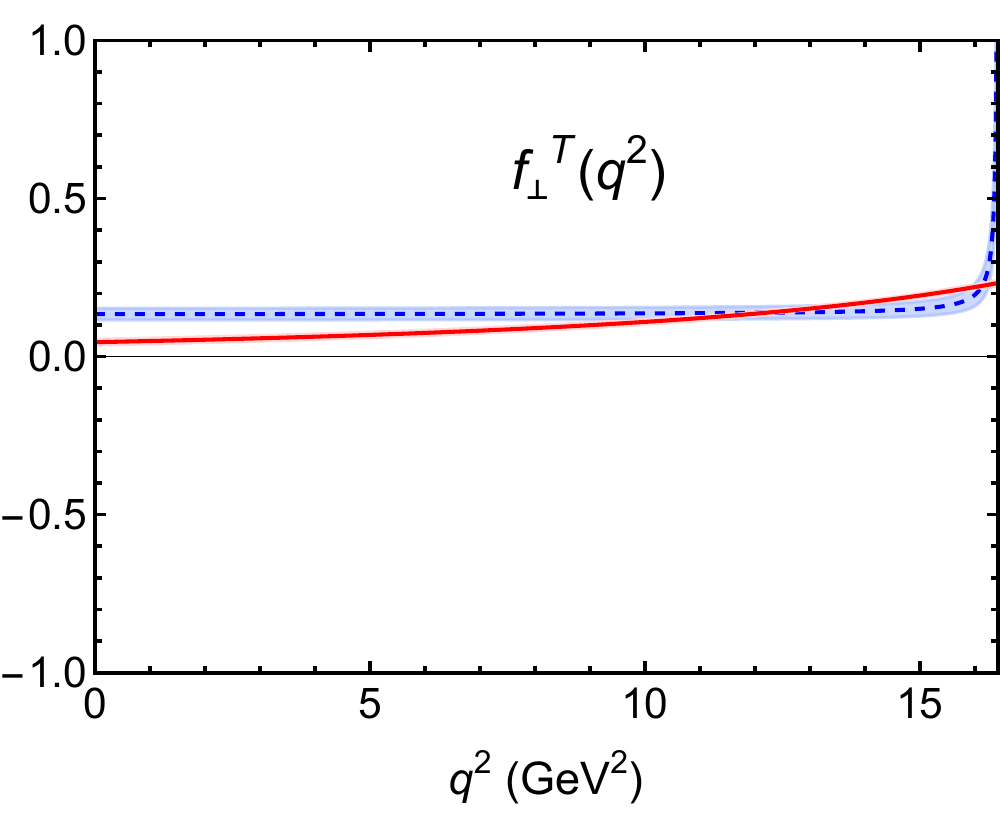}
  \includegraphics[width=54mm]{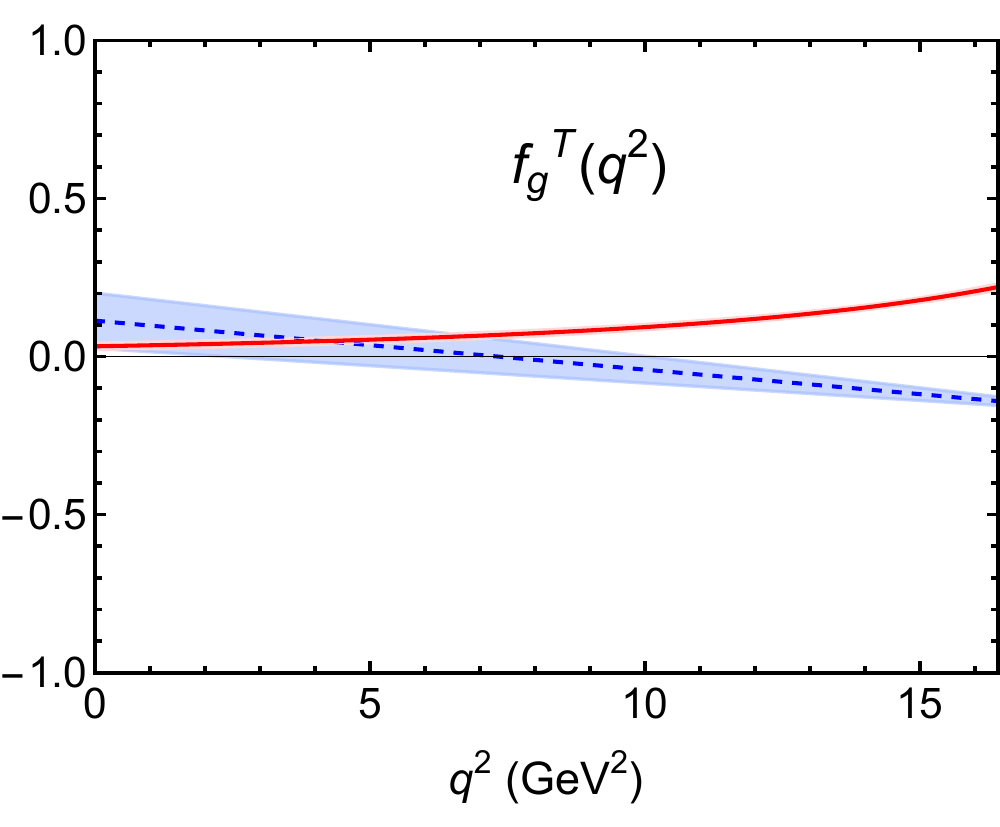}
  \includegraphics[width=54mm]{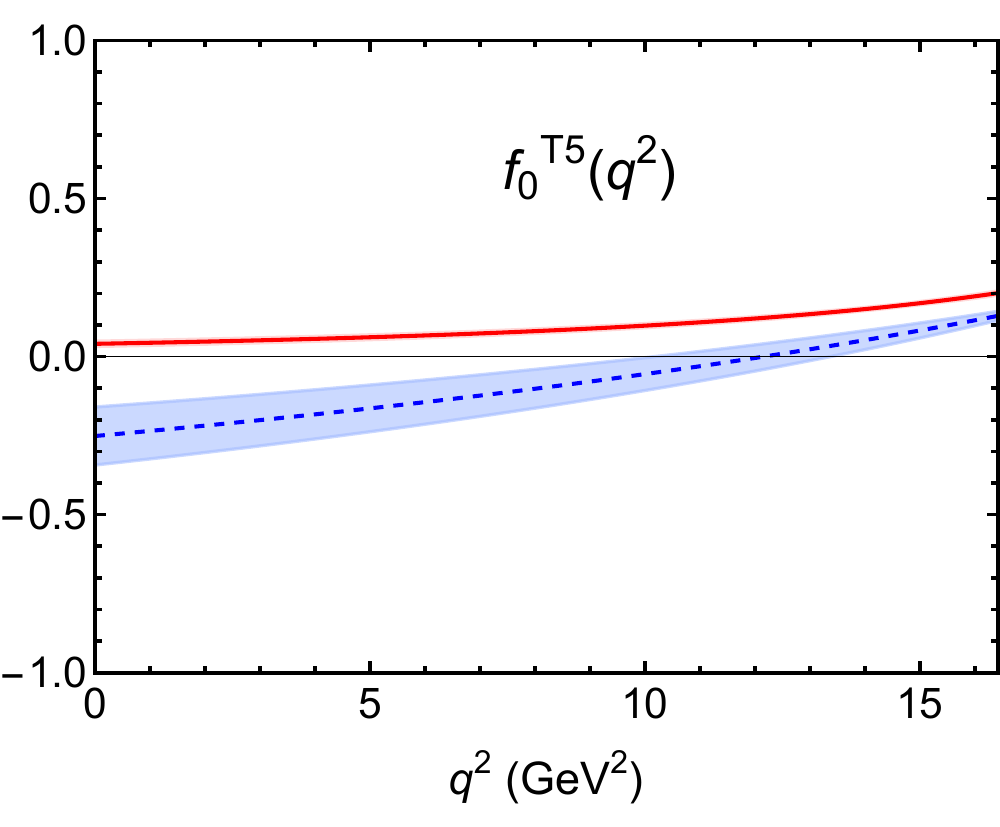}\\
  \includegraphics[width=54mm]{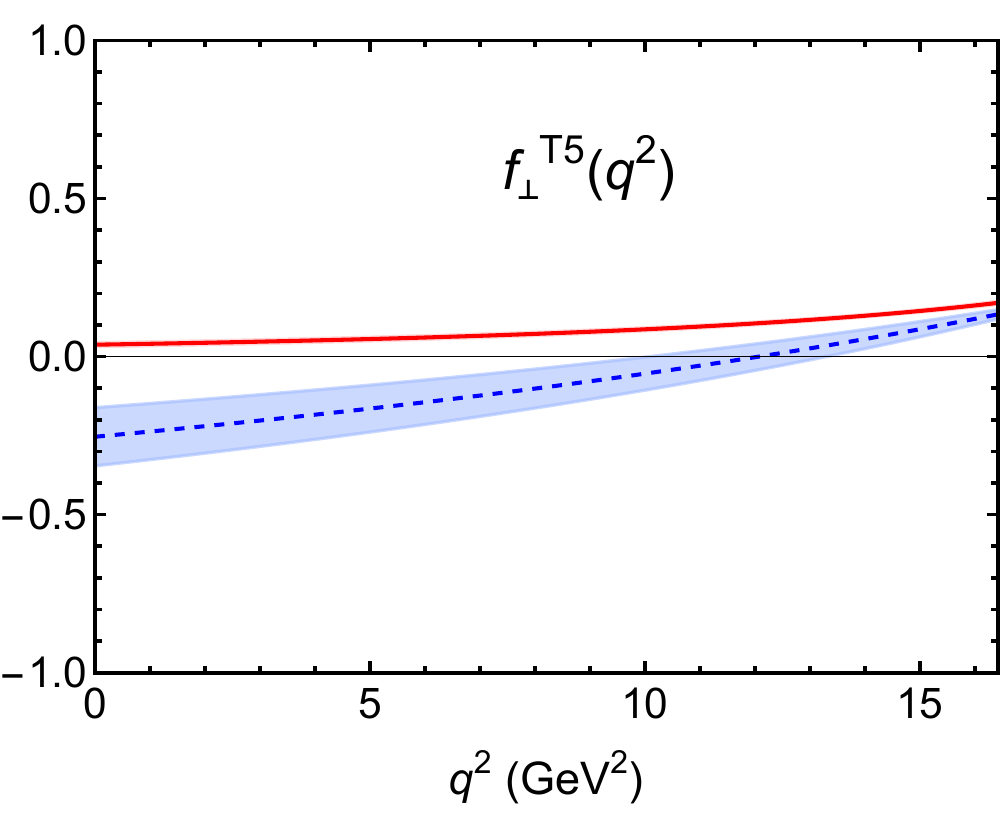}
  \includegraphics[width=54mm]{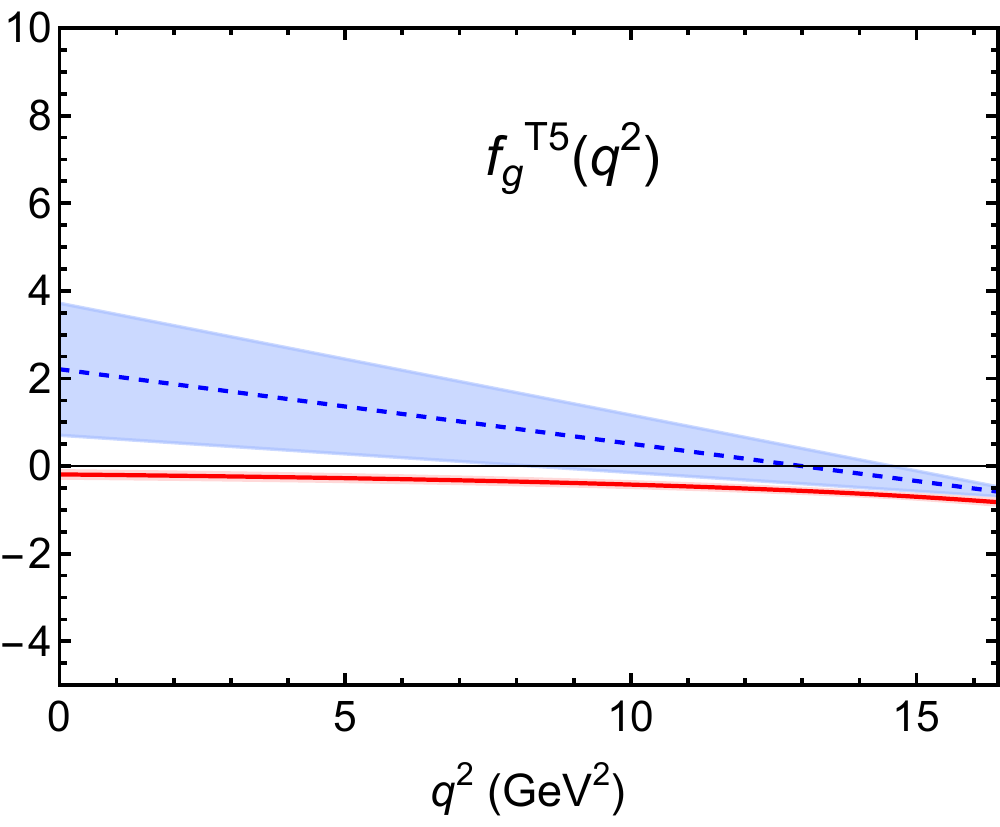}
  \end{tabular}
  \caption{The $q^2$ dependence of the form factors of the vector, axial-vector, tensor, and pseudotensor type currents of the $\Lambda_b\to\Lambda^\ast$ transition, where the red solid curves are central values, and the light red bands are the corresponding errors. The units of the form factors are neglected here.}
\label{fig:ffs}
\end{figure*}

To determine the parameters $a_{0}^{f}$, $a_{1}^{f}$, and $a_{2}^{f}$, we numerically calculate 24 points for each form factor by Eqs.~(\ref{eq:ffs05})-(\ref{eq:ffs08}) from $q^2=-q_{\text{max}}^2$ to $q^2=-0.01\ \text{GeV}^2$ in the spacelike region, and then fit them using Eq.~\eqref{eq:fitness} and Eq.~\eqref{eq:fitness2} with the MINUIT program.
The extrapolated parameters for the form factors of $\Lambda_b\to\Lambda^\ast$ are collected in Table~\ref{tab:ffs}. The $q^2$ dependence of the concerned form factors is shown in Fig.~\ref{fig:ffs}.

{However, as discussed earlier, the less singular behaviors of $f_{0,\bot}^{V}$ and $f_{0,\bot}^{T}$ in the small-recoil limit are also not forbidden. Therefore, in this work, we also use the formula in Eq.~\eqref{eq:fitness} to perform the extrapolation of the form factors $f_{0,\bot}^{V}$ and $f_{0,\bot}^{T}$ again. This extrapolation scheme gives different results at the $q^{2}=q_{\text{max}}^{2}$ point for the four form factors, but has no effect on other form factors compared with the previous scheme. For clarification, we compare our results of the form factors in the $q^{2}=q_{\text{max}}^{2}$ point with the two different extrapolation schemes in Table~\ref{tab:ffscomparison}. Finally, it should be emphasized that there is no established procedure for the extrapolation. The experimental measurement of $\Lambda_b\to\Lambda(1520)\ell^{+}\ell^{-}$ by the LHCb Collaboration can test the different extrapolation schemes.}

\begin{table*}[htbp]\centering
\caption{The form factors of the $\Lambda_b\to\Lambda^\ast$ transition in the standard light-front quark model.}
\label{tab:ffs}
\renewcommand\arraystretch{1.05}
\begin{tabular*}{160mm}{c@{\extracolsep{\fill}}ccccc}
\toprule[1pt]
\toprule[0.5pt]
Parameter &Value &Parameter &Value &Parameter &Value\\
\midrule[0.5pt]
$a_0^{f_t^V}$     &$(0.1041\pm0.0036)~\text{GeV}^{-1}$   &$a_1^{f_t^V}$   &$(-0.4493\pm0.0375)~\text{GeV}^{-1}$   &$a_2^{f_t^V}$   &$(0.5425\pm0.0954)~\text{GeV}^{-1}$\\
$a_0^{f_0^V}$     &$(0.0850\pm0.0037)~\text{GeV}^{-1}$   &$a_1^{f_0^V}$   &$(-0.2465\pm0.0386)~\text{GeV}^{-1}$   &$a_2^{f_0^V}$   &$(0.0637\pm0.0984)~\text{GeV}^{-1}$\\
$a_0^{f_\bot^V}$  &$(0.1538\pm0.0046)~\text{GeV}^{-1}$   &$a_1^{f_\bot^V}$ &$(-0.7505\pm0.0478)~\text{GeV}^{-1}$  &$a_2^{f_\bot^V}$ &$(1.0292\pm0.1210)~\text{GeV}^{-1}$\\
$a_0^{f_g^V}$     &$0.0223\pm0.0001$   &$a_1^{f_g^V}$   &$-0.0807\pm0.0003$   &$a_2^{f_g^V}$   &$0.0798\pm0.0031$\\
$a_0^{f_t^A}$     &$(0.1052\pm0.0026)~\text{GeV}^{-1}$   &$a_1^{f_t^A}$   &$(-0.5337\pm0.0263)~\text{GeV}^{-1}$   &$a_2^{f_t^A}$   &$(0.7542\pm0.0665)~\text{GeV}^{-1}$\\
$a_0^{f_0^A}$     &$(0.0878\pm0.0028)~\text{GeV}^{-1}$   &$a_1^{f_0^A}$   &$(-0.3647\pm0.0293)~\text{GeV}^{-1}$   &$a_2^{f_0^A}$   &$(0.4197\pm0.0747)~\text{GeV}^{-1}$\\
$a_0^{f_\bot^A}$  &$(0.0804\pm0.0022)~\text{GeV}^{-1}$   &$a_1^{f_\bot^A}$ &$(-0.3619\pm0.0227)~\text{GeV}^{-1}$  &$a_2^{f_\bot^A}$ &$(0.4573\pm0.0578)~\text{GeV}^{-1}$\\
$a_0^{f_g^A}$     &$0.0441\pm0.0023$   &$a_1^{f_g^A}$   &$-0.2012\pm0.0240$   &$a_2^{f_g^A}$   &$0.2596\pm0.0605$\\
$a_0^{f_0^T}$     &$(-0.0178\pm0.0003)~\text{GeV}^{-1}$   &$a_1^{f_0^T}$   &$(0.5398\pm0.0037)~\text{GeV}^{-1}$    &$a_2^{f_0^T}$   &$(-1.4719\pm0.0098)~\text{GeV}^{-1}$\\
$a_0^{f_\bot^T}$  &$(0.0565\pm0.0032)~\text{GeV}^{-1}$   &$a_1^{f_\bot^T}$ &$(-0.0233\pm0.0331)~\text{GeV}^{-1}$  &$a_2^{f_\bot^T}$ &$(-0.3596\pm0.0853)~\text{GeV}^{-1}$\\
$a_0^{f_g^T}$     &$(0.0851\pm0.0034)~\text{GeV}$   &$a_1^{f_g^T}$   &$(-0.4603\pm0.0334)~\text{GeV}$   &$a_2^{f_g^T}$   &$(0.6616\pm0.0805)~\text{GeV}$\\
$a_0^{f_0^{T5}}$  &$(0.0923\pm0.0026)~\text{GeV}^{-1}$   &$a_1^{f_0^{T5}}$ &$(-0.4516\pm0.0272)~\text{GeV}^{-1}$  &$a_2^{f_0^{T5}}$ &$(0.6337\pm0.0687)~\text{GeV}^{-1}$\\
$a_0^{f_\bot^{T5}}$ &$(0.0790\pm0.0020)~\text{GeV}^{-1}$ &$a_1^{f_\bot^{T5}}$ &$(-0.3482\pm0.0206)~\text{GeV}^{-1}$ &$a_2^{f_\bot^{T5}}$ &$(0.4288\pm0.0527)~\text{GeV}^{-1}$\\
$a_0^{f_g^{T5}}$  &$(-0.3839\pm0.0276)~\text{GeV}$  &$a_1^{f_g^{T5}}$ &$(1.6524\pm0.2814)~\text{GeV}$  &$a_2^{f_g^{T5}}$  &$(-2.1223\pm0.6945)~\text{GeV}$\\
\bottomrule[0.5pt]
\bottomrule[1pt]
\end{tabular*}
\end{table*}

As shown in Eqs.~(\ref{eq:ffs01})-(\ref{eq:ffs04}), we need eight (axial-)vector and six (pseudo-)tensor form factors to describe the matrix elements in question. The number can apparently be reduced in the heavy quark limit $m_b\to\infty$. We speak separately of  two different kinematic situations, i.e., the outgoing $\Lambda^\ast$ acts softly (the low-recoil limit) and acts energetically (the large-recoil limit). Accordingly, two effective theories, namely heavy quark effective theory (HQET) and soft-collinear effective theory (SCET), are developed to exploit the behaviors of the form factors.

In the low-recoil limit, where HQET is valid~\cite{Isgur:1989vq,Isgur:1990yhj,Isgur:1990pm,Mannel:1990vg}, the weak transition matrix element can be re-expressed by two Isgur-Wise functions as~\cite{Mannel:1990vg,Das:2020cpv,Bordone:2021bop}
\begin{equation}
\langle\Lambda^\ast(p^\prime)\vert\bar{s}\Gamma b\vert\Lambda_b(p)\rangle=
\bar{u}_{\Lambda}^\alpha(p^\prime)\upsilon_{\alpha}\left[\zeta_1(\omega)+\slashed{\upsilon}\zeta_2(\omega)\right]\Gamma u_{\Lambda_b}(p).
\end{equation}
Here, $\Gamma$ is an arbitrary Dirac structure, and $\omega=\upsilon\cdot\upsilon^\prime=(m_{\Lambda_b}^2+m_{\Lambda^\ast}^2-q^2)/(2m_{\Lambda_b}m_{\Lambda^\ast})$, where $\upsilon=p/m_{\Lambda_b}$ and $\upsilon^\prime=p^\prime/m_{\Lambda^\ast}$ represent the four velocities of the bottom baryon and hyperon, respectively. The eight form factors are derived as two independent form factors $\zeta_1(\omega)$ and $\zeta_2(\omega)$. In the low-recoil limit this gives $q^2\to q_{\text{max}}^2\equiv(m_{\Lambda_b}-m_{\Lambda^\ast})^2$ (or $\omega\to1$). With
slightly different definitions of the form factors in Refs.~\cite{Boer:2014kda,Feldmann:2011xf,Mannel:2011xg},we have~\cite{Descotes-Genon:2019dbw}
\begin{equation}
\begin{split}
f_{t}^{V}(q_{\text{max}}^2)& \simeq f_{0}^{A}(q_{\text{max}}^2) \simeq f_{\bot}^{A}(q_{\text{max}}^2) \\
& \simeq f_{0}^{T5}(q_{\text{max}}^2) \simeq f_{\bot}^{T5}(q_{\text{max}}^2) \simeq \big{[}\zeta_1(1)+\zeta_2(1)\big{]}/m_{\Lambda_b},\\
f_{0}^{V}(q_{\text{max}}^2)& \simeq f_{t}^{A}(q_{\text{max}}^2) \simeq f_{\bot}^{V}(q_{\text{max}}^2) \\
& \simeq f_{0}^{T}(q_{\text{max}}^2) \simeq f_{\bot}^{T}(q_{\text{max}}^2) \simeq \big{[}\zeta_1(1)-\zeta_2(1)\big{]}/m_{\Lambda_b},\\
f_{g}^{V}(q_{\text{max}}^2)& \simeq f_{g}^{A}(q_{\text{max}}^2) \simeq f_{g}^{T}(q_{\text{max}}^2) \simeq f_{g}^{T5}(q_{\text{max}}^2) \simeq 0,
\label{eq:HQEF}
\end{split}
\end{equation}
while in the large-recoil limit where SCET is valid, we have~\cite{Mannel:2011xg,Feldmann:2011xf,Wang:2011uv}
\begin{equation}
\langle\Lambda^\ast(p^\prime)\vert\bar{s}\Gamma b\vert\Lambda_b(p)\rangle=
\bar{u}_{\Lambda}^{\alpha}(p^\prime)\upsilon_{\alpha}\left[\zeta(\omega)\right]\Gamma u_{\Lambda_b}(p),
\end{equation}
where $\zeta(\omega)$ is the only remaining form factor. This gives, in the large-recoil limit, $q^2\to0$ (or $\omega\to(m_{\Lambda_b}^2+m_{\Lambda^\ast}^2)/(2m_{\Lambda_b}m_{\Lambda^\ast})$),
\begin{equation}
\begin{split}
f_{t}^{V}(0) &\simeq f_{0}^{V}(0) \simeq f_{\bot}^{V}(0) \simeq f_{t}^{A}(0) \simeq f_{0}^{A}(0) \simeq f_{\bot}^{A}(0) \\
&\simeq f_{0}^{T}(0) \simeq f_{\bot}^{T}(0) \simeq f_{0}^{T5}(0) \simeq f_{\bot}^{T5}(0) \\
&\simeq \zeta\big{(}\frac{m_{\Lambda_b}^2+m_{\Lambda^\ast}^2}{2m_{\Lambda_b}m_{\Lambda^\ast}}\big{)}\big{/}m_{\Lambda_b},
\label{eq:SCEF}
\end{split}
\end{equation}
and four $f_g$ form factors will disappear. {From Fig.~\ref{fig:ffs}, we can see that apart from the $f_{g}^{T(T5)}(q^2)$, which deviates from the predictions, the remaining calculated form factors are consistent with the requirements of HQET and SCET.}

In addition, Bordone has completed the heavy quark expansion (HQE) calculation of the $\Lambda_b\to\Lambda^{*}$ form factors beyond the leading order~\cite{Bordone:2021bop}. {At the zero-recoil limit, the HQE predicts the ratios of the form factors, which are independent of the Isgur-Wise functions, as \cite{Bordone:2021bop}
\begin{equation}
\begin{split}
\frac{F_{1/2,0}}{F_{1/2,\bot}}=&\frac{F_{1/2,0}}{F_{3/2,\bot}}=-2\frac{m_{\Lambda_{b}}-m_{\Lambda^{*}}}{m_{\Lambda_{b}}+m_{\Lambda^{*}}}=-1.15,\\
\frac{F_{1/2,\bot}}{F_{3/2,\bot}}=&1,~~\frac{T_{1/2,0}}{T_{1/2,\bot}}=-2\frac{m_{\Lambda_{b}}+m_{\Lambda^{*}}}{m_{\Lambda_{b}}-m_{\Lambda^{*}}}=-3.48,\\
\frac{T_{1/2,0}}{T_{3/2,\bot}}=&\frac{2m_{\Lambda^{*}}}{m_{\Lambda_{b}}-m_{\Lambda^{*}}}=0.74,~~
\frac{T_{1/2,\bot}}{T_{3/2,\bot}}=-\frac{m_{\Lambda^{*}}}{m_{\Lambda_{b}}+m_{\Lambda^{*}}}=-0.21.
\end{split}
\end{equation}
Note that the form factor base used in Ref.~\cite{Bordone:2021bop} is different from ours. By using the conversions collected in Appendix B of Ref.~\cite{Bordone:2021bop} and Eq.~\eqref{eq:relations}, we can get our results of these ratios as:
\begin{equation}
\begin{split}
\frac{F_{1/2,0}^{\small{\text{this~work}}}}{F_{1/2,\bot}^{\small{\text{this~work}}}}=&\frac{f_{0}^{V}}{f_{\bot}^{V}}=0.521\pm0.026,\\
\frac{F_{1/2,\bot}^{\small{\text{this~work}}}}{F_{3/2,\bot}^{\small{\text{this~work}}}}=&-\frac{s_{-}}{m_{\Lambda^{*}}}\frac{f_{\bot}^{V}}{f_{g}^{V}}=-33.8\pm0.9,\\
\frac{F_{1/2,0}^{\small{\text{this~work}}}}{F_{3/2,\bot}^{\small{\text{this~work}}}}=&-\frac{s_{-}}{m_{\Lambda^{*}}}\frac{f_{0}^{V}}{f_{g}^{V}}=-17.6\pm0.7,\\
\frac{T_{1/2,0}^{\small{\text{this~work}}}}{T_{1/2,\bot}^{\small{\text{this~work}}}}=&\frac{f_{0}^{T}}{f_{\bot}^{T}}=-0.63\pm0.04,\\
\frac{T_{1/2,0}^{\small{\text{this~work}}}}{T_{3/2,\bot}^{\small{\text{this~work}}}}=&s_{-}\frac{f_{0}^{T}}{f_{g}^{T}}=-2.56\pm0.09,\\
\frac{T_{1/2,\bot}^{\small{\text{this~work}}}}{T_{3/2,\bot}^{\small{\text{this~work}}}}=&s_{-}\frac{f_{\bot}^{T}}{f_{g}^{T}}=4.09\pm0.28.
\end{split}
\end{equation}
Our results are very different from those of the HQE.}

In addition, we also compare our results of the form factors with the NRQM~\cite{Mott:2015zma} and the LQCD~\cite{Meinel:2021mdj} at $q^2=0$ and $q^2=q_{\text{max}}^2$ endpoints in Table~\ref{tab:ffscomparison}. Until now, less work has been done on the $\Lambda_b\to\Lambda(1520)$ transition, so more theoretical work is needed to validate these form factors.

\begin{table*}[htbp]\centering
\caption{Theoretical predictions for the form factors of $\Lambda_b\to\Lambda(1520)$ at the endpoints of $q^2=0$ and $q^2=q_{\text{max}}^2$ using different approaches.}
\label{tab:ffscomparison}
\renewcommand\arraystretch{1.20}
\begin{threeparttable}
\begin{tabular*}{172mm}{c@{\extracolsep{\fill}}cccccccc}
\toprule[1pt]
\toprule[0.5pt]
                    &$\text{This\ work}$    &$\text{NRQM}$~\cite{Mott:2015zma}   &$\text{LQCD}$~\cite{Meinel:2021mdj}
&                    &$\text{This\ work}^{[a]}$  &$\text{This\ work}^{[b]}$   &$\text{NRQM}$~\cite{Mott:2015zma}   &$\text{LQCD}$~\cite{Meinel:2021mdj}\\
\midrule[0.5pt]
$f_{t}^{V}(0)$      &$0.051\pm0.007$         &$0.0029$            &$-0.1523\pm0.0530$
&$f_{t}^{V}(q_{\text{max}}^{2})$         &$0.244\pm0.008$  &$0.244\pm0.008$   &$\infty$   &$0.1726\pm0.0138$\\
$f_{0}^{V}(0)$      &$0.051\pm0.007$         &$0.0029$            &$0.0714\pm0.0078$
&$f_{0}^{V}(q_{\text{max}}^{2})$         &$\infty$         &$0.336\pm0.008$   &$\infty$ &$\infty$\\
$f_{\bot}^{V}(0)$   &$0.067\pm0.009$         &$0.0042$            &$0.1093\pm0.0151$
&$f_{\bot}^{V}(q_{\text{max}}^{2})$      &$\infty$         &$0.416\pm0.011$   &$\infty$ &$\infty$\\
$f_{g}^{V}(0)$      &$0.0123\pm0.0001$         &$-0.0002$             &$-0.0385\pm0.0138$
&$f_{g}^{V}(q_{\text{max}}^{2})$         &$0.0582\pm0.0002$ &$0.0582\pm0.0002$  &$0.0323$ &$0.0481\pm0.0028$\\
$f_{t}^{A}(0)$      &$0.044\pm0.005$         &$0.0031$            &$0.0705\pm0.0060$
&$f_{t}^{A}(q_{\text{max}}^{2})$         &$0.293\pm0.006$  &$0.293\pm0.006$   &$\infty$ &$0.1695\pm0.0145$\\
$f_{0}^{A}(0)$      &$0.044\pm0.005$         &$0.0031$            &$-0.1283\pm0.0471$
&$f_{0}^{A}(q_{\text{max}}^{2})$         &$0.197\pm0.006$  &$0.197\pm0.006$    &$0.1791$ &$0.1449\pm0.0109$\\
$f_{\bot}^{A}(0)$   &$0.038\pm0.004$         &$0.0033$            &$-0.1260\pm0.0471$
&$f_{\bot}^{A}(q_{\text{max}}^{2})$      &$0.182\pm0.005$  &$0.182\pm0.005$    &$0.1637$ &$0.1430\pm0.0109$\\
$f_{g}^{A}(0)$      &$0.020\pm0.004$         &$0.0004$            &$0.0086\pm0.0830$
&$f_{g}^{A}(q_{\text{max}}^{2})$         &$0.100\pm0.005$  &$0.100\pm0.005$    &$0.0532$ &$0.0415\pm0.0145$\\
$f_{0}^{T}(0)$      &$0.0294\pm0.0007$         &$0.0038$            &$0.0986\pm0.0151$
&$f_{0}^{T}(q_{\text{max}}^{2})$         &$\infty$   &$0.0026\pm0.0021$   &$\infty$ &$\infty$\\
$f_{\bot}^{T}(0)$   &$0.046\pm0.006$         &$0.0030$            &$0.0690\pm0.0077$
&$f_{\bot}^{T}(q_{\text{max}}^{2})$      &$\infty$   &$0.246\pm0.007$   &$\infty$ &$\infty$\\
$f_{g}^{T}(0)$      &$0.032\pm0.006$         &$-0.0041$             &$0.0130\pm0.0376$
&$f_{g}^{T}(q_{\text{max}}^{2})$         &$0.234\pm0.008$  &$0.234\pm0.008$   &$-0.1509$ &$-0.1506\pm0.0127$\\
$f_{0}^{T5}(0)$     &$0.040\pm0.005$         &$0.0032$            &$-0.1207\pm0.0500$
&$f_{0}^{T5}(q_{\text{max}}^{2})$        &$0.211\pm0.005$  &$0.211\pm0.005$   &$0.1861$ &$0.1412\pm0.0127$\\
$f_{\bot}^{T5}(0)$  &$0.038\pm0.004$         &$0.0030$            &$-0.1333\pm0.0471$
&$f_{\bot}^{T5}(q_{\text{max}}^{2})$     &$0.178\pm0.004$  &$0.178\pm0.004$   &$0.1897$ &$0.1476\pm0.0127$\\
$f_{g}^{T5}(0)$     &$-0.19\pm0.05$        &$0.0072$            &$1.167\pm0.688$
&$f_{g}^{T5}(q_{\text{max}}^{2})$        &$-0.86\pm0.06$   &$-0.86\pm0.06$   &$-0.3268$ &$-0.5826\pm0.0918$\\
\bottomrule[0.5pt]
\bottomrule[1pt]
\end{tabular*}
\begin{tablenotes}
\footnotesize
\item[a] These results, listed in the fourth column, are obtained by the first extrapolation scheme. Here, the $f_{t,g}^{V}$, $f_{t,0,\bot,g}^{A}$, $f_{g}^{T}$ and $f_{0,\bot,g}^{T5}$ are extrapolated by Eq.~\eqref{eq:fitness}, while the $f_{0,\bot}^{V}$ and $f_{0,\bot}^{T}$ are extrapolated by Eq.~\eqref{eq:fitness2}.
\item[b] These results, shown in the fifth column, are from the second extrapolation scheme. Here, all form factors are extrapolated by Eq.~\eqref{eq:fitness}.
\end{tablenotes}
\end{threeparttable}
\end{table*}

\subsection{The branching ratio and angular observables}

With the above preparations, we will present our numerical results. The baryon and lepton masses used in our calculation are taken from the PDG~\cite{ParticleDataGroup:2020ssz}, as well as $\tau_{\Lambda_b}=1.470$ ps. We also use $\mathcal{B}_{\Lambda^\ast}\equiv\mathcal{B}\big{(}\Lambda^\ast\to N\bar{K}\big{)}=45\%$~\cite{ParticleDataGroup:2020ssz}. To compare with the experimental data, we examine a number of angular observables, including the $CP$-averaged normalized angular coefficients, the differential branching ratios, the lepton's forward-backward asymmetry $(A_{FB}^{\ell})$, and the transverse ($F_{T}$) and longitudinal ($F_{L}$) polarization fractions of the dilepton system.

First, we examine the $CP$-averaged normalized angular distributions
\begin{equation}
S_{i}=\frac{L_{i}+\bar{L}_{i}}{d(\Gamma+\bar{\Gamma})/dq^2},
\end{equation}
where the angular distributions $L_{i}$ and the differential decay width $d\Gamma/dq^2$ are defined in Eqs.~\eqref{eq:AbgularDistribution2} and \eqref{eq:branchingrate}, respectively. For the $CP$-conjugated mode, the corresponding expression for the angular decay distribution should be written as
\begin{equation}
\frac{d^4\bar{\Gamma}}{dq^2 d\cos\theta_{\Lambda^\ast} d\cos\theta_\ell d\phi}
=\frac{3}{8\pi}\sum_{i}\bar{L}_{i}(q^2)f_{i}(q^2,\theta_{\ell},\theta_{\Lambda^{*}},\phi),
\end{equation}
where $\bar{L}_{i}(q^2)$ can be obtained by doing the full conjugation for all weak phases in $L_{i}(q^2)$. We should also do the substitutions as
\begin{equation}
\begin{split}
L_{1c,2c}&\to-\bar{L}_{1c,2c},~~L_{1cc,1ss,2cc,2ss}\to\bar{L}_{1cc,1ss,2cc,2ss},\\
L_{3ss}&\to\bar{L}_{3ss},~~~L_{4ss}\to-\bar{L}_{4ss},\\
L_{5s}&\to-\bar{L}_{5s},~~~L_{5sc}\to\bar{L}_{5sc},\\
L_{6s}&\to\bar{L}_{6s},~~~L_{6sc}\to-\bar{L}_{6sc},
\end{split}
\end{equation}
where the minus sign is a result from the operations of $\theta_{\ell}\to\theta_{\ell}-\pi$ and $\phi\to-\phi$. The differential decay width of the conjugated mode is
\begin{equation}
\frac{d\bar{\Gamma}}{dq^2}=\frac{1}{3}(\bar{L}_{1cc}+2\bar{L}_{1ss}+2\bar{L}_{2cc}+4\bar{L}_{2ss}+2\bar{L}_{3ss}).
\end{equation}

\begin{figure*}[htbp]\centering
  \begin{tabular}{ccc}
  \includegraphics[width=56mm]{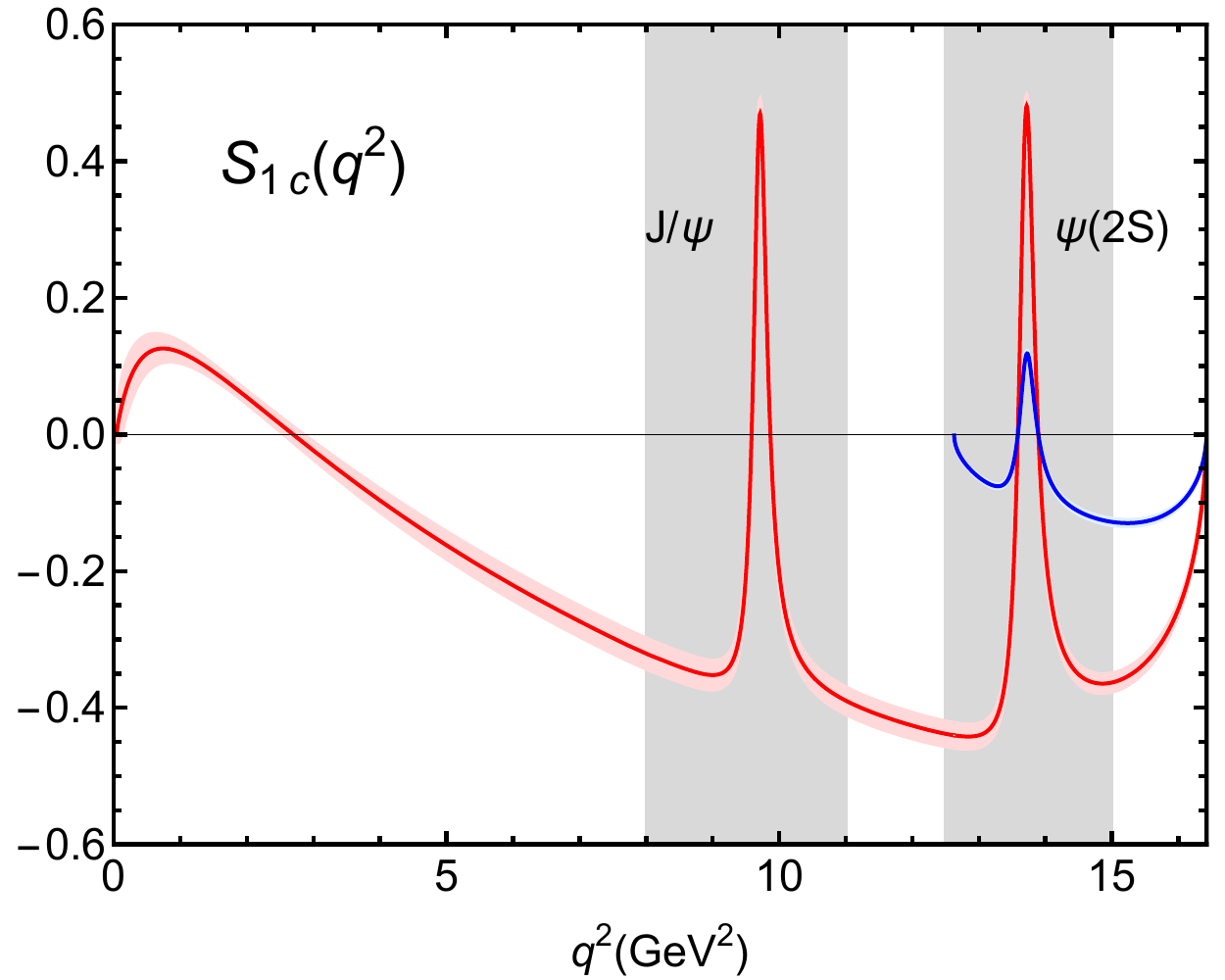}
  \includegraphics[width=56mm]{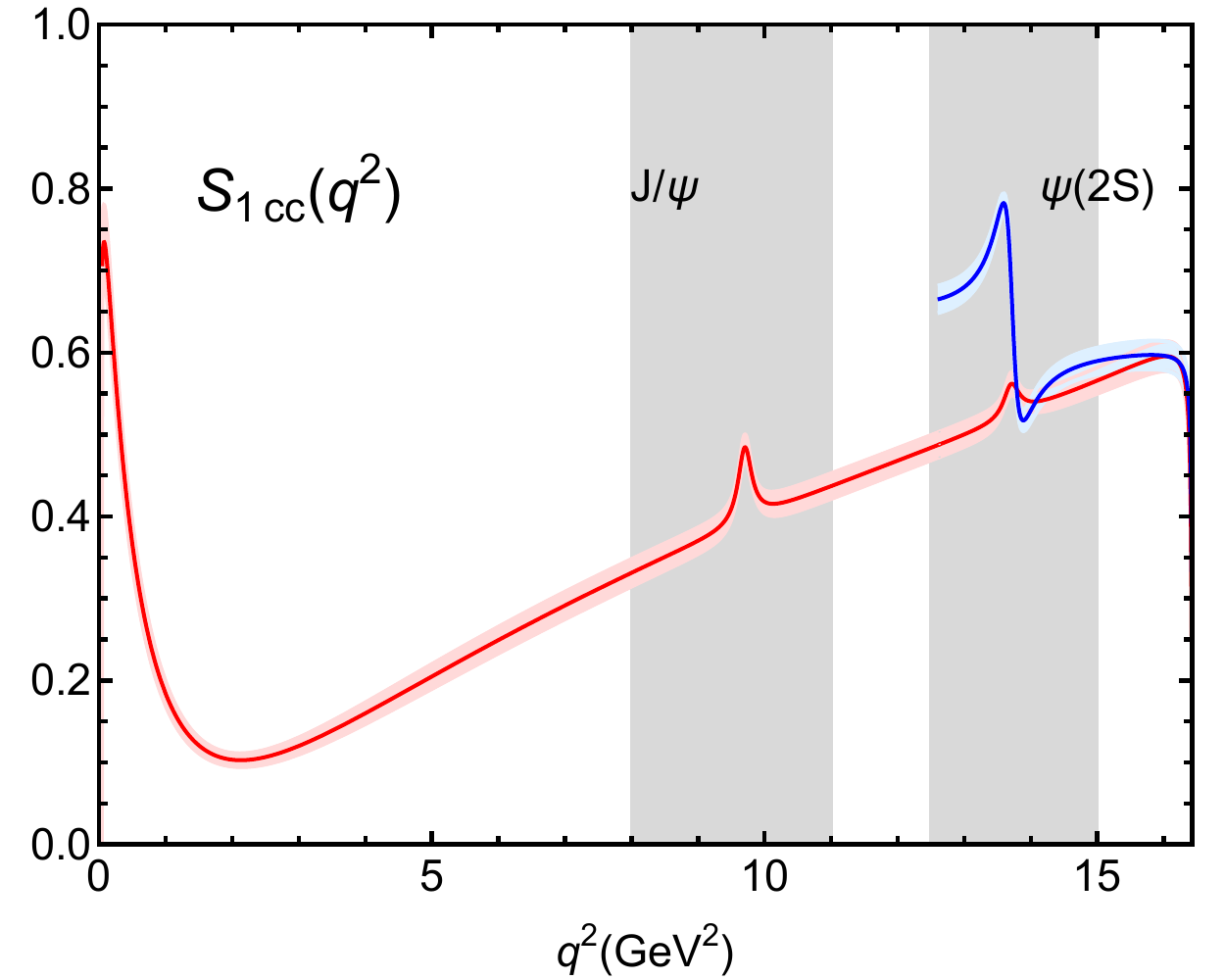}
  \includegraphics[width=56mm]{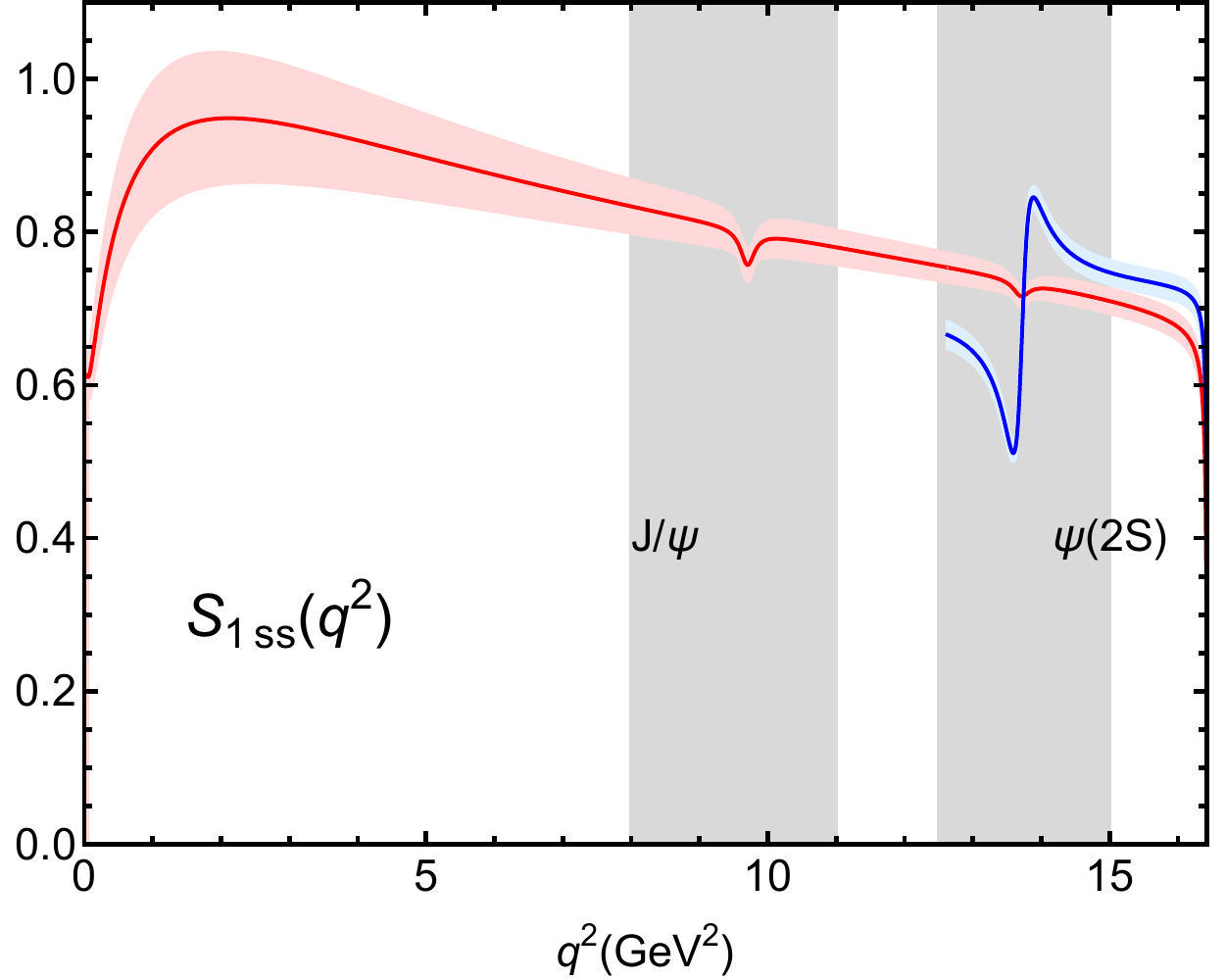}\\
  \includegraphics[width=56mm]{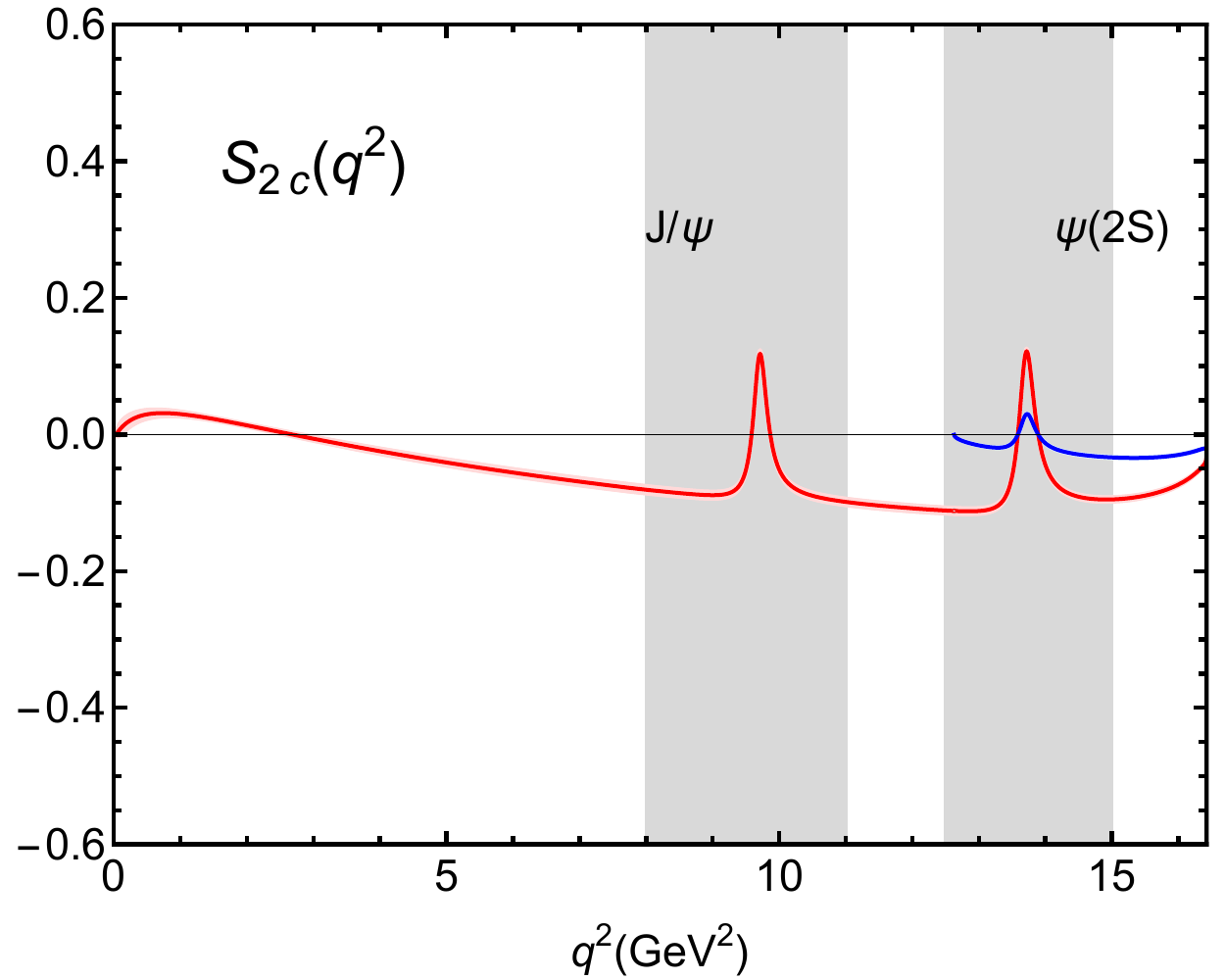}
  \includegraphics[width=56mm]{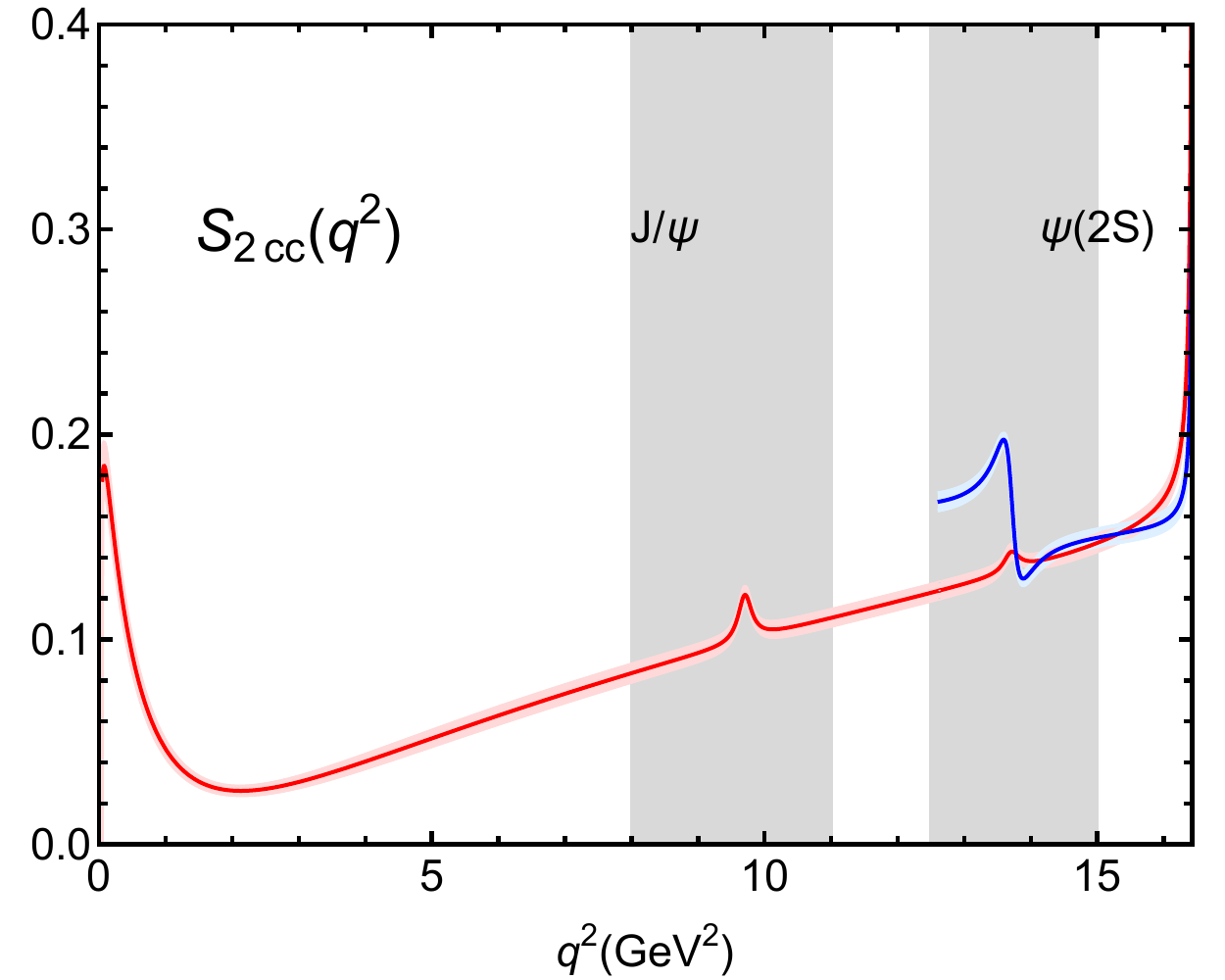}
  \includegraphics[width=56mm]{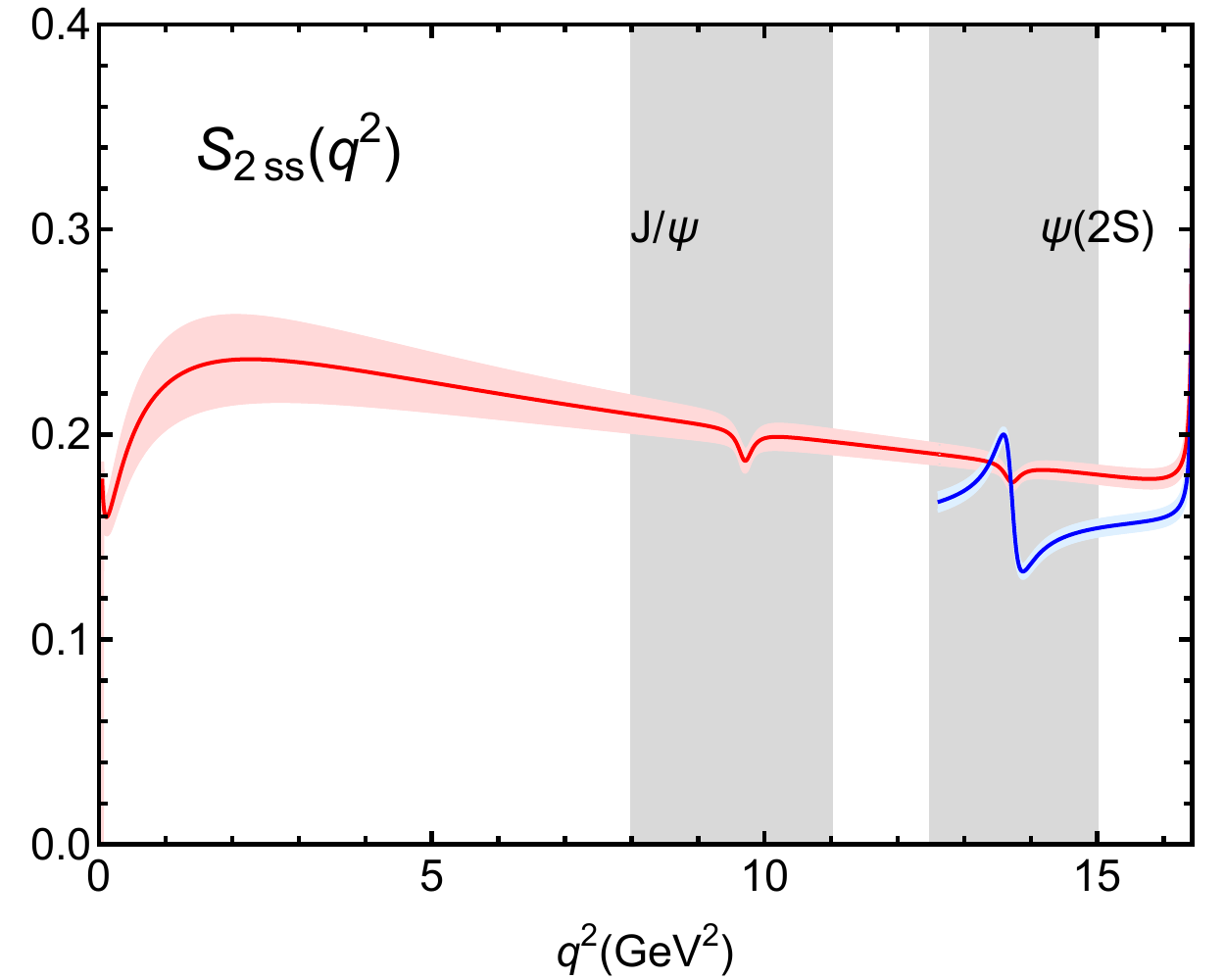}\\
  \includegraphics[width=56mm]{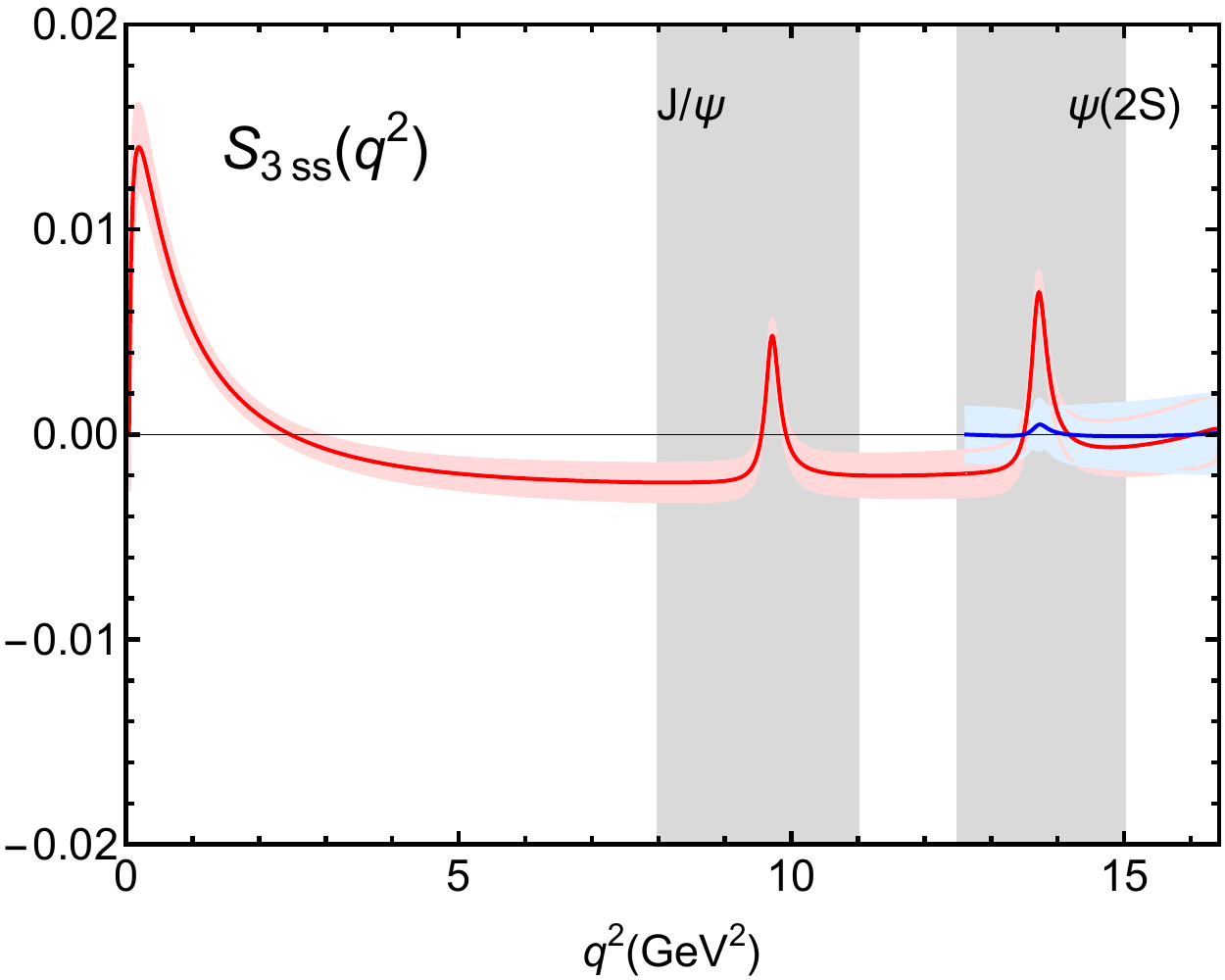}
  \includegraphics[width=56mm]{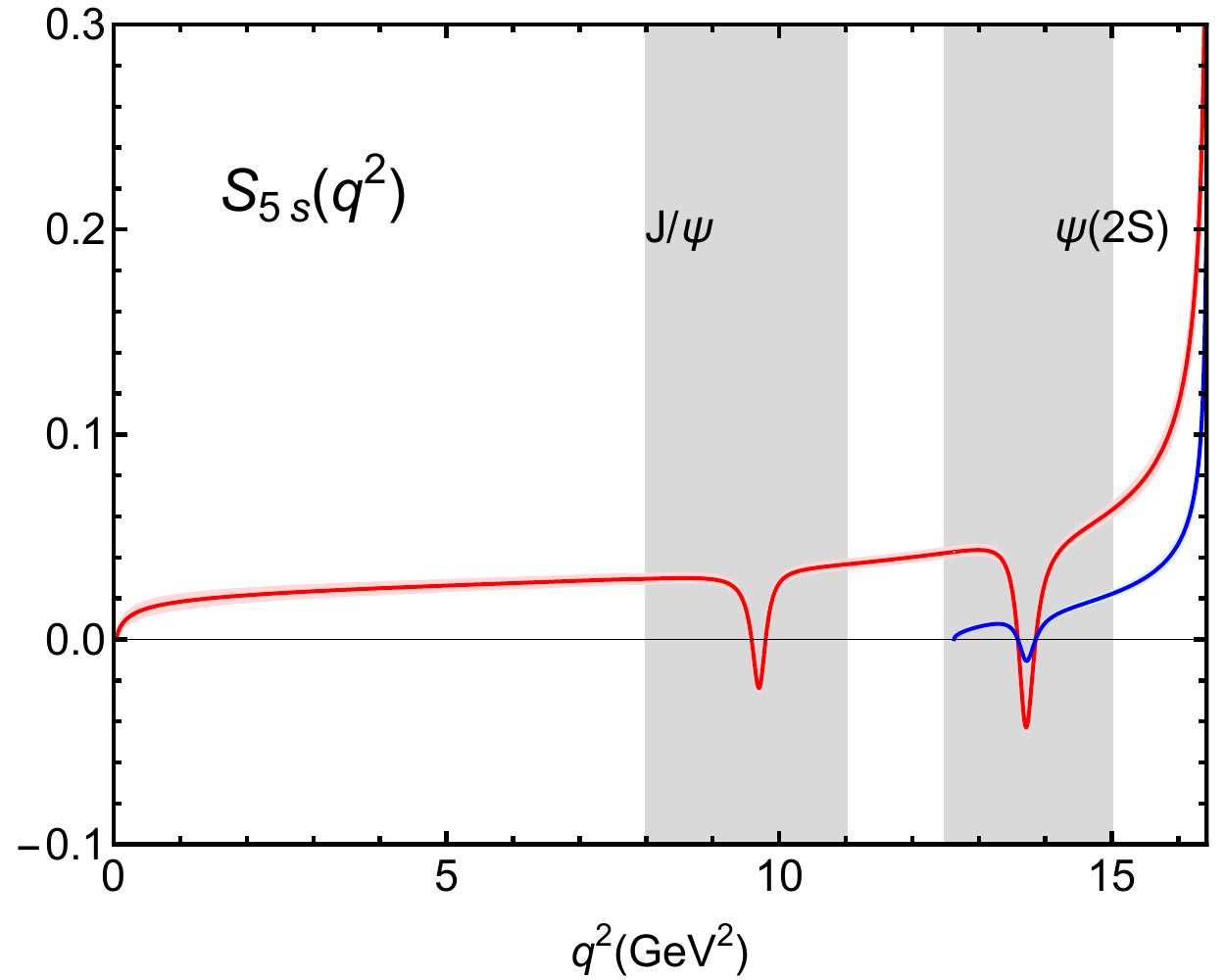}
  \includegraphics[width=56mm]{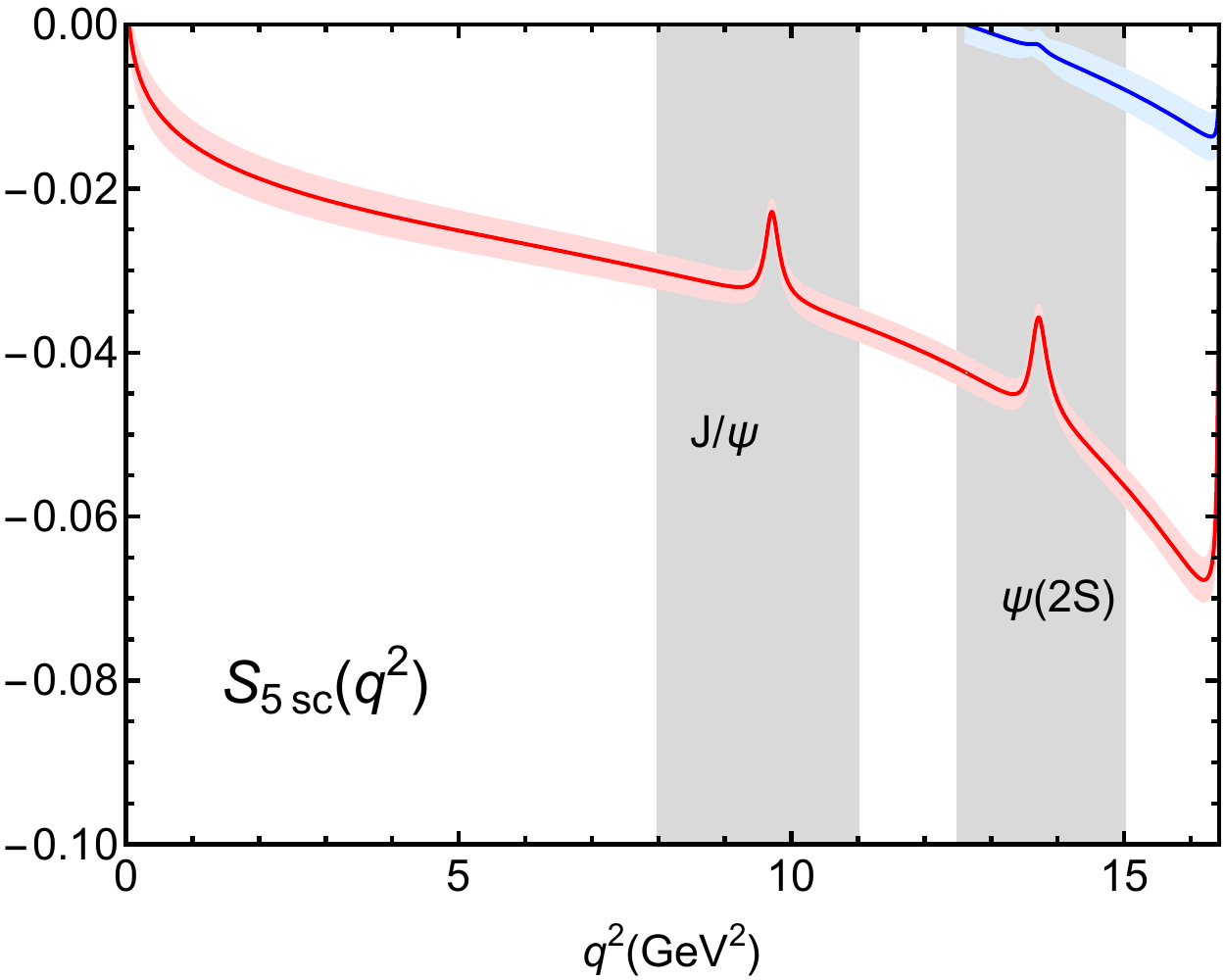}
  \end{tabular}
  \caption{The $q^2$ dependence of the normalized angular coefficients $S_{1c}$, $S_{1cc}$, $S_{1ss}$, $S_{2c}$, $S_{2cc}$, $S_{2ss}$, $S_{3ss}$, $S_{5s}$, and $S_{5sc}$. Here, the red curve and the blue curve are our results of the $\mu$ and $\tau$ channels, respectively, and the concomitant shadows are corresponding errors.}
\label{fig:angular}
\end{figure*}

In Fig.~\ref{fig:angular}, we present our results for the $q^2$ dependent normalized angular coefficients. Since the $e$ channel shows similar behavior to the $\mu$ channel, we only present the results of the $\mu$ and the $\tau$ channels here. These angular distributions are important physical observables, and  can be checked by future experiments.

\begin{figure}[htbp]\centering
  \includegraphics[width=60mm]{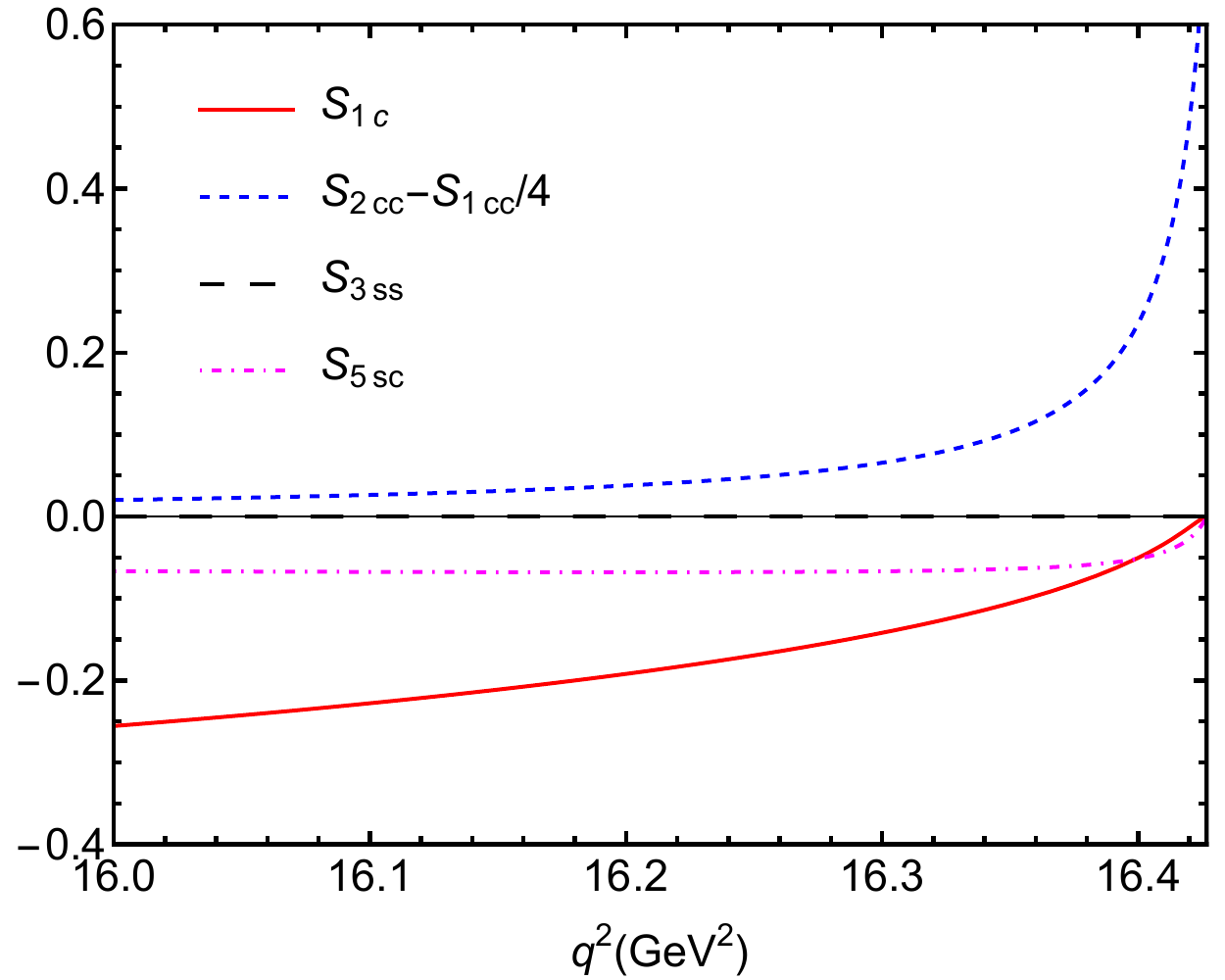}\\
  \caption{The behaviors of the normalized angular coefficients $S_{1c}$, $S_{2cc}-S_{1cc}/4$, $S_{3ss}$, and $S_{5sc}$ in the low-recoil region with $m_{\ell}=0$.}
  \label{fig:S}
\end{figure}

At the low-recoil endpoint for $q^2\to(m_{\Lambda_b}-m_{\Lambda^{*}})^2$, Descotes-Genon and Novoa-Brunet predicted \cite{Descotes-Genon:2019dbw}
\begin{equation*}
\begin{split}
S_{1c}&\to0,~~S_{2cc}-S_{1cc}/4\to3/8,\\
S_{3ss}&\to-1/4,~~S_{5sc}\to-1/2
\end{split}
\end{equation*}
by neglecting the contribution from the photon pole. In Fig.~\ref{fig:S}, we present the behavior of the normalized angular coefficients $S_{1c}$, $S_{2cc}-S_{1cc}/4$, $S_{3ss}$, and $S_{5sc}$ in the low-recoil region by assuming $m_{\ell}=0$. It is obvious that our result for $S_{1c}$ is strictly consistent with the above prediction, while the $S_{2cc}-S_{1cc}/4$, $S_{3ss}$ and $S_{5sc}$ show apparent deviations.

\begin{figure*}[htbp]\centering
  \begin{tabular}{ccc}
  \includegraphics[width=56mm]{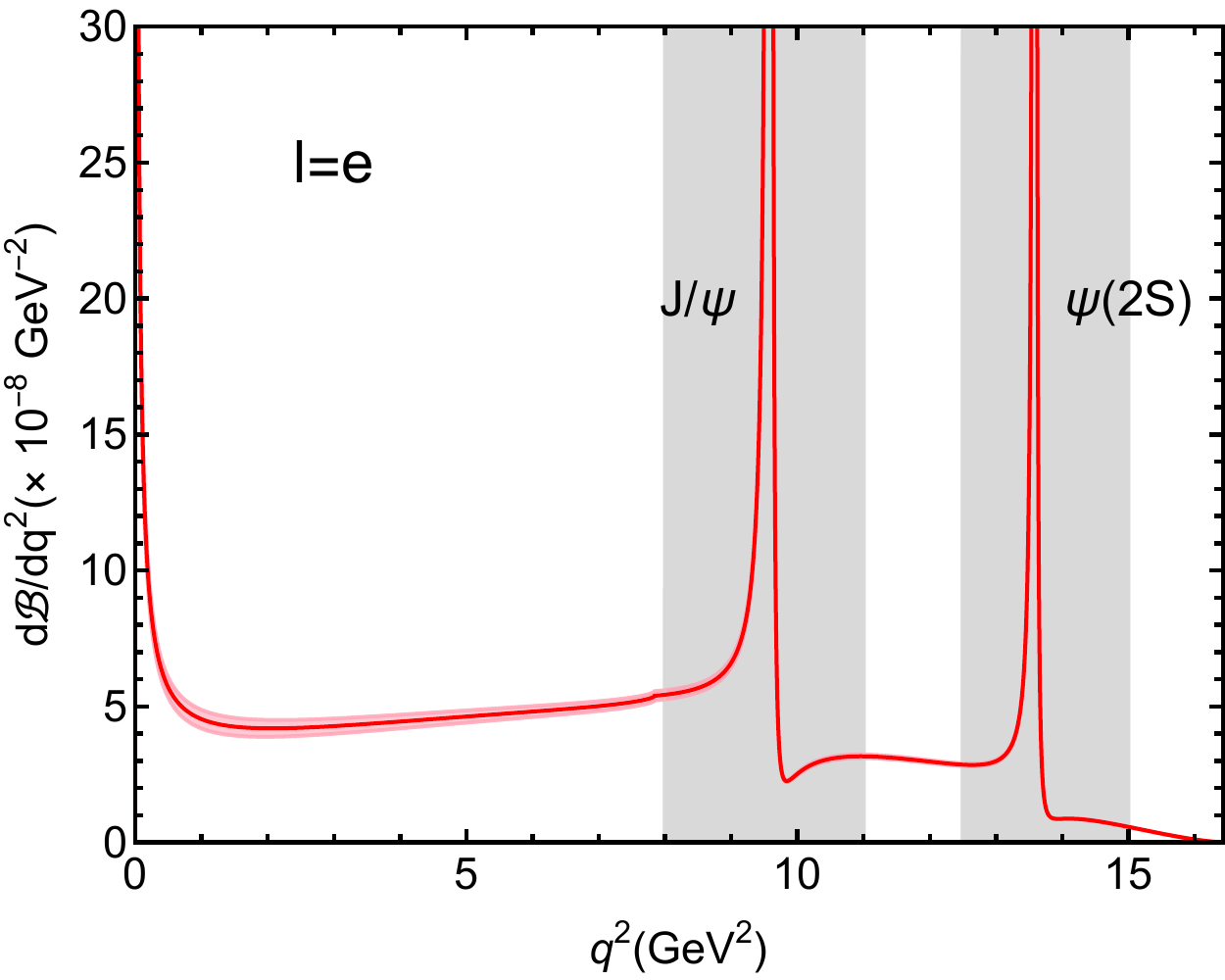}
  \includegraphics[width=56mm]{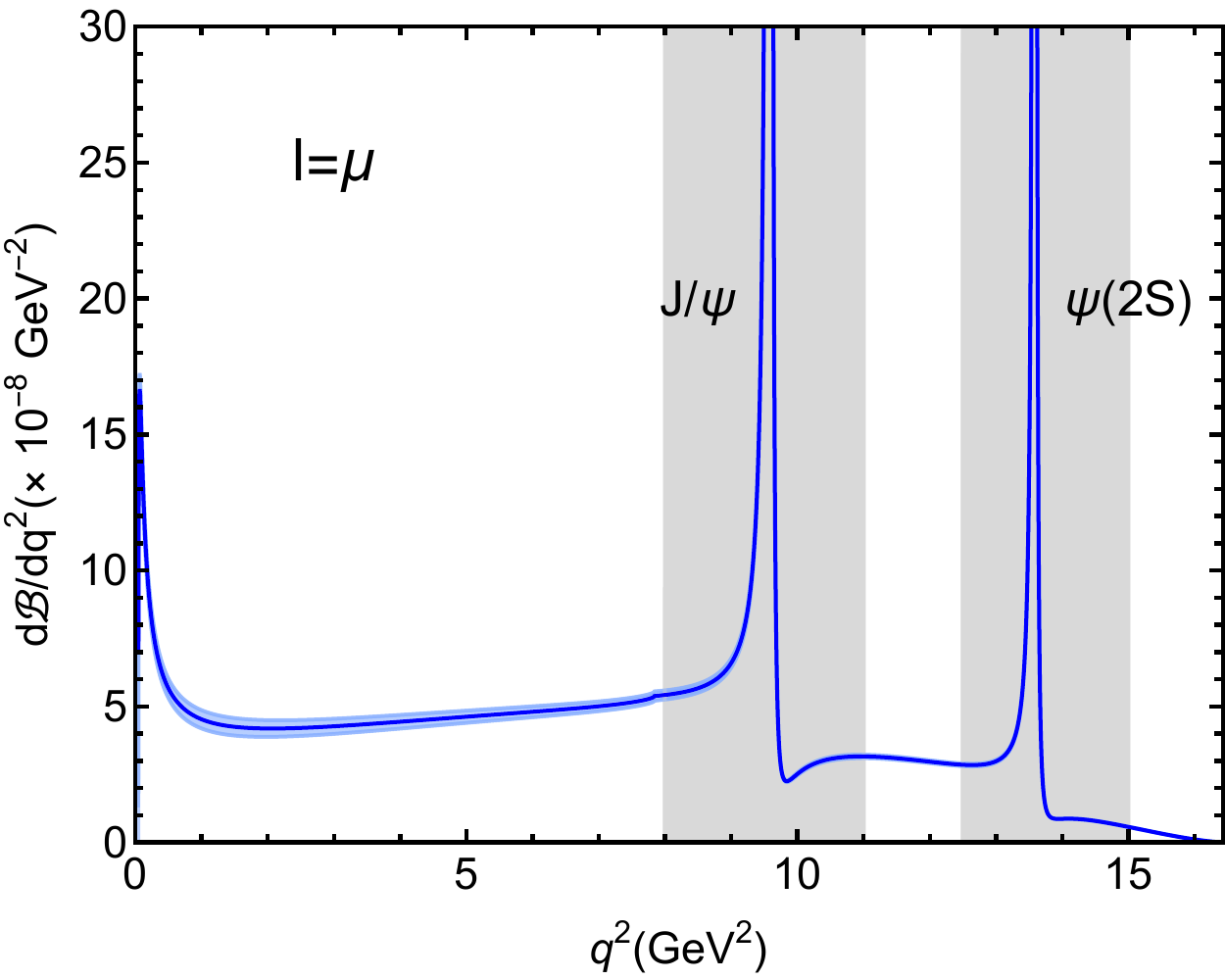}
  \includegraphics[width=56mm]{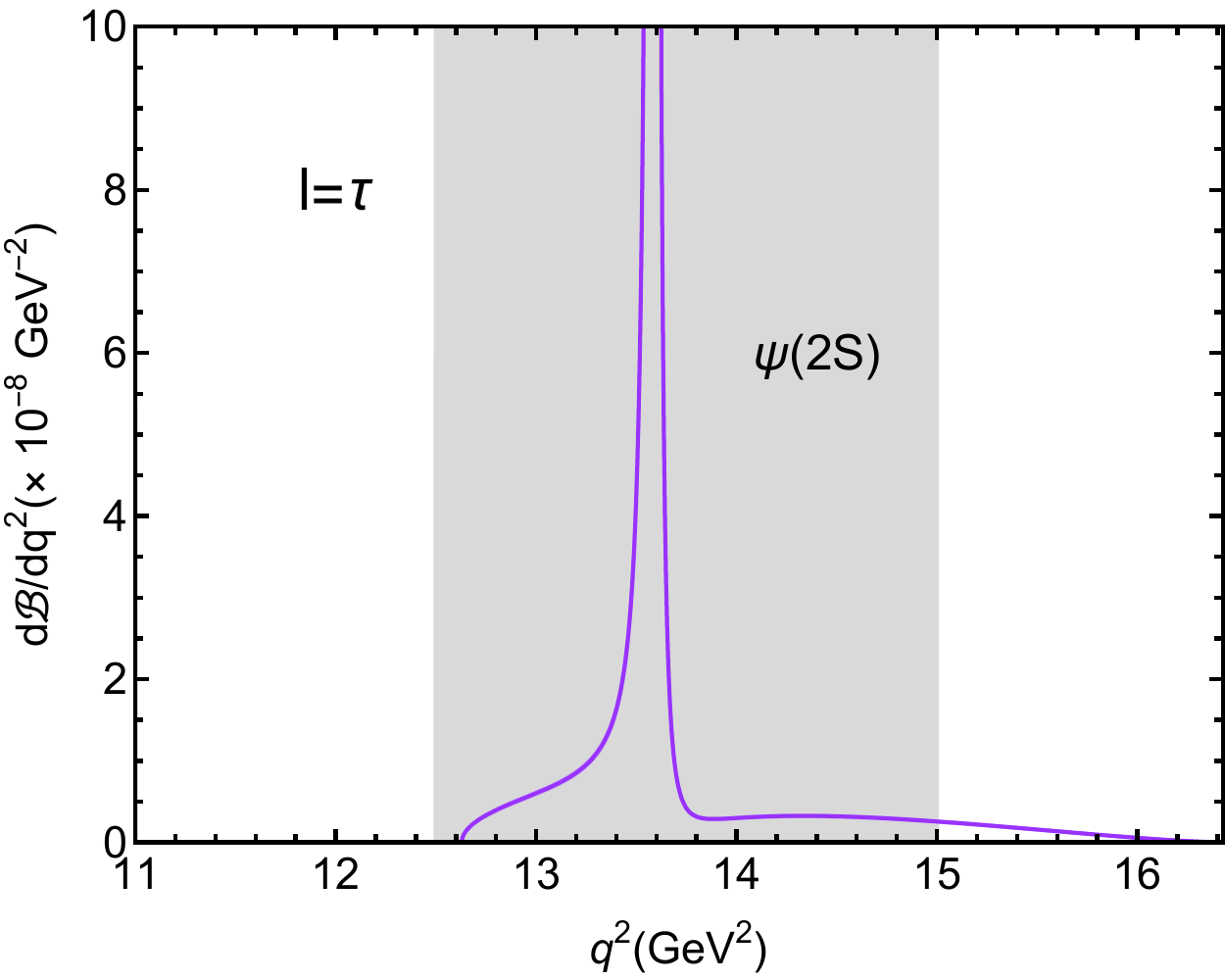}
  \end{tabular}
  \caption{The $q^2$ dependence of the differential branching ratios for $\Lambda_b\to\Lambda^{\ast}(\to N\bar{K})\ell^+\ell^-$ ($\ell=e~(\text{left~panel})$, $\mu~(\text{center~panel})$, $\tau~(\text{right~panel})$), where the red, the blue, and the purple curves are our results from the $e$, $\mu$, and $\tau$ channels, respectively, and the concomitant shadows are the corresponding errors.}
\label{fig:branchingratios}
\end{figure*}

We further evaluate the differential branching ratios using Eq.~\eqref{eq:branchingrate}. The $q^2$ dependence of the differential branching ratios is shown in Fig.~\ref{fig:branchingratios}, where the orange solid curve, the blue dashed curve, and the purple dot-dashed curve are our results for the $e$, $\mu$, and $\tau$ channels, respectively. The gray zones in the regions of the dilepton mass squared $8.0<q^2<11.0\ \text{GeV}^2$ and $12.5<q^2<15.0\ \text{GeV}^2$ show the contributions from the charmonium resonances $J/\psi$ and $\psi(2S)$, respectively.

Recently, the LHCb collaboration measured the ``nonresonant" contributions, which are different from the ``resonant" contribution from $\Lambda_b^0\to pK^{-}J\psi(\to\ell^+\ell^-)$, to $\mathcal{B}(\Lambda_b^0\to pK^{-}e^+e^-)$ and $\mathcal{B}(\Lambda_b^0\to pK^{-}\mu^+\mu^-)$ decays as
\begin{eqnarray*}
\mathcal{B}(\Lambda_b^0\!\to\!pK^{-}e^+e^-)\!&\!=\!&\!(3.1\!\pm0.4\!\pm0.2\!\pm0.3^{+0.4}_{-0.3})\times10^{-7},\\
\mathcal{B}(\Lambda_b^0\!\to\!pK^{-}\mu^+\mu^-)\!&\!=\!&\!(2.65\!\pm0.14\!\pm0.12\!\pm0.29^{+0.38}_{-0.23})\times10^{-7},
\end{eqnarray*}
in the region of $0.1\leq q^2 \leq6$ $\text{GeV}^{2}/c^4$ and $m(pK^-)<2600\ \text{MeV}/c^2$~\cite{LHCb:2019efc}. Assuming $\mathcal{B}(\Lambda(1520)\to pK^{-})=\mathcal{B}(\Lambda(1520)\to n\bar{K}^{0})$, we calculate
\begin{eqnarray*}
\mathcal{B}(\Lambda_b^0\to\Lambda^{\ast}(\to pK^{-})e^+e^-)_{0.1\leq q^2\leq6\ \text{GeV}^2}\\
=(1.618\pm0.108)\times10^{-7},\\
\mathcal{B}(\Lambda_b^0\to\Lambda^{\ast}(\to pK^{-})\mu^+\mu^-)_{0.1\leq q^2\leq6\ \text{GeV}^2}\\
=(1.610\pm0.106)\times10^{-7}.
\end{eqnarray*}
This indicates that the contribution from $\Lambda(1520)$ is significant. We find from the PDG~\cite{ParticleDataGroup:2020ssz} that $\Lambda(1600)$, $\Lambda(1670)$, and other hyperons can also decay to the $N\bar{K}$ final state. Their contributions need to be carefully studied. Further studies with more excited $\Lambda$ hyperons will make a difference to the $\Lambda_b^0\to pK^{-}\ell^+\ell^-$ decays.

\begin{figure*}[htbp]\centering
  \begin{tabular}{ccc}
  \includegraphics[width=56mm]{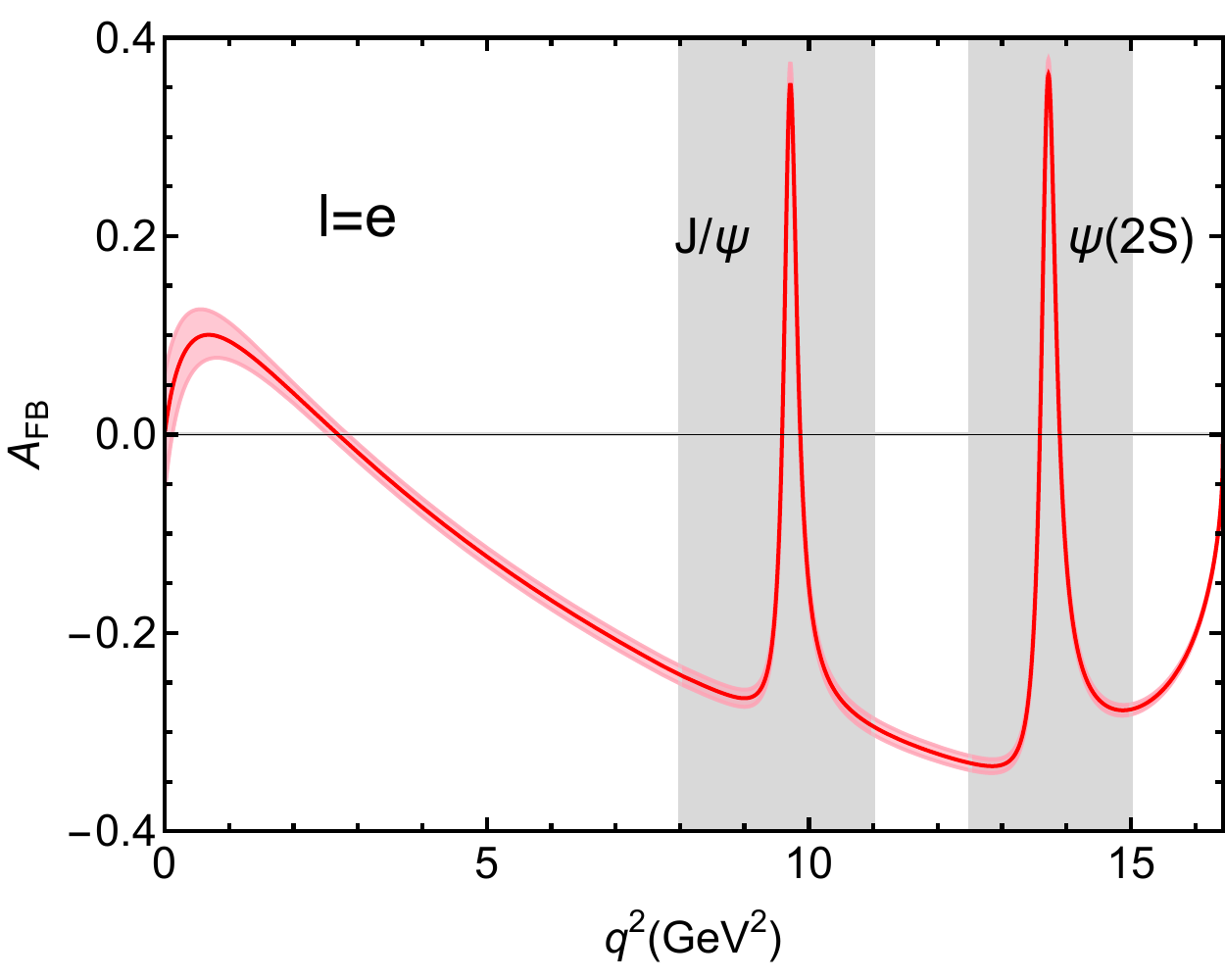}
  \includegraphics[width=56mm]{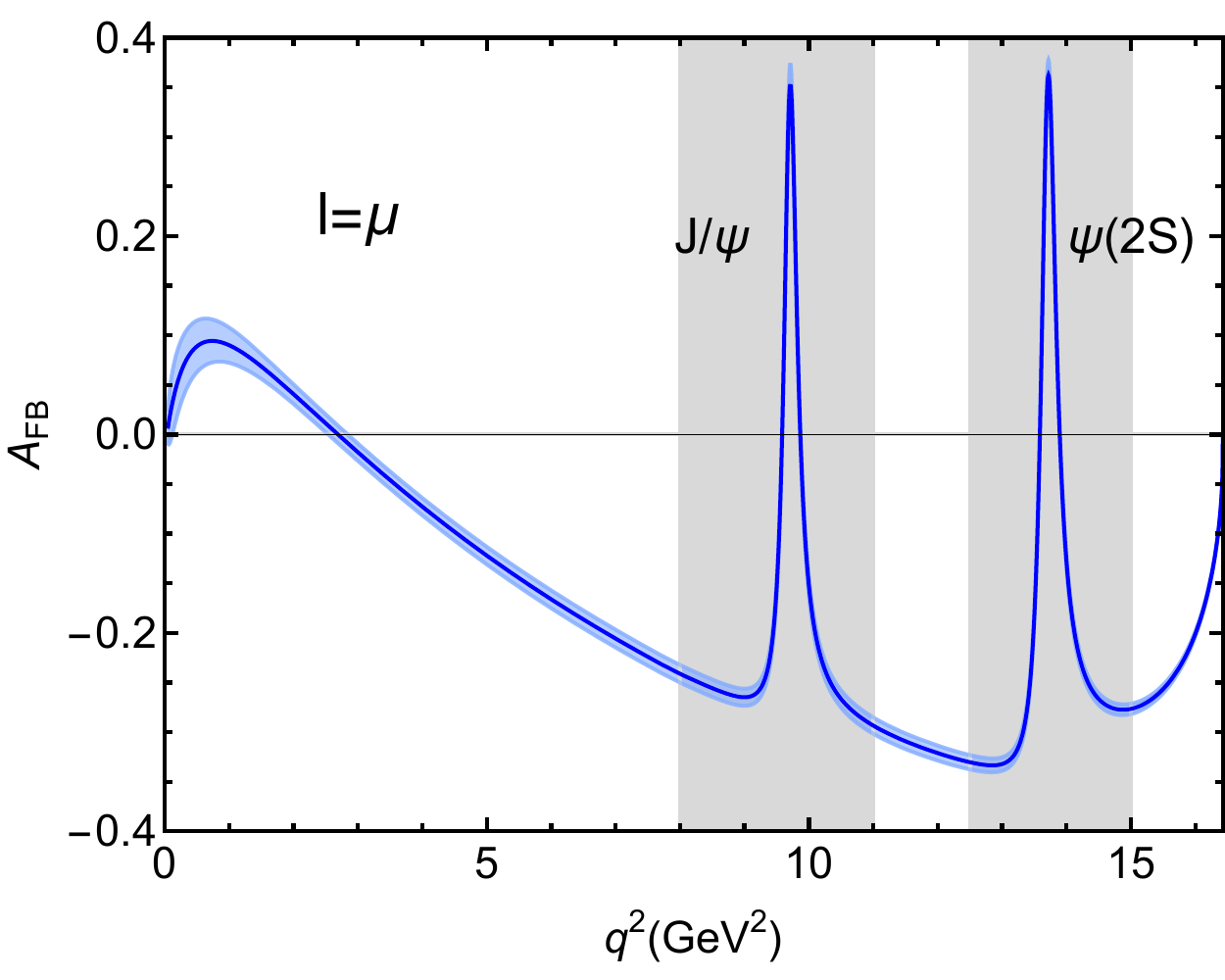}
  \includegraphics[width=56mm]{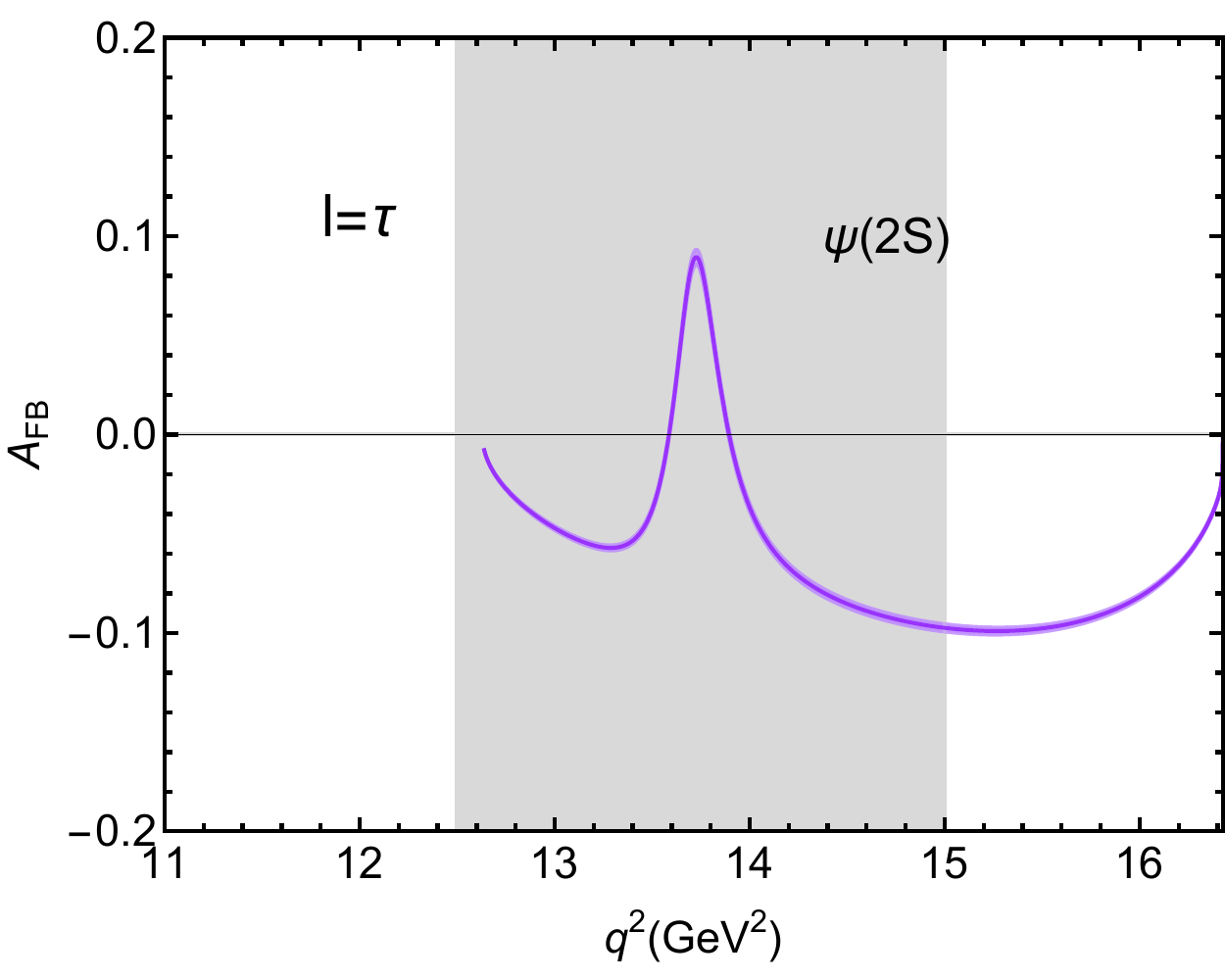}
  \end{tabular}
  \caption{The $q^2$ dependence of the lepton-side forward-backward asymmetry parameter $(A_{FB}^{\ell})$ for $\Lambda_b\to\Lambda^{*}(\to N\bar{K})\ell^+\ell^-$ ($\ell=e~(\text{left~panel})$, $\mu~(\text{center~panel})$, $\tau~(\text{right~panel})$), where the red, blue, and purple curves are our results from the $e$, $\mu$, and $\tau$ channels, respectively, and the shadows are the  corresponding errors.}
\label{fig:AFB}
\end{figure*}

\begin{figure*}[htbp]\centering
  \begin{tabular}{ccc}
  \includegraphics[width=56mm]{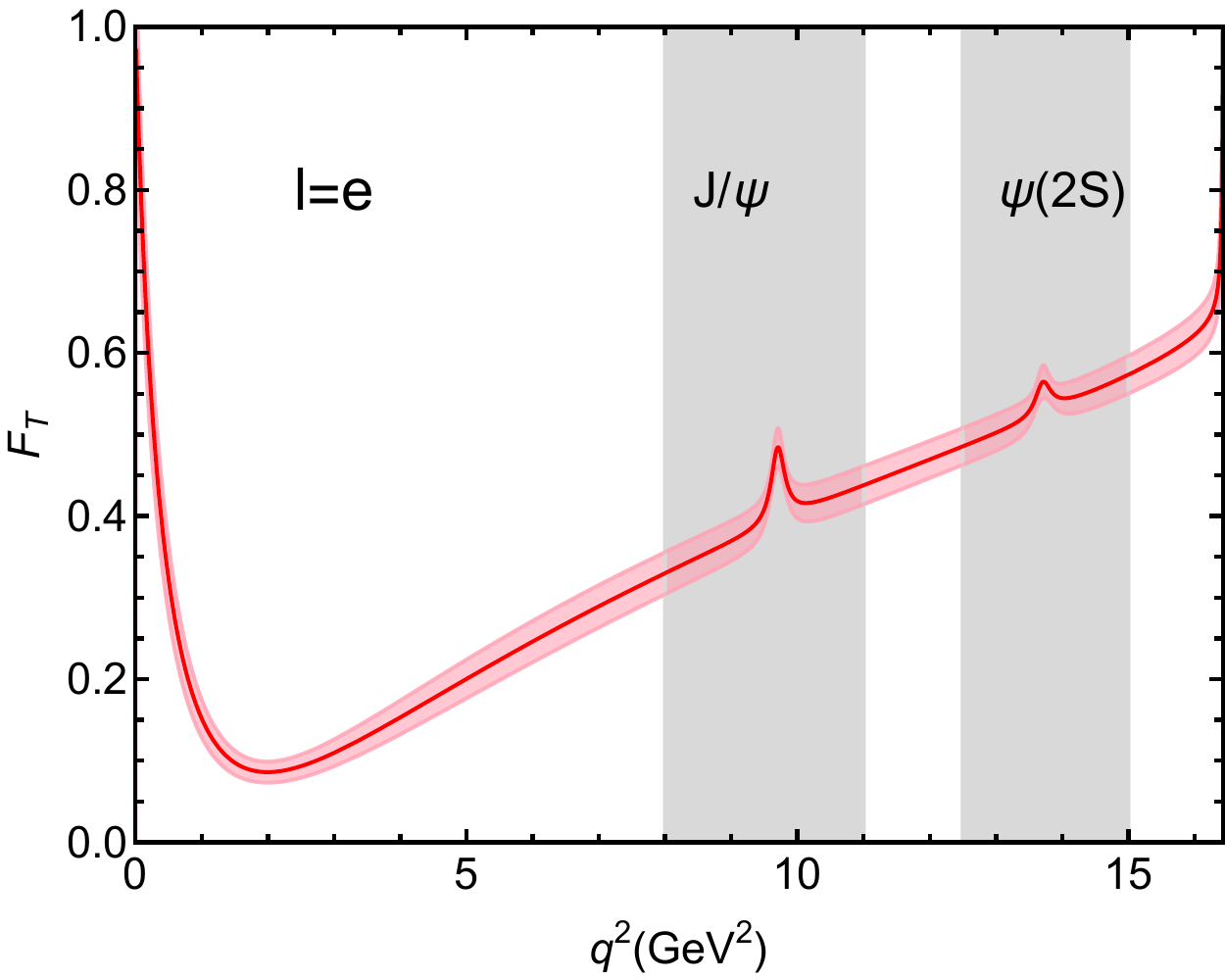}
  \includegraphics[width=56mm]{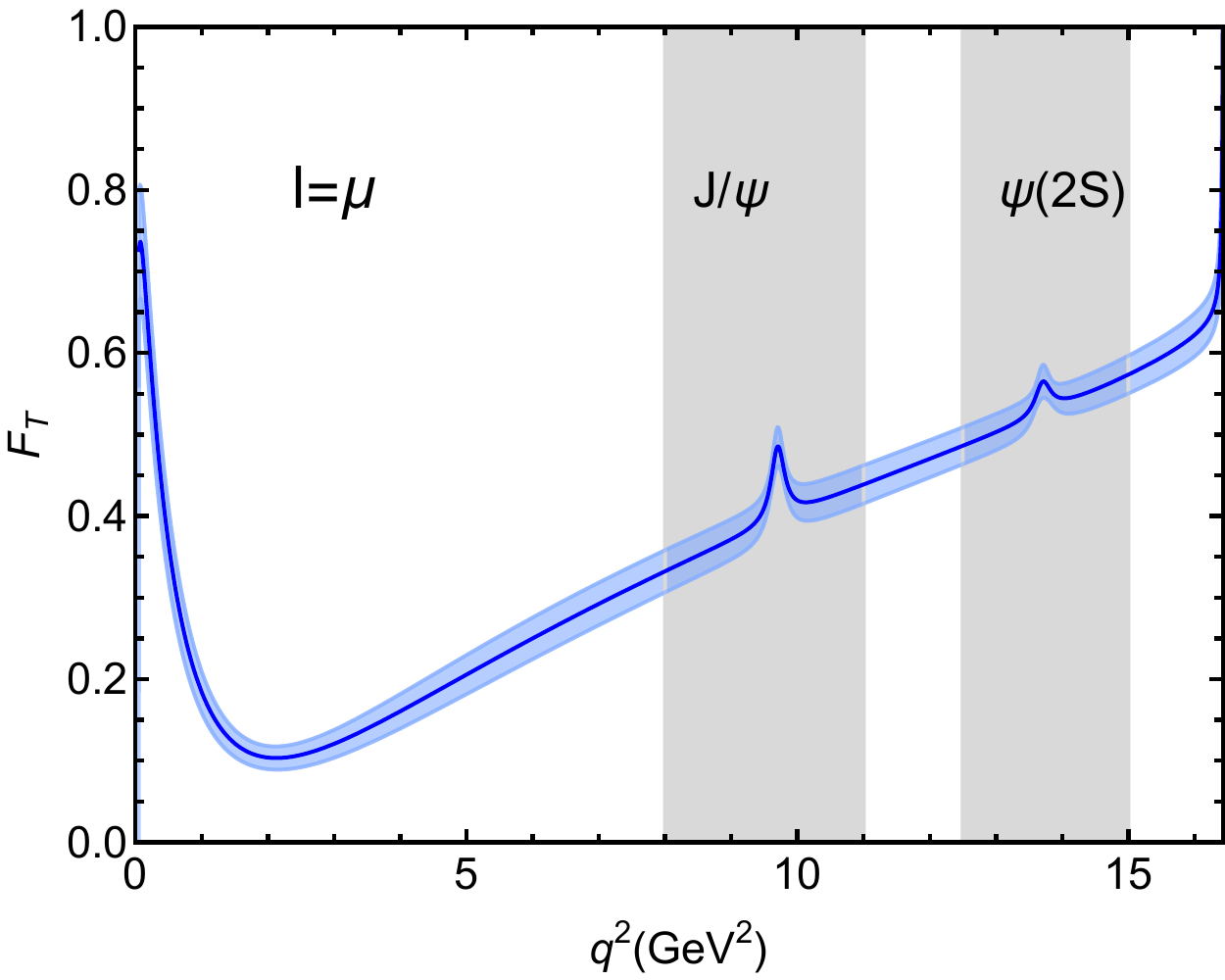}
  \includegraphics[width=56mm]{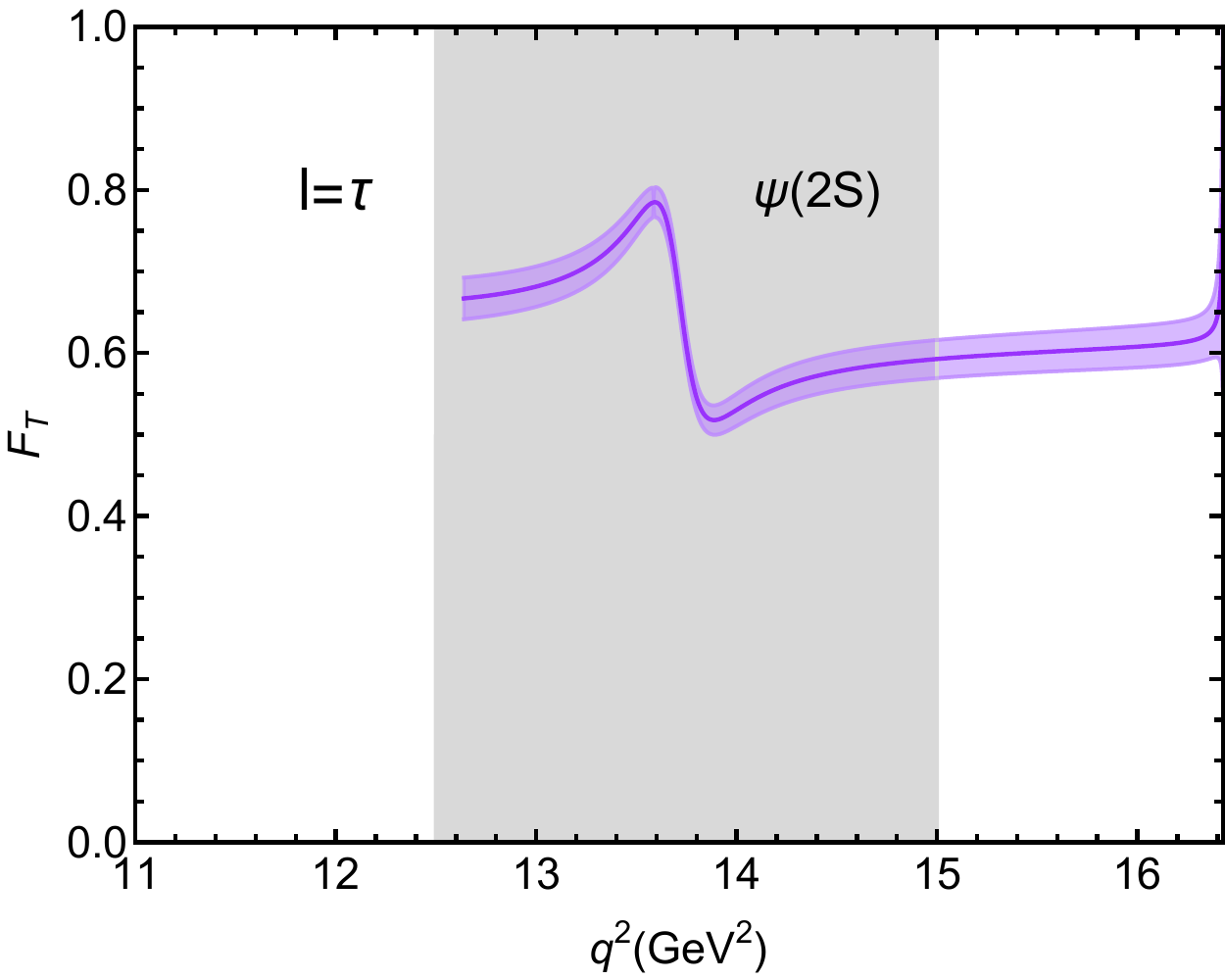}
  \end{tabular}
  \caption{The $q^2$ dependence of the transverse polarization fractions ($F_{T}$) for $\Lambda_b\to\Lambda^{*}(\to N\bar{K})\ell^+\ell^-$ ($\ell=e~(\text{left~panel})$, $\mu~(\text{center~panel})$, $\tau~(\text{right~panel})$), where the red, blue, and purple curves are our results of the $e$, $\mu$, and $\tau$ channels, respectively, and the shadows are the corresponding errors.}
\label{fig:FT}
\end{figure*}

\begin{figure*}[htbp]\centering
  \begin{tabular}{ccc}
  \includegraphics[width=56mm]{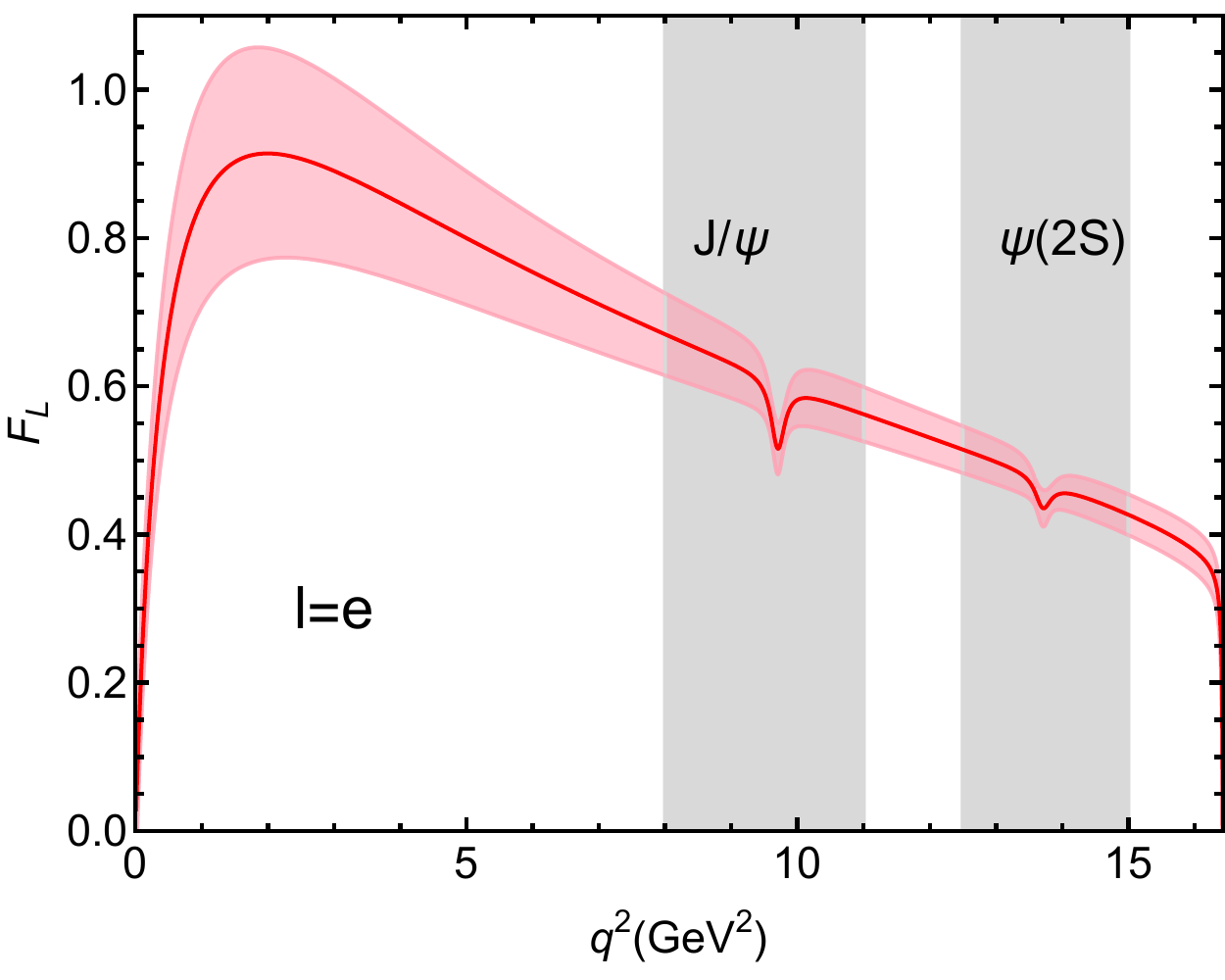}
  \includegraphics[width=56mm]{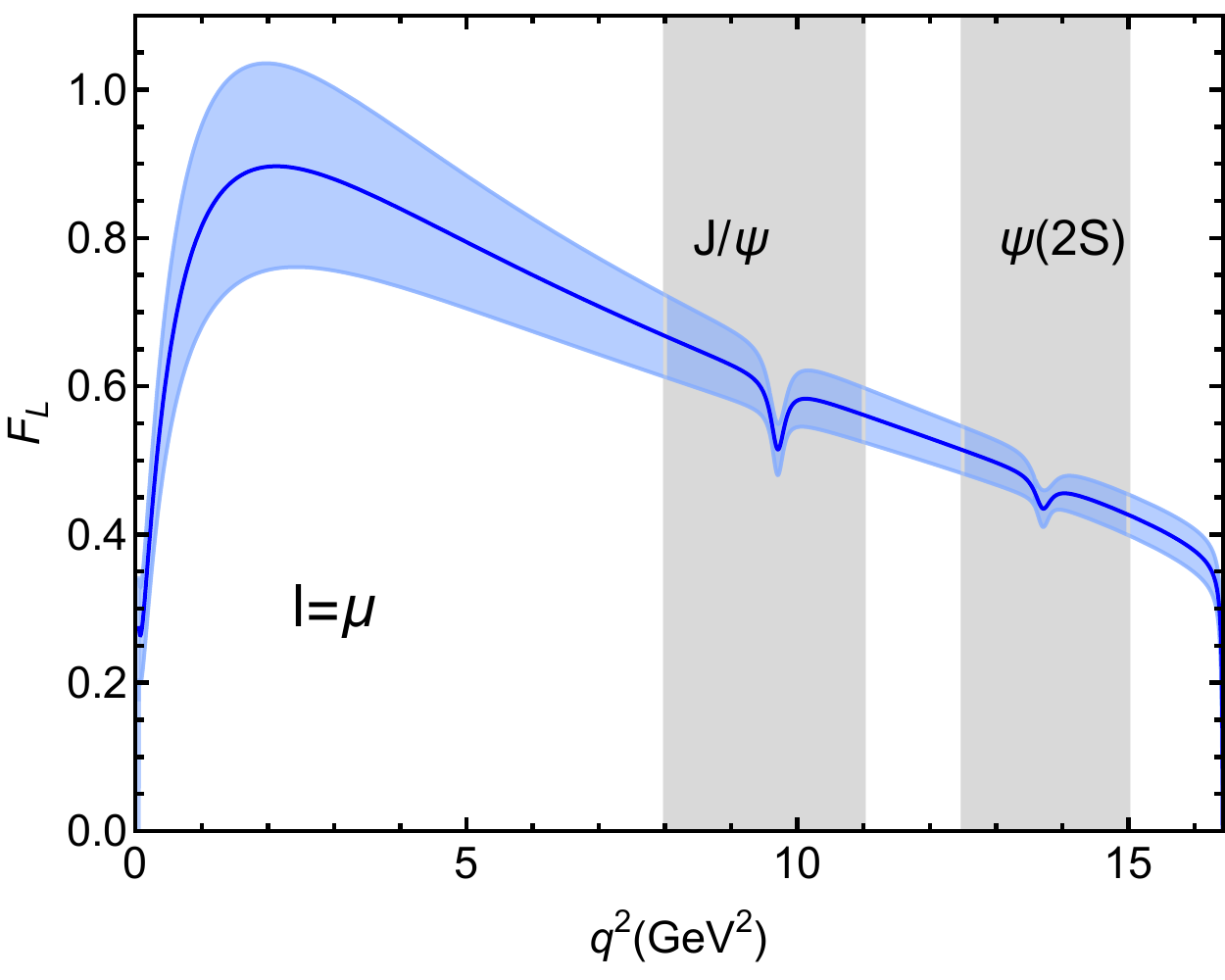}
  \includegraphics[width=56mm]{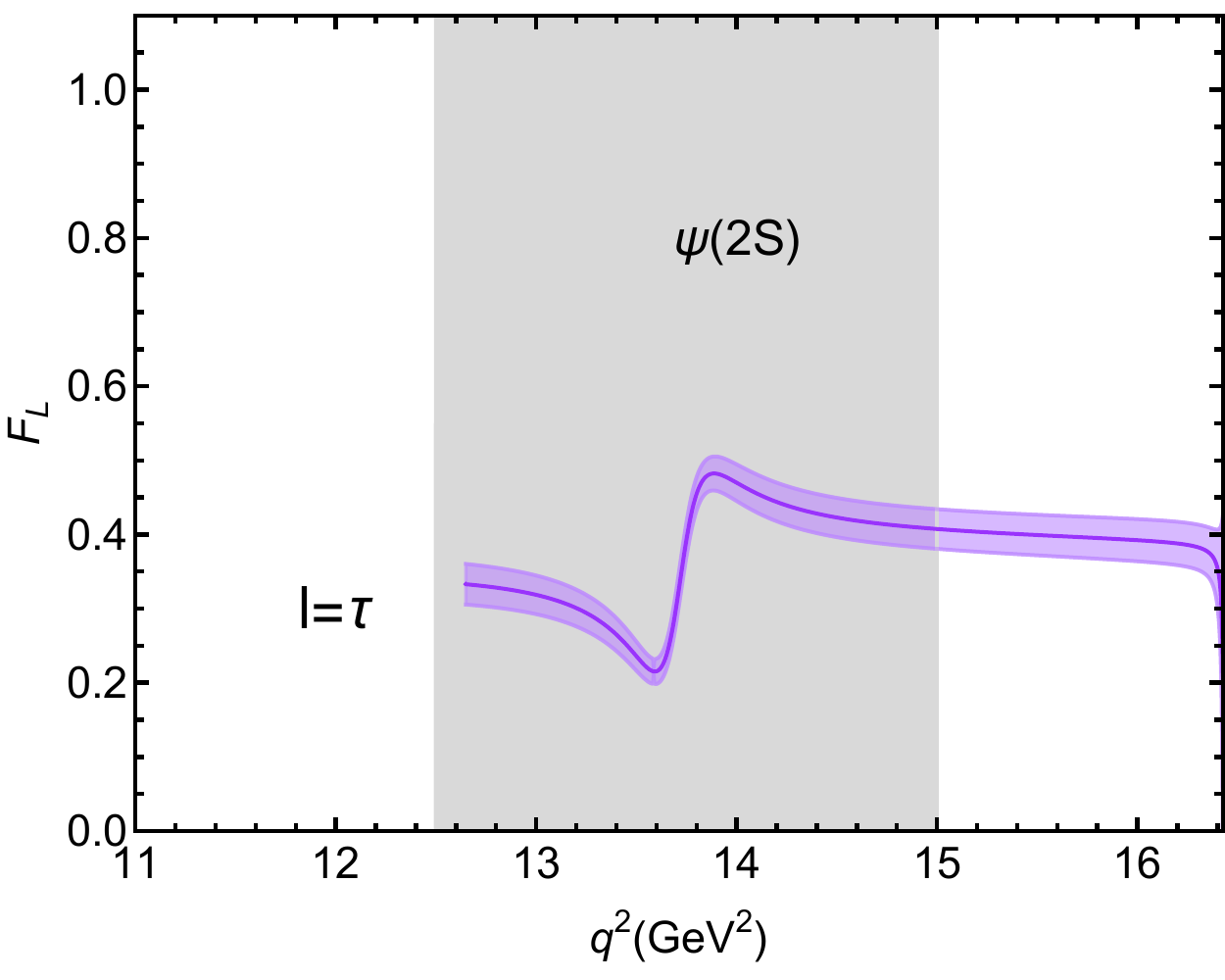}
  \end{tabular}
  \caption{The $q^2$ dependence of the longitudinal polarization fractions ($F_{L}$) for $\Lambda_b\to\Lambda^{*}(\to N\bar{K})\ell^+\ell^-$ ($\ell=e~(\text{left~panel})$, $\mu~(\text{center~panel})$, $\tau~(\text{right~panel})$), where the red, blue, and purple curves are our results of the $e$, $\mu$, and $\tau$ channels, respectively, and the shadows are the corresponding errors.}
\label{fig:FL}
\end{figure*}

In addition, the $q^2$ dependence of the lepton-side forward-backward asymmetry $(A_{FB}^{\ell})$, and the transverse ($F_{T}$) and longitudinal ($F_{L}$) polarization fractions of the dilepton system are presented in Figs.~\ref{fig:AFB}, \ref{fig:FT} and \ref{fig:FL}, respectively, where we also show the contributions from the charmonium resonances $J/\psi$ and $\psi(2S)$ with gray zones. The averaged values of these angular observables for the $e$ and $\mu$ channels defined in Eq.~\eqref{eq:AveragedPhysicalObservable} in the region of $0.1<q^2<6.0\ \text{GeV}^2$ are presented in Table~\ref{tab:AveragedPhysicalObservable}. The angular distributions provide a rich set of physical observables to study the weak interaction and the structure of $\Lambda(1520)$, and are also important to study the NP effects beyond the SM~\cite{Azizi:2012vy,Yan:2019tgn,Descotes-Genon:2019dbw,Das:2020cpv,Amhis:2020phx}, so we call for the ongoing LHCb experiment to measure them.

\begin{table}[htbp]\centering
\caption{The predictions for the averaged lepton-side forward-backward asymmetry $\langle A_{FB}^{\ell}\rangle$, the averaged transverse polarization fraction $\langle F_T\rangle$, and the averaged longitudinal polarization fraction $\langle F_L\rangle$ in the region of $0.1<q^2<6.0\ \text{GeV}^2$.}
\label{tab:AveragedPhysicalObservable}
\renewcommand\arraystretch{1.05}
\begin{tabular*}{86mm}{c@{\extracolsep{\fill}}ccc}
\toprule[0.7pt]
\toprule[0.7pt]
Channels  &$\langle A_{FB}^{\ell}\rangle$  &$\langle F_T\rangle$  &$\langle F_L\rangle$\\
\midrule[0.7pt]
$\ell=e$    &$-0.030\pm0.012$          &$0.208\pm0.055$          &$0.792\pm0.231$\\
$\ell=\mu$   &$-0.032\pm0.010$          &$0.221\pm0.057$          &$0.779\pm0.225$\\
\bottomrule[0.7pt]
\bottomrule[0.7pt]
\end{tabular*}
\end{table}

\section{Summary}
\label{sec6}
With the accumulation of experimental data in the LHCb Collaboration, the experimental exploration of rare decays $b\to s\ell^+\ell^-$ ($\ell$=$e$, $\mu$, $\tau$) in the baryon sector, especially
the $P$-wave final state $\Lambda_b\to\Lambda(1520)\ell^+\ell^-$, will attract more attention. Given this opportunity, in this work we focus on the quasi-four-body decay $\Lambda_b\to\Lambda(1520)(\to N\bar{K})\ell^+\ell^-$, where the angular coefficients, the differential branching ratio, and several angular observables, including the lepton-side forward-backward asymmetry $(A_{FB}^{\ell})$, and the transverse and longitudinal polarization fractions $(F_{T(L)})$ are investigated.

To describe the weak process, we have worked in the helicity formula, where the relevant weak transition form factors are obtained through the three-body light-front quark model. Our main advantage is the improved treatment of the spatial wave functions of the involved baryons, where a semirelativistic potential model is applied to solve the numerical spatial wave functions of the baryons assisted by the GEM. Thus, we emphasize that our study of the rare decay $\Lambda_b\to\Lambda(1520)(\to N\bar{K})\ell^+\ell^-$ is supported by the baryon spectroscopy. Our results of the form factors are comparable with the predictions of HQET and SCET, and also with  the calculations by the LQCD approach. These form factors will be useful for the study of the corresponding weak decays.

Overall, we have systematically investigated  the $\Lambda_b\to\Lambda(1520)(\to N\bar{K})\ell^+\ell^-$ ($\ell=e,\mu,\tau$) processes in the framework of the three-body light-front quark model based on the Gaussian expansion method. We believe that the present work can serve as an essential step toward strong dynamics on the beauty baryon decays. We expect that under the considerable progress on the experimental side, the above predictions could be tested by future LHCb experiments.

\appendix

\section{The differential decay width}
\label{app01}

\setcounter{equation}{0}
\renewcommand{\theequation}{A.\arabic{equation}}

The differential decay width for the quasi-four-body decay $\Lambda_b\to\Lambda^{*}(\to N\bar{K})\ell^+\ell^-$ is
\begin{equation}
d\Gamma=\frac{|\mathcal{M}|^2}{2m_{\Lambda_b}}d\Phi_4(p;k_1,k_2,q_1,q_2),
\end{equation}
where $d\Phi_4$ is the four-body phase space given by
\begin{equation}
\begin{split}
d\Phi_4&(p;k_1,k_2,q_1,q_2)\\
=&(2\pi)^4\delta^4\Big{(}p\!-\!k_1\!-\!k_2\!-q_1\!-\!q_2\Big{)}
\prod_{i=1}^2\!\frac{d^3\vec{k}_i}{(2\pi)^32E_{k_i}}
\prod_{j=1}^2\!\frac{d^3\vec{q}_j}{(2\pi)^32E_{q_j}}\\
=&\frac{dk^2}{2\pi}\frac{dq^2}{2\pi}d\Phi_2(k;k_1,k_2)d\Phi_2(q;q_1,q_2)d\Phi_2(p;k,q),
\end{split}
\end{equation}
where the two-body phase spaces are written as
\begin{equation}
\begin{split}
\int\!d\Phi_2(k;k_1,k_2)\!=&\frac{1}{32\pi^2}\frac{\sqrt{\lambda(k^2,k_1^2,k_2^2)}}{k^2}\!\int_{-1}^{1}\!d\cos\theta_{\Lambda^\ast}\!\int_{0}^{2\pi}\!d\phi,\\
\int\!d\Phi_2(q;q_1,q_2)\!=&\frac{1}{32\pi^2}\frac{\sqrt{\lambda(q^2,q_1^2,q_2^2)}}{q^2}\!\int_{-1}^{1}\!d\cos\theta_{\ell}\times(2\pi),\\
\int\!d\Phi_2(p;k,q)\!=&\frac{1}{32\pi^2}\frac{\sqrt{\lambda(p^2,k^2,q^2)}}{p^2}\times 2 \times (2\pi).
\end{split}
\end{equation}
As shown in Fig.~\ref{fig:kinematics}, three angles are defined (i) the angle $\theta_{\Lambda^\ast}$ is defined as the angle that the nucleon makes with the $+z$ axis in the $(N\bar{K})$ center of mass system, (ii) the angle $\theta_{\ell}$ is defined as the angle made by the $\ell^-$ with the $+z$ axis in the $(\ell^+\ell^-)$ cms, and (iii) the angle $\phi$ between the two decay planes, respectively.

The decay width of the concerned decay $\Lambda_b\to\Lambda^\ast(\to N\bar{K})\ell^+\ell^-$ is expressed as
\begin{equation}
\begin{split}
\int\!d\Phi_4\frac{|\mathcal{M}|^2}{2m_{\Lambda_b}}\!=&
\frac{2}{(32\pi^2)^3}\!\int\!dk^2
\frac{\sqrt{\lambda(k^2,k_1^2,k_2^2)}}{k^2}
\frac{\sqrt{\lambda(q^2,q_1^2,q_2^2)}}{q^2}\\
&\times\frac{\sqrt{\lambda(p^2,k^2,q^2)}}{p^2}
(dq^2d\cos\theta_{\Lambda^\ast}d\cos\theta_{\ell}d\phi)\\
&\times\frac{|\mathcal{M}|^2}{2m_{\Lambda_b}}.
\end{split}
\end{equation}
We also take into account the width of $\Lambda^\ast$ to modify its propagator, but treat it as narrow ($\Gamma_{\Lambda^\ast}\ll m_{\Lambda^\ast}$) state\footnote{Checking the PDG~\cite{ParticleDataGroup:2020ssz}, we notice that $m_{\Lambda(1520)}=1519\ \text{MeV}$ and $\Gamma_{\Lambda(1520)}=16\ \text{MeV}$,  indicating that it is reasonable to take the narrow-width approximation.}. This gives~\cite{Yan:2019tgn},
\begin{equation}
\begin{split}
\int\!d\Phi_{4}\frac{|\mathcal{M}|^2}{2m_{\Lambda_b}}
=&\int\!d\Phi_4 \frac{|\mathcal{M}|^2}{2m_{\Lambda_b}}\frac{1}{(k^2-m_{\Lambda^\ast}^2)^2}(k^2-m_{\Lambda^\ast}^2)^2\\
\stackrel{\Gamma_{\Lambda^\ast}\ll m_{\Lambda^\ast}}{\longrightarrow} &\int\!d\Phi_4 \frac{|\mathcal{M}|^2}{2m_{\Lambda_b}}\frac{(k^2-m_{\Lambda^\ast}^2)^2}{(k^2-m_{\Lambda^\ast}^2)^2+(m_{\Lambda^\ast}\Gamma_{\Lambda^\ast})^2}\\
=&\int\!d\Phi_4
\frac{|\mathcal{M}|^2}{2m_{\Lambda_b}}\frac{(k^2-m_{\Lambda^\ast}^2)^2}{m_{\Lambda^\ast}^3\Gamma_{\Lambda^\ast}}\frac{\Gamma_{\Lambda^\ast}/m_{\Lambda^\ast}}{\frac{\Gamma_{\Lambda^\ast}^2}{m_{\Lambda^\ast}^2}+(\frac{k^2}{m_{\Lambda^\ast}^2}-1)^2}\\
\stackrel{\Gamma_{\Lambda^\ast}\ll m_{\Lambda^\ast}}{\longrightarrow} &\int\!d\Phi_4
\frac{|\mathcal{M}|^2}{2m_{\Lambda_b}}(k^2-m_{\Lambda^\ast}^2)^2\frac{\pi}{m_{\Lambda^\ast}^3\Gamma_{\Lambda^\ast}}\delta\Big{(}\frac{k^2}{m_{\Lambda^\ast}^2}\!-\!1\Big{)}\\
=&\int\!d\Phi_4
\frac{|\mathcal{M}|^2}{2m_{\Lambda_b}}(k^2-m_{\Lambda^\ast}^2)^2\frac{\pi}{m_{\Lambda^\ast}\Gamma_{\Lambda^\ast}}\delta(k^2-m_{\Lambda^\ast}^2),\\
\end{split}
\end{equation}
with the properties of the Dirac delta function
\begin{equation}
\begin{split}
\lim_{\epsilon\to0}\frac{\epsilon}{\epsilon^2+x^2}&=\pi\delta(x),\\ \delta\Big{(}\frac{k^2}{m_{\Lambda^\ast}^2}\!-\!1\Big{)}&=m_{\Lambda^\ast}^2\delta(k^2-m_{\Lambda^\ast}^2)
\end{split}
\end{equation}
applied.

Following the above discussion, we can finally obtain
\begin{equation}
\begin{split}
\int\!d\Phi_4\frac{|\mathcal{M}|^2}{2m_{\Lambda_b}}\!=&
\frac{1}{2^{15}\pi^{5}m_{\Lambda_b}m_{\Lambda^\ast}\Gamma_{\Lambda^\ast}}
\int\!dq^2\ d\cos\theta_{\Lambda}\ d\cos\theta_{\ell}\ d\phi \\
&\times\frac{\sqrt{\lambda(k^2,k_1^2,k_2^2)}}{k^2}
\frac{\sqrt{\lambda(q^2,q_1^2,q_2^2)}}{q^2}
\frac{\sqrt{\lambda(p^2,k^2,q^2)}}{p^2}\\
&\times(k^2-m_{\Lambda^\ast})^2|\mathcal{M}|^2\Big|_{k^2=m_{\Lambda^\ast}^2},
\label{appeq:decayrate}
\end{split}
\end{equation}
where $\lambda(x,y,z)=x^2+y^2+z^2-2xy-2xz-2yz$ is the kinematic triangle  K\"{a}ll\'{e}n function.

\section{The kinematic conventions}
\label{app02}

\setcounter{equation}{0}
\renewcommand{\theequation}{B.\arabic{equation}}

\begin{figure}[htbp]\centering
  \includegraphics[width=60mm]{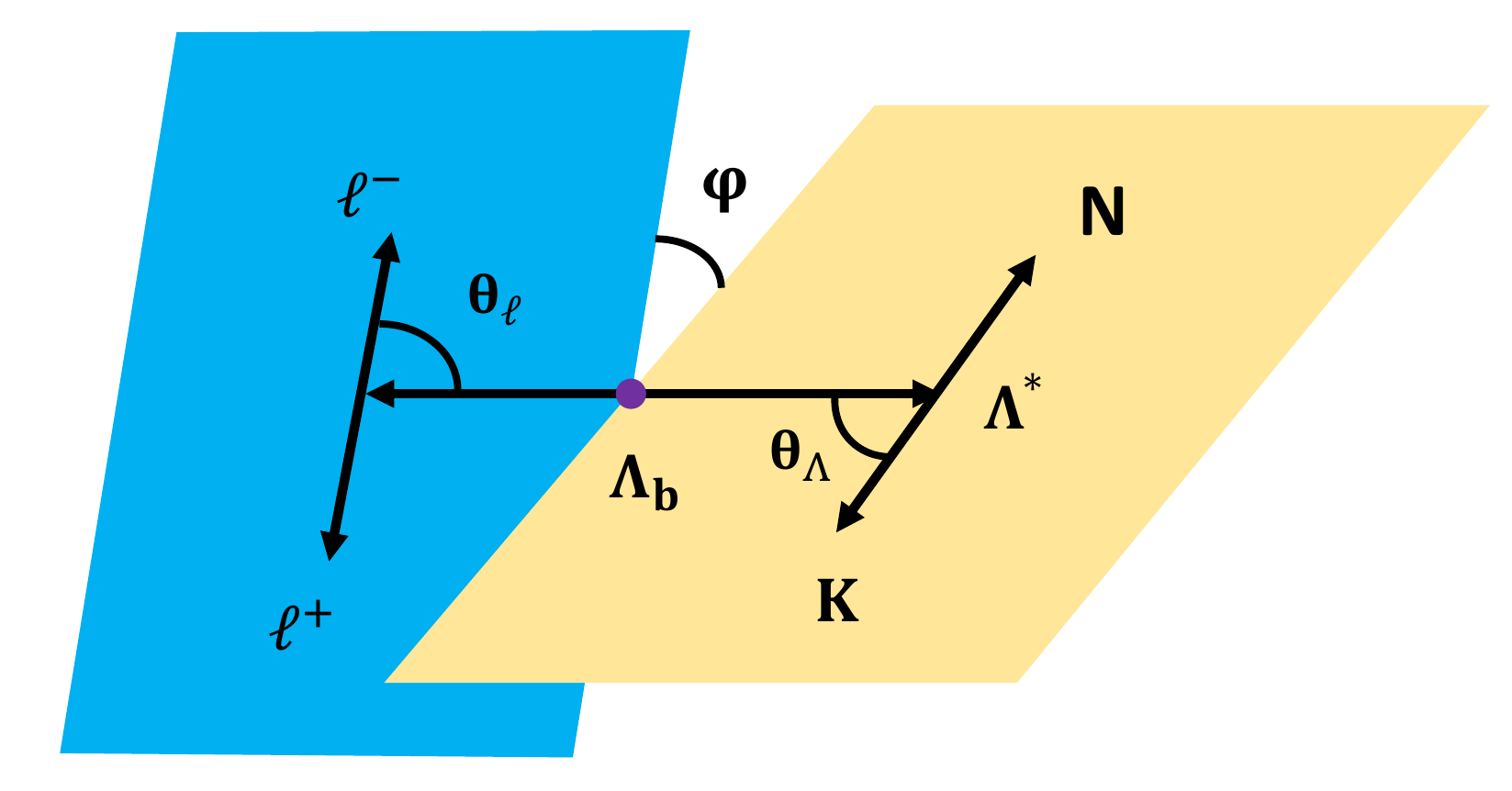}\\
  \caption{Kinematics of the four-body $\Lambda_b\to\Lambda^\ast(\to N\bar{K})\ell^+\ell^-$ decay, where the angles are defined in the corresponding rest frames.}
  \label{fig:kinematics}
\end{figure}

In this paper, we assign the particle momenta and spin variables for the hadrons in the $\Lambda_b\to\Lambda^*(\to N\bar{K})\ell^+\ell^-$ process according to:
\begin{equation}
\begin{split}
\Lambda_b(p,s_{\Lambda_b}) & \to\Lambda^\ast(k,s_{\Lambda^\ast})\ell^{-}(q_1,s_{\ell})\nu_{\ell}(q_2,s_{\nu}),\\
\Lambda^\ast(k,s_{\Lambda^\ast}) & \to N(k_1,s_{N})\bar{K}(k_2),
\end{split}
\end{equation}
as shown in Fig.~\ref{fig:kinematics}. Here we have some relations like $q^\mu=q_1^\mu+q_2^\mu$, $k^\mu=k_1^\mu+k_2^\mu$, and $p^\mu=k^\mu+q^\mu$.

In the following, we will introduce some kinematic conventions that are useful for the calculation of the involved helicity amplitudes.

\subsection{Some conventions in $\mathbf{\Lambda_b}$ rest frame}
\label{app02.1}

In the $\Lambda_b$ rest frame, we have the four-momentum of $\Lambda_b$, $\Lambda^\ast$ and the vector boson as
\begin{equation}
\begin{split}
p^\mu&=(m_{\Lambda_b},0,0,0),\\
k^\mu&=(E_{\Lambda^\ast},0,0,|\vec{p}_{\Lambda^\ast}|),\\
q^\mu&=(q_0,0,0,-|\vec{q}|),
\end{split}
\end{equation}
where
\begin{equation}
\begin{split}
E_{\Lambda^\ast}&=\frac{m_{\Lambda_b}^2+m_{\Lambda^\ast}^2-q^2}{2m_{\Lambda_b}},~~~|\vec{p}_{\Lambda^\ast}|=|\vec{q}|=\frac{\sqrt{s_+s_-}}{2m_{\Lambda_b}},\\
q_0&=\frac{m_{\Lambda_b}^2-m_{\Lambda^\ast}^2+q^2}{2m_{\Lambda_b}},~~~s_{\pm}=(m_{\Lambda_b}\pm m_{\Lambda^\ast})^2-q^2.
\end{split}
\end{equation}
We have the following solutions for the Dirac spinor of $\Lambda_b$ for different $s_{\Lambda_b}$ as
\begin{equation}
u_{\Lambda_b}(-1/2)=\begin{pmatrix} \sqrt{2m_{\Lambda_b}}\\0\\0\\0\\ \end{pmatrix},~~
u_{\Lambda_b}(+1/2)=\begin{pmatrix} 0\\\sqrt{2m_{\Lambda_b}}\\0\\0\\ \end{pmatrix},
\end{equation}
and the solutions for the Rarita-Schwinger spinor $u_{\Lambda^\ast,\alpha}(s_{\Lambda^\ast})$ for different $s_{\Lambda^\ast}$ as~\cite{Descotes-Genon:2019dbw}
\begin{equation}
\begin{split}
u_{\Lambda^\ast}\Big{(}\!-\!\frac{3}{2}\!\Big{)}\!=&\frac{1}{2\sqrt{m_{\Lambda_b}}}\!
\begin{pmatrix}
0&0&0&0\\
0&\sqrt{s_+}&0&-\sqrt{s_-}\\
0&-i\sqrt{s_+}&0&i\sqrt{s_-}\\
0&0&0&0\\
\end{pmatrix},\\
u_{\Lambda^\ast}\Big{(}\!-\!\frac{1}{2}\!\Big{)}\!=&\frac{\sqrt{s_-s_+}}{4\!\sqrt{3}m_{\Lambda^\ast}m_{\Lambda_b}^{3/2}}\!
\begin{pmatrix}
0&2\!\sqrt{s_+}&0&-2\!\sqrt{s_-}\\
\frac{2m_{\Lambda^\ast}m_{\Lambda_b}}{\sqrt{s_-}}&0&\frac{2m_{\Lambda^\ast}m_{\Lambda_b}}{\sqrt{s_+}}&0\\
\frac{-2im_{\Lambda^\ast}m_{\Lambda_b}}{\sqrt{s_-}}&0&\frac{-2im_{\Lambda^\ast}m_{\Lambda_b}}{\sqrt{s_+}}&0\\
0&\frac{s_-+s_+}{\sqrt{s_-}}&0&-\frac{s_-+s_+}{\sqrt{s_+}}\\
\end{pmatrix},\\
u_{\Lambda^\ast}\Big{(}\!+\!\frac{1}{2}\!\Big{)}\!=&\frac{\sqrt{s_-s_+}}{4\!\sqrt{3}m_{\Lambda^\ast}m_{\Lambda_b}^{3/2}}\!
\begin{pmatrix}
2\!\sqrt{s_+}&0&2\!\sqrt{s_-}&0\\
0&\frac{-2m_{\Lambda^\ast}m_{\Lambda_b}}{\sqrt{s_-}}&0&\frac{2m_{\Lambda^\ast}m_{\Lambda_b}}{\sqrt{s_+}}\\
0&\frac{-2im_{\Lambda^\ast}m_{\Lambda_b}}{\sqrt{s_-}}&0&\frac{2im_{\Lambda^\ast}m_{\Lambda_b}}{\sqrt{s_+}}\\
\frac{s_-+s_+}{\sqrt{s_-}}&0&\frac{s_-+s_+}{\sqrt{s_+}}&0\\
\end{pmatrix},\\
u_{\Lambda^\ast}\Big{(}\!+\!\frac{3}{2}\!\Big{)}\!=&\frac{1}{2\sqrt{m_{\Lambda_b}}}\!
\begin{pmatrix}
0&0&0&0\\
-\sqrt{s_+}&0&-\sqrt{s_-}&0\\
-i\sqrt{s_+}&0&-i\sqrt{s_-}&0\\
0&0&0&0\\
\end{pmatrix},
\end{split}
\end{equation}
where the column and row notations correspond to the spinor indices $\alpha$ and the vector indices, respectively. In addition, the polarization vectors for the virtual vector boson alone on the $-z$ axis in the $\Lambda_b$ rest frame are expressed as
\begin{equation}
\begin{split}
\epsilon^\mu(t)&=\frac{1}{\sqrt{q^2}}(q_0,0,0,-|\vec{q}|),\\
\epsilon^\mu(0)&=\frac{1}{\sqrt{q^2}}(-|\vec{q}|,0,0,q_0),\\
\epsilon^\mu(\pm)&=\frac{1}{\sqrt{2}}(0,\mp1,-i,0),
\end{split}
\end{equation}
where we use $t$ and $0$ to distinguish the two $\lambda_W=0$ states ($0$ for $J=1$ and $t$ for $J=0$), and $\pm$ to represent $\lambda_W=\pm$ for $J=1$, respectively.

\subsection{Some conventions in the dilepton rest frame}
\label{app02.2}

In the dilepton rest frame, we have the four-momentum of the vector bosons and leptons as
\begin{equation}
\begin{split}
q^\mu&=(\sqrt{q^2},0,0,0),\\
q_{\ell^-}^\mu&=(E_\ell,|\vec{q}_\ell|\sin\theta_\ell,0,|\vec{q}_\ell|\cos\theta_\ell),\\
q_{\ell^+}^\mu&=(E_\ell,-|\vec{q}_\ell|\sin\theta_\ell,0,-|\vec{q}_\ell|\cos\theta_\ell),
\end{split}
\end{equation}
where $|\vec{q}_\ell|=\sqrt{q^2}\beta_\ell/2$ and $E_\ell=\sqrt{q^2}/2$. The Dirac spinors for $\ell^-$ and $\ell^+$ in Dirac representation are
\begin{equation}
\begin{split}
u_{\ell^-}(\vec{q}_{\ell},s_{\ell^-})&=\begin{pmatrix}\sqrt{E_\ell+m_{\ell}}\ \chi\left(\vec{q}_\ell,s_{\ell^-}\right) \\ 2s_{\ell^-}\sqrt{E_\ell-m_{\ell}}\ \chi\left(\vec{q}_\ell,s_{\ell^-}\right)\end{pmatrix},\\
v_{\ell^+}(-\vec{q}_{\ell},s_{\ell^+})&=\begin{pmatrix}\sqrt{E_\ell-m_{\ell}}\ \xi\left(-\vec{q}_\ell,s_{\ell^+}\right) \\ -2s_{\ell^+}\sqrt{E_\ell+m_{\ell}}\ \xi\left(-\vec{q}_\ell,s_{\ell^+}\right)\end{pmatrix},
\end{split}
\end{equation}
respectively, where
\begin{equation}
\begin{split}
&\chi(\vec{q}_\ell,\frac{1}{2})=\xi(-\vec{q}_\ell,\frac{1}{2})=\begin{pmatrix}\cos\frac{\theta_\ell}{2} \\ \sin\frac{\theta_\ell}{2}\end{pmatrix},\\
&\chi(\vec{q}_\ell,-\frac{1}{2})=-\xi(-\vec{q}_\ell,-\frac{1}{2})=\begin{pmatrix}-\sin\frac{\theta_\ell}{2} \\ \cos\frac{\theta_\ell}{2}\end{pmatrix}.
\end{split}
\end{equation}
In addition, the polarization vectors of the virtual vector boson in the dilepton rest frame are written as
\begin{equation}
\begin{split}
\bar{\epsilon}^\mu(t)&=(1,0,0,0),\\
\bar{\epsilon}^\mu(0)&=(0,0,0,1),\\
\bar{\epsilon}^\mu(\pm)&=\frac{1}{\sqrt{2}}(0,\mp1,-i,0),
\end{split}
\end{equation}
which satisfy the following orthogonality and completeness relations~\cite{Descotes-Genon:2019dbw,Yan:2019tgn,Das:2020cpv}
\begin{eqnarray}
\bar{\epsilon}^{\ast\mu}(m)\bar{\epsilon}_{\mu}(n)&=&\tilde{g}_{mn},\\
\sum_{m,n}\bar{\epsilon}^{\ast\mu}(m)\bar{\epsilon}^{\nu}(n)\tilde{g}_{mn}&=&g^{\mu\nu},
\end{eqnarray}
where $m,n\in\{t,\pm,0\}$, $\tilde{g}_{mn}=\text{diag}(+1,-1,-1,-1)$, and $g^{\mu\nu}=\text{diag}(+1,-1,-1,-1)$.

\subsection{Some conventions in $\mathbf{\Lambda^\ast}$ rest frame}
\label{app02.3}

In the $\Lambda^\ast$ rest frame, we have the following solutions for the Rarita-Schwinger spinor $u_{\Lambda^\ast,\alpha}(s_{\Lambda^\ast})$ with different $s_{\Lambda^\ast}$ as~\cite{Descotes-Genon:2019dbw,Das:2020cpv}
\begin{equation}
\begin{split}
u_{\Lambda^\ast}(-3/2)=&\sqrt{m_{\Lambda^\ast}}
\begin{pmatrix}
0&0&0&0\\
0&1&0&0\\
0&-i&0&0\\
0&0&0&0\\
\end{pmatrix},\\
u_{\Lambda^\ast}(-1/2)=&\sqrt{\frac{m_{\Lambda^\ast}}{3}}
\begin{pmatrix}
0&0&0&0\\
1&0&0&0\\
-i&0&0&0\\
0&2&0&0\\
\end{pmatrix},\\
u_{\Lambda^\ast}(+1/2)=&\sqrt{\frac{m_{\Lambda^\ast}}{3}}
\begin{pmatrix}
0&0&0&0\\
0&-1&0&0\\
0&-i&0&0\\
2&0&0&0\\
\end{pmatrix},\\
u_{\Lambda^\ast}(+3/2)=&\sqrt{m_{\Lambda^\ast}}
\begin{pmatrix}
0&0&0&0\\
-1&0&0&0\\
-i&0&0&0\\
0&0&0&0\\
\end{pmatrix},
\end{split}
\end{equation}
where the column and row notations correspond to the spinor indices $\alpha$ and the vector indices, respectively, and the solutions for the Dirac spinor of the nucleon for different $s_N$ as
\begin{equation}
\begin{split}
u_N(-1/2)=&\frac{1}{2m_{\Lambda^\ast}}
\begin{pmatrix}
-\sqrt{r_+}\sin\frac{\theta_{\Lambda^\ast}}{2}e^{-i\phi}\\
\sqrt{r_+}\cos\frac{\theta_{\Lambda^\ast}}{2}\\
\sqrt{r_-}\sin\frac{\theta_{\Lambda^\ast}}{2}e^{-i\phi}\\
-\sqrt{r_-}\cos\frac{\theta_{\Lambda^\ast}}{2}\\
\end{pmatrix},\\
u_N(+1/2)=&\frac{1}{2m_{\Lambda^\ast}}
\begin{pmatrix}
\sqrt{r_+}\cos\frac{\theta_{\Lambda^\ast}}{2}\\
\sqrt{r_+}\sin\frac{\theta_{\Lambda^\ast}}{2}e^{i\phi}\\
\sqrt{r_-}\cos\frac{\theta_{\Lambda^\ast}}{2}\\
\sqrt{r_-}\sin\frac{\theta_{\Lambda^\ast}}{2}e^{i\phi}\\
\end{pmatrix}.
\end{split}
\end{equation}

\section*{ACKNOWLEDGMENTS}

We would like to thank Prof. Yu-Ming Wang, Prof. Wei Wang, and Dr. Si-Qiang Luo for helpful discussions. This work is supported by the China National Funds for Distinguished Young Scientists under Grant No. 11825503, the National Key Research and Development Program of China under Contract No. 2020YFA0406400, the 111 Project under Grant No. B20063, the National Natural Science Foundation of China under Grant No. 12047501, the Project for top-notch innovative talents of Gansu province, and by the Fundamental Research Funds for the Central Universities under Grant No. lzujbky-2022-it17. J.G. is also supported by the National Natural Science Foundation of China under Grant No.12147118.

\end{document}